\documentclass[
reprint, 
nofootinbib,
amsmath,
amssymb,
aps]{revtex4-2}

\usepackage{graphicx}
\usepackage{dcolumn}
\usepackage{bm}
\usepackage{hyperref}
\usepackage{autobreak}
\usepackage{mathtools}
\usepackage[normalem]{ulem}
\usepackage{xcolor}
\usepackage{pifont}

\newcommand{\pmat}[1]{\begin{pmatrix}
#1
\end{pmatrix}}
\newcommand{\del}[2]{\frac{\partial #1}{\partial #2}}
\newcommand{\rd}[0]{\mathrm{d}}
\newcommand{\tr}[0]{\mathrm{Tr}}

\begin{document}

\preprint{APS/123-QED}

\title{\boldmath VISH$\nu$: a unified solution to five SM shortcomings with a protected electroweak scale}

\author{Alexei H. Sopov}
 \email{sopova@student.unimelb.edu.au}
\author{Raymond R. Volkas}%
 \email{raymondv@unimelb.edu.au}
\affiliation{
ARC Centre of Excellence for Dark Matter Particle Physics,
School of Physics, \\
The University of Melbourne,
Victoria 3010, Australia
}

\begin{abstract}
We propose a Standard Model extension, coined VISH$\nu$ (Variant-axIon Seesaw Higgs $\nu$-trino), that is an $N_{\text{DW}} = 1$ variation of its predecessor, the $\nu$DFSZ model. In accounting for the origin of neutrino masses, dark matter and the baryon asymmetry of the universe, VISH$\nu$ inherits the explanatory power of $\nu$DFSZ while, of course, resolving the strong $CP$ problem. In both models, the electroweak scale is naturally protected from a high seesaw scale that is identified with the Peccei-Quinn (PQ) spontaneous symmetry breaking scale. Through a flavour variant coupling structure, VISH$\nu$ evades a domain wall problem, extending the cosmological reach of the $\nu$DFSZ to include a viable period of inflation. The primary focus of this paper is on the inflationary dynamics of VISH$\nu$ and their naturalness in the sense of radiative stability. We find that non-minimal gravitational couplings, generically developed by the VISH$\nu$ scalar fields, naturally support a viable inflaton field which typically has both a PQ scalar and Higgs component. An axion mass window [$40\mu\text{eV}, \sim 2\text{meV}$] accessible to forthcoming searches, results for the case where PQ symmetry is restored during the (p)reheating phase.
\end{abstract}

\maketitle

\section{\label{sec:sec1}Introduction}

The Standard Model (SM) features both remarkable experimental agreement and considerable phenomenological deficit, predicting neither neutrino masses nor a viable dark matter candidate. Given that sub-Planckian energies were explored in the early universe, cosmological phenomena also lie within its estimation, and while the SM predicts an inflaton field if one admits non-minimal coupling of the Higgs doublet to gravity~\cite{HiggsInflation1}, it cannot generate an acceptable baryon asymmetry of the universe (BAU). 

It may be that these shortcomings point to additional physics that does not exceed the electroweak scale. In the $\nu$MSM~\cite{Asaka:2005an,Asaka:2005pn}, three additional right-handed neutrino singlets with sub-electroweak masses can already support type-I seesaw~\cite{Minkowski:1977sc,Yanagida:1979as,GellMann:1980vs,Mohapatra:1979ia} masses for SM neutrinos, a keV-range dark matter candidate (the lightest neutrino singlet) and a version of ARS leptogenesis~\cite{ARS}, without spoiling Higgs-driven inflation. However, the SM \emph{also} has a strong $CP$ problem (why is $\Bar{\theta} \ll 1$?), for which viable axion~\cite{Peccei:1977hh} solutions require the presence of a new fundamental scale that exceeds the electroweak scale by several orders of magnitude.

Whenever there is such a measurable hierarchy of scales, there is a naturalness issue concerning the stability of the lower mass-scale under large and physically meaningful radiative corrections induced by the higher mass-scale. While its origin, a ``hierarchy problem'', is perhaps answerable only in the far UV, we can avoid introducing \textit{new} naturalness problems through cautious model-building, provided the larger scale is sub-Planckian. Hidden-sector extensions, discussed in Ref.~\cite{Foot:2013hna}, are free of naturalness problems arising from any new non-gravitational interactions. This is because the high-scale physics is assumed to interact with the low-energy SM sector only through a weakly-coupled interaction Lagrangian whose small couplings ($\lambda_{\text{int},i}$) are protected by an enhanced Poincar\'e symmetry manifest in the decoupling limit ($\lambda_{\text{int},i}\rightarrow0$). The electroweak scale is therefore stabilised under the induced corrections by parameters which remain appropriately small under running.\footnote{\label{foonote1}
Of course, to completely address this electroweak naturalness problem, one should also understand the effect of gravitational corrections arising from a Planckian scale. (This is the case for the SM.) Our focus, however, is ensuring that any new high-scale physics we introduce \textit{beyond} the SM does not create naturalness problems of its own: a necessary (but not sufficient) condition for solving the electroweak naturalness problem posed in a given SM extension. This is exactly the restricted sense in which we refer to a ``protected'' electroweak scale.}

An exemplar, the $\nu$DFSZ model~\cite{Volkas1988,nuDFSZ}, is a minimal hidden sector extension with a radiatively stable hierarchy of scales built around the DFSZ axion~\cite{DFSZ,Zhitnitsky:1980tq} and three right-handed neutrinos. The new fundamental scale, $f_{PQ} \sim 10^{10}$\text{--}$10^{11}\, \text{GeV}$, is associated with a spontaneously broken Peccei-Quinn symmetry $U(1)_{\text{PQ}}$ to realise an invisible axion, and also serves as the seesaw neutrino mass scale~\cite{Langacker1986,Shin1987}. In Ref.~\cite{nuDFSZ} a critical density of axionic dark matter and the BAU were shown also to be viable outcomes without requiring unnatural inter-sector couplings. The BAU is generated by the $CP$-violating decays of the lightest right-handed neutrino out-of-equilibrium~\cite{nu2HDM}, as in standard thermal leptogenesis~\cite{Fukugita:1986hr}. Subsequently, the SMASH model~\cite{Salvio:2015cja,Ballesteros:2016euj,Ballesteros:2016xej,Ballesteros:2019tvf,Salvio:2018rv} was proposed, built around the KSVZ axion~\cite{Kim:1979if,Shifman:1979if}, with similar explanatory power to the $\nu$DFSZ (neutrino masses, axion dark matter, BAU, $\Bar{\theta} \rightarrow 0$) but supplemented by an analysis of an early cosmological history (inflation followed by a period of reheating~\cite{Ballesteros:2016euj,Ballesteros:2016xej,Ballesteros:2019tvf,Salvio:2018rv}). However, unlike the SMASH model~\cite{Ballesteros:2016xej}, the $\nu$DFSZ model can preserve the high-scale validity of its Higgs sector, without introducing a destabilising contribution to the Higgs mass parameters~\cite{Oda:2019njo}, and does not require an exotic quark. 

A primary objective of this paper is to extend the original analysis of the $\nu$DFSZ model to also include an early period of inflation~\cite{STAROBINSKY198099,Guth1981}, without reintroducing the naturalness concerns it serves to resolve. This is accomplished by non-minimal couplings to gravity of the scalar sector which, for a general range of initial conditions~\cite{Kaiser:2013sna,DeCross:2015uza}, gives rise to effectively single-field inflation along valleys of the conformally stretched large-field potential in the Einstein frame, producing an excellent fit to CMB data~\cite{Planck:2018jri}. We identify the possible inflaton configurations for a general non-minimally coupled three-scalar model, before specialising to the hidden-sector parameter space where the inflaton field generally has both a singlet scalar and Higgs component. When non-minimal couplings are much larger than unity, inflationary predictions replicate Higgs inflation, while smaller couplings, unscathed by unitarity concerns, are also viable and compatible with both naturalness and stability requirements.

We also remove a well-known cosmological axion domain wall problem~\cite{Sikivie:1982qv}, which afflicts the $\nu$DFSZ model if PQ symmetry is restored after the inflation, using a variant axion. Drawing inspiration from original proposals in Refs.~\cite{Peccei:1986pn,KRAUSS1986189}, we recast the model with a flavour-dependent $U(1)_{PQ}$ charge assignment that realises a unit domain wall number which is cosmologically benign~\cite{Davidson:1983tp,Davidson:1984ik,Geng1989,Geng:1990dv} and leads to generation non-universal quark Yukawa couplings. For the sake of definiteness, we focus on a top-specific Yukawa structure, in which one of two Higgs doublets is responsible for giving a heavier mass to the top quark and no other fermion, gaining a natural explanation for the top mass as a remarkable bonus. This structure leads to interesting phenomenology~\cite{Chiang:2015cba, Chiang:2017fjr} which we briefly review alongside its discovery prospects.

We outline our $\nu$DFSZ-variant, top-specific ``VISH$\nu$'' model\footnote{Though in this paper we single out the top-specific model for definiteness and phenomenological simplicity, we have in mind a family of models, dubbed ``avatars'' of VISH$\nu$, encompassing all unit axion domain wall number $\nu$DFSZ-like models with viable flavour-variant structures.} in Sec. \ref{sec:sec2}. We then arrive at inflationary predictions in Sec. \ref{sec:sec3}, discuss implications for dark matter predictions in Sec. \ref{sec:sec4} and establish a parameter space which is consistent with reviewed naturalness desiderata, as well as high-scale validity and possible unitarity concerns, in Sec. \ref{sec:sec5}. We conclude in Sec. \ref{sec:con} with a summary of our findings. 

\section{\label{sec:sec2}The model and its collider phenomenology}

\subsection{The model}

Preserving the SM gauge group, our model extends the SM fermion content by three right-handed neutrinos singlets $\nu_R \sim (1,1,0)$ and upgrades its scalars to two Higgs electroweak doublets $\Phi_{1,2} \sim (1,2,1)$ with an additional gauge singlet $S \sim (1,1,0)$. To realise an invisible axion solution to the strong $CP$ problem, an anomalous chiral global $U(1)_{\text{PQ}}$ symmetry constrains possible renormalisable terms in the potential and Yukawa sectors. 

Gauge invariance and renormalisability forces two independent choices for non-trivially $U(1)_{\text{PQ}}$ invariant terms in the potential. So that, if we denote as $X_1, X_2$ the PQ charges of $\Phi_1, \Phi_2$, and fix the charge of $S$ to unity:
\begin{equation}\label{potential}
    \begin{split}
        V &= M_{11}^2\Phi^\dagger_1 \Phi_1 + M_{22}^2\Phi^\dagger_2 \Phi_2 + M_{SS}^2 S^* S  
        \\ &\quad + \frac{\lambda_1}{2}(\Phi^\dagger_1 \Phi_1)^2 + \frac{\lambda_2}{2}(\Phi^\dagger_2 \Phi_2)^2 + \frac{\lambda_S}{2}(S^*S)^2 \\ 
        &\quad + \lambda_3 (\Phi^\dagger_1 \Phi_1)(\Phi^\dagger_2 \Phi_2) + \lambda_4 (\Phi^\dagger_1 \Phi_2)(\Phi^\dagger_2 \Phi_1) \\
        &\quad + \lambda_{1S}(\Phi^\dagger_1 \Phi_1)(S^*S) + \lambda_{2S} (\Phi^\dagger_2 \Phi_2)(S^*S) \\
        & + \begin{cases} \kappa \Phi_1^\dagger\Phi_2 S\ + \text{h.c.}\ \ \text{if}\ X_2 - X_1 = 1\ [\text{VISH}\nu]\\
        \epsilon \Phi_1^\dagger\Phi_2 S^2 + \text{h.c.}\ \ \text{if}\ X_2 - X_1 = 2 \ [\nu\text{DFSZ}].
        \end{cases}
    \end{split}
\end{equation}
While the second choice of final term ($X_2 - X_1 = 2$) was considered in both the original DFSZ model and the $\nu$DFSZ papers~\cite{Volkas1988,nuDFSZ}, we adopt the \emph{first} throughout. Our Yukawa sector is determined by the $U(1)_{\text{PQ}}$ charge assignments given in Table \ref{tab:table1}. Implying summation over the generation indices $a = 1,2$ and $j,k = 1,2,3$: 
\begin{equation}
\begin{split}
    - \mathcal{L}_Y &= \overline{q_L}^{j} y_{u1}^{j3}  \widetilde{\Phi}_1 u_R^3 + \overline{q_L}^j y_{u2}^{ja}  \widetilde{\Phi}_2 u_R^a   + \overline{q_L}^{j} y_{d}^{jk} \Phi_2 d_R^k \\
    &\quad + \overline{l_L}^j y_{e}^{jk}  \Phi_2 e_R^k  + \overline{l_L}^{j} y_{\nu}^{jk}  
    \widetilde{\Phi}_2 \nu_R^k \\
    &\quad + \frac{1}{2} \overline{(\nu_R)^c}^j y_{N}^{jk} S \nu_R^k   +
    \text{h.c.},
\end{split}
\end{equation}
where, of the right-handed fermions, $\Phi_1$ Yukawa-couples exclusively to $u^3_R$ and we work hereafter in a basis where $y_N$ is real and diagonal. The potential (\ref{potential}) is minimised when the scalar fields develop the real-valued vacuum expectation values (vevs), for $i=1,2$:
\begin{equation}\label{vevs}
\langle S \rangle = \frac{v_S}{\sqrt{2}},\quad \langle \Phi_i \rangle = \frac{1}{\sqrt{2}} \pmat{0 \\ v_i},
\end{equation}
where $v = \sqrt{v_1^2 + v_2^2} \simeq 246$ GeV and, since we require $v_S \gg v_1, v_2$ for an ``invisible'' axion, $v_S \simeq \sqrt{-2 M_{SS}^2/\lambda_S} \sim 10^{10}$\text{--}$10^{11}\ \text{GeV}$. Integrating out $S$ leads to the following quadratic terms in the Higgs potential:
\begin{equation}\label{higgspotential}
    V \supset m_{11}^2\Phi^\dagger_1 \Phi_1 + m_{22}^2\Phi^\dagger_2 \Phi_2 
    - m^2_{12}(\Phi^\dagger_1\Phi_2+ \Phi_1\Phi_2^\dagger), 
\end{equation}
where $m^2_{ii} = M^2_{ii} + \lambda_{iS}\langle S \rangle^2$ and $m^2_{12} = \kappa \langle S \rangle$. Together with a weakly-coupled axion ($f_A \sim v_{S}$), the low energy model is the $\nu$2HDM~\cite{nu2HDM} up to $O(v/v_S)$ corrections and top-specific Yukawa couplings.

\begin{table*}
\caption{\label{tab:table1} $U(1)_{\text{PQ}}$ charge assignments for the VISH$\nu$ model. In order to decouple the $U(1)_{\text{PQ}}$ current from the Goldstone boson eaten by the $Z$ boson, it is sufficient to impose $X_1 = (-\cot^2 \beta) X_2$~\cite{DFSZ}, where $X_1, X_2$ denote the PQ charges of $\Phi_1, \Phi_2$.}
\begin{ruledtabular}
\begin{tabular}{ccccccccccc}
 Field: & $q_L$ & $u^{a}_R$ & $u^3_R$ & $d_R$ & $l_L$ & $e_R$ & $\nu_R$ & $\Phi_1$ & $\Phi_2$ & $S$ \\ 
 Charge: & 0 & $-\sin^2 \beta$ & $\cos^2 \beta$ & $\sin^2 \beta$ & $\sin^2 \beta - \frac{1}{2}$ & $2\sin^2 \beta - \frac{1}{2}$ & $- \frac{1}{2}$ & $\cos^2 \beta$ & $-\sin^2 \beta $ & 1
\end{tabular}
\end{ruledtabular}
\end{table*}

Values for the mass parameters, $m_{ij}$, impact two important predictions of the model. These are: the viability of thermal hierarchical leptogenesis, and the protection of the electroweak scale from physically meaningful corrections induced by the PQ scale. For instance, we see that \emph{tree-level} naturalness already requires that $\lambda_{iS} \langle S \rangle^2$ to not exceed $M_{ii}^2$. In Ref.~\cite{nu2HDM} it was shown that both predictions can be additionally met if one of the Higgs vevs, namely $v_2$, is slightly smaller than the other ($\sim$ few~GeV). This is due to the convergence of a Vissani-like naturalness bound, ensuring $\nu_R$-induced corrections to the electroweak Higgs masses are controlled, and a Davidson-Ibarra bound, ensuring sufficient $CP$ asymmetry 
during leptogenesis, under the assumption of perturbative validity.\footnote{We direct the reader to Fig. 5 of Ref.~\cite{nu2HDM} for a more detailed representation of the bounds on $v_2$, and their dependence on the lightest neutrino mass.} To realise the observed value $v \simeq 246$ GeV, this forces $v_1$ to a comparatively larger value ($\sim 10^2$ GeV) than $v_2$. As $\Phi_1$ is responsible only for giving mass to the top quark, an explanation for its large mass in comparison to other SM fermions is an immediate consequence.\footnote{This remarkable bonus hints suggestively at an inroad to the flavour puzzle, since richer VISH$\nu$ flavour-structures are also possible.} We require that $|m^2_{12}|/m^2_{22} \ll 1$ to protect the resulting small hierarchy in Higgs vevs from radiative corrections. As discussed in Ref.~\cite{nuDFSZ}, the ensuing values of the vevs, $v_i$, mean that it is typical for the mass parameters to take the values $m^2_{11} \simeq - (88\ \text{GeV})^2$ and $m_{22} \sim 10^3$ GeV together with $m^2_{22}/\tan^2 \beta \ll |m^2_{11}|$, where $\tan \beta \equiv v_2/v_1$. Taken together, we find that the consistency of VISH$\nu$ predictions with our naturalness criteria require that $|\lambda_{1S}| \lesssim 10^{-18}$, $|\lambda_{2S}| \lesssim 10^{-16}$ and $|\kappa| \ll O(0.1)$ keV, where the final condition replaces its analogue $|\epsilon| \ll 10^{-18}$ in the $\nu$DFSZ model~\cite{nuDFSZ}.

Observe that the model manifests two decoupling limits which render weak inter-sector interactions technically natural in the Poincar\'{e}-protection sense, as will be fully discussed in Sec. \ref{sec:sec5}. The heavy sector $\{ S, \nu_R \}$ is decoupled from the Higgs (and SM) sector in the limit $\lambda_{1S}, \lambda_{2S}, \kappa, y_\nu \rightarrow 0$, while $\{ S\}$ can itself be decoupled by $\lambda_{1S}, \lambda_{2S}, \kappa, y_N \rightarrow 0$. Additionally, there is an enhanced $U(1)$ symmetry in the limits $y_N, y_\nu, \kappa \rightarrow 0$ which individually protect small values of $y_N, y_\nu$ and $\kappa$. Necessary and sufficient bounded-from-below conditions for the potential (\ref{potential}) were derived in Ref.~\cite{Klimenko:1984qx}. For $\widetilde{\lambda} = \max \{\lambda_3, \lambda_{3} + \lambda_{4}\}$, they describe the parameter space: $\{\lambda_S, \lambda_i > 0,\ \lambda_{iS} > -\sqrt{\lambda_{i}\lambda_S},\ \widetilde{\lambda}> - \sqrt{\lambda_{1}\lambda_{2}},\ \lambda_{1S} \geq - \lambda_{2S}\sqrt{\lambda_1/\lambda_2} \} \cup \{\lambda_S, \lambda_i > 0,\ \sqrt{\lambda_2\lambda_S} \geq \lambda_{2S} > -\sqrt{\lambda_{2}\lambda_{S}},\ -\lambda_{2S}\sqrt{\lambda_1/\lambda_2} \geq \lambda_{1S} > - \sqrt{\lambda_1\lambda_S},\ \lambda_S \widetilde{\lambda} > \lambda_{1S}\lambda_{2S} - \sqrt{(\lambda^2_{1S}-\lambda_1\lambda_S)(\lambda^2_{2S}-\lambda_2\lambda_S)}\}$. For the very weak interactions coupling the Higgs and $S$ sectors, the conditions diminish to those of a general two Higgs doublet model~\cite{Deshpande1978,Klimenko:1984qx,Maniatis:2006fs,nu2HDM} with couplings limited by $U(1)_{\text{PQ}}$:
\begin{equation}\label{AppxBoundedFromBelow}
    \lambda_1 > 0,\ \lambda_2 > 0,\ \widetilde{\lambda}  \gtrsim - \sqrt{\lambda_1\lambda_2},
\end{equation}
together with the requirement that $\lambda_S > 0$, up to small corrections. To ensure that (\ref{vevs}) is the global minimum of the potential, we also require that $\lambda_1\lambda_2-(\lambda_{3}+\lambda_4)^2>0$, which strengthens the final condition of (\ref{AppxBoundedFromBelow}).

\subsection{A brief summary of collider phenomenology}

We now mention some interesting phenomenological differences between VISH$\nu$ and the $\nu$DFSZ model born from the top-specific flavour structure, referring the reader to Refs.~\cite{nuDFSZ,nu2HDM} for the natural explanations of neutrino masses and the BAU, which remain the same, and Sec. \ref{sec:sec4} for an updated discussion of axionic dark matter production. 

Unlike the standard DFSZ model, flavour-variant models generically feature neutral Higgs-induced flavour-changing processes at tree level. We now explain how VISH$\nu$ produces a SM-like mass eigenstate $h$ in a natural alignment limit, and how a top-specific structure abides by current experimental constraints. Since $\Phi_1$ couples only to the right-handed top-quark while $\Phi_2$ couples to right-handed up- and charm-quarks, there are tree-level FCNC decays of the top quark via mixing of the $CP$-even neutral Higgs states. Hence, the collider signature of our flavour structure is the decay $t\rightarrow ch$ (or $t \rightarrow uh$), where $h$ is the observed Higgs~\cite{Chiang:2015cba,Chiang:2017fjr,Hou:2020chc}.

Upper limits on the observed (expected) branching ratios of each decay have recently been made more stringent by the CMS collaboration~\cite{CMS:2021cqc}:
\begin{equation}\label{FCNCConstraints}
\begin{split}
    &\mathcal{B}(t \rightarrow hc) < 0.073\ (0.051)\ \%,\\
    &\mathcal{B}(t \rightarrow hu) < 0.019\ (0.031)\ \%,
\end{split}
\end{equation}
with $\sqrt{s} = $ 13 TeV and 137 fb$^{-1}$ data, representing no significant excess above the predicted background. For the case where $t \rightarrow hc$ is dominant (which we expect anyway from mass-mixing~\cite{Hou:2020chc}), this corresponds to the constraint on model parameters~\cite{Chiang:2017fjr}:
\begin{equation}\label{ModelConstraintFromFCNC}
    a^2 \sin^2 \rho < 0.023,
\end{equation}
where $a = (\tan \beta + \cot \beta)\cos(\beta-\alpha)$, $\alpha$ is the rotation angle which diagonalises the neutral Higgs scalars and $\rho$ is a $t-c$ mixing angle defined in Ref.~\cite{Chiang:2017fjr}. 

In addition to the usual decoupling suppression ($v_1/m^2_{22} \ll 1$)~\cite{nu2HDM}, the quantity $\cos (\beta - \alpha)$ is also suppressed\footnote{See, for example, Eq.~15 of Ref.~\cite{nu2HDM}.} in VISH$\nu$ by the approximate $U(1)$ symmetry realised for $m^2_{12}/m^2_{22} \ll 1$. Note that, in the full theory, this corresponds to the natural limit $\kappa \rightarrow 0$ discussed above. As a consequence, the light $CP$-even neutral scalar \emph{naturally} resembles the SM Higgs. Since the product (\ref{ModelConstraintFromFCNC}) is suppressed in the alignment limit, $\cos (\beta - \alpha) \rightarrow 0$, $t-c$ mixing does not have to be unnaturally small to remain compatible with exclusion limits. In fact, while permitting in-principle tree-level FCNC decays associated to the \emph{SM-like} Higgs, it follows that flavour-variants of the $\nu$DFSZ may automatically arrange their suppression. 

An additional way to access the FCNC top-netural-Higgs interaction, without alignment suppression, is in the top-associated production of the heavy scalar bosons $H, A$ and $H^+$~\cite{Hou:2020chc} via the channels: $cg \rightarrow tH$ or $tA$~\cite{KOHDA2018379} as well as $cg \rightarrow bH^+$~\cite{Ghosh2020}, which can be searched for at the LHC and its high-luminosity upgrade~\cite{Hou:2020chc}.

\section{Inflation}
\label{sec:sec3}

A scalar inflaton field which is non-minimally coupled to gravity remains among the best fits to CMB data~\cite{Planck:2018jri}, and may be identified with the SM Higgs field~\cite{HiggsInflation1}. However, there also exist realistic models, such as invisible axion or multi-Higgs doublet models, which employ an extended scalar sector. Compellingly, non-minimal gravitational couplings of a multi-scalar sector generically give rise to one or more attractors for inflation trajectories which replicate the successful inflationary predictions of the single-field model~\cite{Kaiser:2013sna}, with some exceptional cases~\cite{Schutz:2013fua}. This behaviour has, for example, motivated analyses of an inflation phase in non-minimally coupled versions of the two Higgs doublet~\cite{Gong:2012ri,Modak:2020fij,Nakayama2015} and KSVZ models~\cite{Ballesteros:2016euj,Ballesteros:2016xej,Ballesteros:2019tvf}.

In this section, we extend each of these analyses to include a third scalar non-minimally coupled to gravity, with couplings limited by our $U(1)_{\text{PQ}}$ symmetry. Since this simply restricts to a $Z_2$ symmetry on the quartic potential relevant in the inflation regime, our analysis is quite general.\footnote{In particular, Table \ref{tab:table2} may be useful for inflation model-builders.} In Sec. \ref{VISHnuInflation}, we discuss implications of weak inter-sector couplings and present fits to CMB observations in Sec. \ref{largexifit} and Sec. \ref{smallxifit}. Of course, these findings are independent of our flavour structure, and so constitute a general analysis of both non-minimally coupled VISH$\nu$ and cubic DFSZ models for inflation, the results of which are summarised in Figure \ref{Figure1}, Figure \ref{Figure2} and Table \ref{tab:table3}.

\subsection{VISH$\nu$ in the Einstein frame}

If they respect the $U(1)_{\text{PQ}}$ symmetry, mass-dimension four non-minimal couplings of the VISH$\nu$ scalar fields to gravity will generically arise in curved spacetime, since they are generated radiatively, and are required for renormalisation~\cite{Chernikov:1968zm,CALLAN197042,Birrell1980,Bunch_1980,Bunch_1980_2,Birrell:1982ix,Nelson1982,Ford1982,Parker1984,Parker1985,Buchbinder2017,faraoni2004cosmology}. These modify the Einstein-Hilbert action as follows:
\begin{equation}\label{InflationAction}
    \frac{\mathcal{L}^{\mathcal{J}}}{\sqrt{-g^{\mathcal{J}}}} \supset
      \left(\frac{M_P^2}{2} + \xi_1 \Phi_1^\dagger \Phi_1 + \xi_2 \Phi_2^\dagger \Phi_2 + \xi_S S^\dagger S\right) R^\mathcal{J},
\end{equation}
where $\xi_1, \xi_2, \xi_S$ are real dimensionless couplings which we will assume to be non-negative, $M_P$ is the reduced Planck mass and $R^\mathcal{J}$ is the Ricci scalar in the Jordan frame. Since the non-minimal couplings are diagonal, we parameterise the neutral scalars as a modulus which, at large values, plays a role in driving inflation, and a phase:
\begin{equation}\label{fieldparam}
    \Phi^0_1 = \frac{\rho_1}{\sqrt{2}} e^{i \vartheta_1 / v_1},\ \Phi^0_2 = \frac{\rho_2}{\sqrt{2}} e^{i \vartheta_2 / v_2},\ S = \frac{\sigma}{\sqrt{2}} e^{i \vartheta_S / v_S}
\end{equation}
(we set the charged Higgs components to zero for the inflation analysis~\cite{Gong:2012ri,Modak:2020fij,Lee:2021rzy}). We will assume the inflation phase begins with a displacement of one or more of the modulus fields from their minimum which satisfies a \emph{large-field} regime ($\xi_1 \rho_1^2 + \xi_2 \rho_2^2 + \xi_S \sigma^2 \gg M_P^2$).

To analyse the predicted inflation, it is standard to work in the Einstein frame ($\mathcal{E}$), which is related to Jordan frame ($\mathcal{J}$) of (\ref{InflationAction}) by a local rescaling of the spacetime metric:
\begin{equation}\label{WeylXfm}
\begin{split}
    &g^\mathcal{J}_{\mu\nu} \rightarrow g^\mathcal{E}_{\mu\nu} = \Omega^2 (\rho_1, \rho_2, \sigma) g^\mathcal{J}_{\mu\nu}, \\
    &\text{where}\quad \Omega^2(\rho_1, \rho_2, \sigma)  \equiv 1 + \frac{\xi_1 \rho_1^2 + \xi_2 \rho_2^2 + \xi_S \sigma^2}{M_P^2}.
\end{split}
\end{equation}
The effect of the Weyl transformation is both to restore a minimal coupling to gravity and flatten out the potential for large values of the modulus fields, but this necessarily comes at the expense of non-canonical kinetic terms~\cite{Kaiser2010}, so that the gravi-scalar action becomes:
\begin{equation}\label{EinsteinframeExpression}
    \frac{\mathcal{L}^{\mathcal{E}}}{\sqrt{-g^{\mathcal{E}}}} \supset \frac{M_P^2}{2} R^\mathcal{E} - \frac{1}{2} \sum_{I,J} \mathcal{G}^\mathcal{E}_{IJ} \partial_{\mu}  \varphi^I \partial^\mu \varphi^J - V^\mathcal{E}(\varphi^I),
\end{equation}
where $\varphi^I = (\rho_ 1, \rho_2, \sigma)$ and $\mathcal{G}^\mathcal{E}_{IJ}$ is the non-trivial metric induced by (\ref{WeylXfm}) on the scalar field space.\footnote{We discuss implications of terms with non-renormalisable mass dimension in Sec. \ref{sec:sec5}.} It is:
\begin{equation}\label{inducedfieldmetric}
    \mathcal{G}^{\mathcal{E}}_{IJ} = \frac{\delta_{IJ}}{\Omega^2}  + \frac{3M_P^2}{2} \del{\log \Omega^2}{\varphi^I} \del{\log \Omega^2}{\varphi^J} .
\end{equation}
In the large-field regime ($\Omega^2 \gg 1$) the Einstein frame scalar potential is:
\begin{equation}\label{EinsteinFramePot1}
\begin{split}
    V^\mathcal{E}(\varphi^I) &= \Omega^{-4}(\varphi^I) V^\mathcal{J} (\varphi^I) \\ 
    &= \frac{M_P^{4}}{8} \frac{\lambda_i \rho_i^4  + 2\lambda_{34} \rho_1^2 \rho_2^2 + 2\lambda_{iS} \rho_i^2 \sigma^2 + \lambda_S  \sigma^4}{(\xi_i \rho_i^2 + \xi_S \sigma^2)^2}\\
    &\quad \times \left[1 - \mathcal{O}\left(\frac{M_P^2}{\xi_i \rho_i^2 + \xi_S \sigma^2}\right) \right]^2,
    \end{split}
\end{equation}
where $\lambda_{34} = \lambda_3 + \lambda_4$ and we will compress sums over $i = 1,2$ when what we mean is obvious.

\subsection{Inflation model with three non-minimally coupled scalars}
\label{valleyorientations}

We will now outline the inflation scenarios for a model of three real scalar fields, $\chi^I = (\chi_1, \chi_2, \chi_3)$, with non-minimal gravitational couplings, $\xi_k = (\xi_1,\xi_2,\xi_3)$, and which respects independent $\mathbb{Z}_2$ transformations of the components of $\chi^I$. In other words, we will analyse the scalar potential which, in the large-field regime, is:
\begin{equation}\label{threescalarinflationpotential}
    V(\chi^I) \simeq \frac{M^4_P}{8}\frac{\sum_{ij} \lambda_{ij}(\chi_i\chi_j)^2}{\left[\sum_k \xi_k(\chi_k)^2\right]^2} \left[1 - \mathcal{O}\left(\frac{M_P^2}{\sum_k \xi_k(\chi_k)^2}\right) \right]^2,
\end{equation}
and which is obtained in the Einstein frame by a Weyl transformation that induces a field-space metric in direct analogy to (\ref{WeylXfm}) and (\ref{inducedfieldmetric}).\footnote{Of course, (\ref{threescalarinflationpotential}) is merely a generalisation of (\ref{EinsteinFramePot1}). We do this so that the calculations of this subsection apply now to an arbitrary parameter space, $\{\xi_I, \lambda_{IJ}\}$, with the minimal requirements: $\xi_I \geq 0$, and $\sum_{I,J} \lambda_{IJ} (\chi^I\chi^J)^2 \geq 0$. We reserve Sec. \ref{VISHnuInflation} for the hidden-sector parameter space of VISH$\nu$.}

Unless there is an internal $O(3)$ symmetry of the field space, which forces both the equality of all quartic couplings $\lambda_{ij}$ and all non-minimal couplings $\xi_k$, (\ref{threescalarinflationpotential}) describes a manifold with higher-dimensional analogues of ridges and valleys~\cite{Kaiser:2012ak}. At large field-values, and with respect to a radial cross-section, we will denote as ``hyper-valleys'' features which are local minima, reserving ``ridge'' for features which either resemble a local maximum or are saddle-shaped. For general large-field initial displacements $\chi_0^I$, and a broad range of initial velocities $\Dot{\chi}_0^I$, the hyper-valleys correspond to effectively single-field attractors for inflaton trajectories due to Hubble friction~\cite{Kaiser:2013sna}. After settling into a hyper-valley, the field trajectory: satisfies the single field slow-roll approximation, exhibits a negligible turn rate during the inflation, and orthogonal directions can acquire an effective mass which generally exceeds the inflationary Hubble scale, so that both non-Gaussianities and isocurvature pertubations are negligible~\cite{Kaiser:2013sna,Kaiser:2012ak,Schutz:2013fua,DeCross:2015uza}.\footnote{As explained in Ref.~\cite{Schutz:2013fua}, there are exceptions to this general situation which can arise when the curvature near a ridge is sufficiently weak. When this is the case, severely finely-tuned initial field trajectories can either drive a period of inflation along a ridge which is incompatible with observations, or persist there for a time, before falling into a valley at a later stage. This leads, for example, to the generation of sizeable isocurvature modes~\cite{Schutz:2013fua}, which are generally suppressed for the effectively single-field trajectories and initial conditions we are interested in.}

To determine the possible inflation scenarios, we thus identify the hyper-valleys of the manifold (\ref{threescalarinflationpotential}) compatible with a given parameter space. We do this by characterising the curvature near the extrema of the surface at large field displacement (\ref{largesurfacesphere}). While the existence of ridges and hyper-valleys will be independent of the parametrisation we use for the underlying field space, we can simplify the problem using spherical coordinates:
\begin{equation}\label{sphericalxfm}
    \chi_1 = r \cos\phi \sin \theta\quad \chi_2 = r \sin \phi \sin \theta\quad \chi_3 = r\cos\theta
\end{equation}
where $r^2 = \sum_k \chi_k^2$. The azimuthal and polar angles have the definitions when $\chi_1 \neq 0$:
\begin{equation}
    \tan \phi = \frac{\chi_2}{\chi_1},\quad  \cos \theta = \frac{\chi_3}{r}, 
\end{equation}
and we focus on the first octant without loss of generality: $\phi \in [0,\pi/2]$, $\theta \in [0,\pi/2]$ and $r > 0$. In the large-field regime, the leading order part of the potential (\ref{threescalarinflationpotential}) is therefore independent of the radial variable (with $c_{\psi}, s_{\psi}  \equiv \cos \psi, \sin \psi$ and $\lambda_{ij} = \lambda_{ji}$):
\begin{equation}\label{largesurfacesphere}
\begin{split}
    V \simeq \Lambda[\phi, \theta] =\ &\frac{M_P^4}{8} [s_\theta^2 (\xi_1 c_\phi^2 + \xi_2 s_\phi^2) + \xi_3 c^2_\theta]^{-2} \\
    &\times \left\{(\lambda_1 c_\phi^4 + 2\lambda_{12}c_\phi^2 s_\phi^2 + \lambda_2 s_\phi^4) s^4_\theta \right.\\
    &\quad \quad \left.
    + 2[(\lambda_{23} - \lambda_{13})c^2_{\phi} + \lambda_{13}]c^2_\theta s^2_\theta \right.\\
    &\quad \quad \left. + \lambda_3 c_\theta^4 \right\}
\end{split}
\end{equation}
and, when extremised, represents a false vacuum energy density ($\Lambda_{\text{inf}}$) responsible for a de Sitter expansion, to leading order. 

\begin{table*}
\caption{\label{tab:table2} We supply the locations $(\overline{\phi}, \overline{\theta})$ of the critical points of the false vacuum energy density which is the leading order contribution of the inflation potential (\ref{largesurfacesphere}). The parameters $\kappa_{ij}$ and $A_{ij}$ are defined in (\ref{paramstocompress}), and we have indicated which components in the $\chi_k$ basis take non-trivial values at each location. For the existence of each critical point, we use ``$\checkmark$'' for when this is guaranteed in general and derive the non-trivial conditions using restrictions on the codomain of the cosine function, namely that it is real-valued and does not exceed unity. For the conditions which must be satisfied to ensure a hyper-valley (a valley of the 3-manifold) at the point $(\overline{\phi}, \overline{\theta})$, we require the positivity of both principal curvatures which, for a critical point, constitute the eigenvalues of the Hessian matrix of double derivatives. For brevity, we omit the existence conditions for the case where a principal curvature vanishes. The $\overline{\phi}$ value for the critical point along the $\chi_3$ axis is degenerate.}
\begin{ruledtabular}
%\centering
\def\arraystretch{1.8}
\begin{tabular}{ccccc}
    $\cos^2 \overline{\phi}$ & $\cos^2 \overline{\theta}$ & Non-zero components & Existence of critical point & Hyper-valley conditions \\
    \hline 
    $1$ & $0$ & $\chi_1$ & $\checkmark$ & $\kappa_{12},\kappa_{13}>0$ \\
    0 & 0 & $\chi_2$ & $\checkmark$ & $\kappa_{21},\kappa_{23}>0$  \\
    \text{deg.} & 1 & $\chi_3$ & $\checkmark$ & $\kappa_{31},\kappa_{32}>0$ \\
    $\dfrac{\kappa_{21}}{\kappa_{12} + \kappa_{21}}$ & $0$ & $\chi_1,\chi_2$ & $\kappa_{12}\kappa_{21} > 0$ & $\kappa_{12}, \kappa_{21}, A_{12} < 0$ \\[1.2ex]
    $1$ &  $\dfrac{\kappa_{13}}{\kappa_{13} + \kappa_{31}}$  & $\chi_1,\chi_3$ & $\kappa_{13}\kappa_{31} > 0$ & $\kappa_{13}, \kappa_{31}, A_{13} < 0$ \\[1.2ex]
    $0$ &  $\dfrac{\kappa_{23}}{\kappa_{23} + \kappa_{32}}$  & $\chi_2,\chi_3$ & $\kappa_{23}\kappa_{32} > 0$ & $\kappa_{23}, \kappa_{32}, A_{23} < 0$ \\[1.5ex]
    $\dfrac{A_{23}}{A_{23}+A_{13}}$ & $\dfrac{A_{12}}{A_{12}+ A_{23}+ A_{13}}$ &  $\chi_1,\chi_2,\chi_3$  &  $A_{13}A_{23},\ A_{12}(A_{23}+A_{13}) > 0$  & $A_{12}, A_{23}, A_{13}> 0$  \\[1.2ex]
\end{tabular}
\end{ruledtabular}
\end{table*}

In a general parameter space, we identify seven possible critical points of the surface $(\phi, \theta, \Lambda[\phi, \theta])$. Despite a non-trivial field-space metric, they merely occur when both $\partial_\phi \Lambda = 0$ and  $\partial_\theta \Lambda = 0$, as the function $\Lambda$ is a field-space scalar. Their locations are given in Table \ref{tab:table2}. To compress expressions, we introduce the parameters, for $i \neq j \neq k$ and no implied sums:
\begin{equation}\label{paramstocompress}
\begin{split}
    &\kappa_{ij} \equiv \lambda_{ij}\xi_i - \lambda_i \xi_j,\quad
    \gamma_{ij} = \gamma_{ji} \equiv  \lambda_{i}\lambda_{j} - \lambda^2_{ij} > 0,\\
    & A_{ij} = A_{ji}\equiv \xi_k\gamma_{ij} + \lambda_{ik}\kappa_{ji} + \lambda_{jk}\kappa_{ij}
\end{split}
\end{equation}
which have geometric roles. The parameter $\kappa_{ij}$ is proportional to one of the pair of principal curvatures at the critical point in the $\chi_i$-direction, namely that which is associated to the $\chi_j$-direction, while $\gamma_{ij}  > 0$ forbids unwanted minima of the Jordan frame potential in the $\chi_i\chi_j$-plane. The $A_{ij}$ arise in the products $\kappa_{31}A_{13}$, $\kappa_{32}A_{23}$ and $(\kappa_{12}+\kappa_{21})A_{12}$ which respectively fix the sign of the principal curvature associated to the $\chi_{1}$, $\chi_2$ and $\chi_{3}$ directions at the critical point misaligned in the $\chi_1\chi_3$-, $\chi_2\chi_3$- and $\chi_1\chi_2$-planes. In fact, the relevant curvature data for all seven critical point factorises into the $\kappa_{ij}$ and $A_{ij}$, and so their signs are sufficient to determine when each realises a local minimum (or maximum/saddle point) of the leading order potential, and hence a hyper-valley (or ridge) of the full potential. We supply this information in the final column of Table \ref{tab:table2} and, hence, exhaust the possible attractor orientations for a general parameter space $\{\lambda_{ij},\xi_k\}$.

 In Sec. \ref{VISHnuInflation}, we show how the results of this subsection can be used identify the hyper-valleys of (\ref{EinsteinFramePot1}) by restricting the parameter space to realise the hidden-sector structure, as this is the parameter space which is relevant to the VISH$\nu$ model, and the remainder of our discussion. Having established the possible inflation scenarios, it remains to analyse the effectively single-field inflationary dynamics within each hyper-valley and demonstrate their reproduction of observed features of the CMB power spectrum. Although our discussion will focus on the VISH$\nu$ model, these results are generally independent of the parameter space, with the exception of the scalar power spectrum amplitude constraint ($A_s$), and are complementary to previous studies. In Sec. \ref{largexifit}, we assume that $\xi_k \gg 1$ whenever a hyper-valley has a component in the $\chi_k$ direction (we call this the ``large $\xi_k$'' regime) and these inflationary predictions are degenerate with the Higgs inflation model~\cite{HiggsInflation1}. In Sec. \ref{smallxifit}, we use the parametric freedom of our extended scalar sector to explore representative limiting cases that continue the parameter space outside the large $\xi_k$ regime, i.e. ``small $\xi_k$'', and this regime can avoid unitarity concerns. As we soon demonstrate, a simplification achieved in either context is the suppression of the kinetic mixing of the field components induced by the field-space metric in a convenient basis. In Appendix \ref{sec:AppendixA} we compare the masses of canonically normalised fluctuations orthogonal to the hyper-valleys with the inflationary Hubble scale. 

\subsection{Inflation scenarios in VISH$\nu$}
\label{VISHnuInflation} 

Let us now discuss the possible orientations of hyper-valleys specific to the case of the large-field VISH$\nu$ potential (\ref{EinsteinFramePot1}), for which we require $\lambda_{iS} \ll \lambda_{i}, \lambda_S, \lambda_{34}$ to realise the hidden sector structure (see Sec. \ref{sec:hiddensectorphil}). To do this, we merely identify the components of $\varphi^I = (\rho_1,\rho_2,\sigma)$ with the components of $\chi^I$ in the preceding section. Correspondingly, we will substitute ``$S$'' for the ``$3$'' index of the results we have obtained, and identify the coefficients of (\ref{threescalarinflationpotential}) with (\ref{EinsteinFramePot1}) in the obvious way, e.g. $\lambda_{34} = \lambda_{12}$. 

In general, the effect that decoupling $S$ from the Higgs sector has on the curvature of the inflation potential is encoded by the set of parameters $\{\kappa_{1S}$, $\kappa_{S1}$, $\kappa_{2S}$, $\kappa_{S2}\}$ which all become negative, and $A_{12}$, now positive. When this is the case, initial trajectories will quickly develop a $\sigma$ component, but, as the $\sigma$ axis forms a ridge ($\kappa_{S1},\kappa_{S2} < 0$), there are at most three possible hyper-valleys where the trajectories can stabilise.  We will denote the corresponding inflation scenarios by the fields which give rise to a non-zero component in the inflaton field parameterising the radial time-evolution. Hence, for the \textit{general} situation where are all non-minimal couplings are sizeable, there are three candidate scenarios to consider: $\Phi_1S$-Inflation can occur if $A_{1S}<0$ ($\kappa_{12} > 0$), $\Phi_2S$-Inflation can occur if $A_{2S} < 0$ ($\kappa_{21} > 0$), while $\Phi_1\Phi_2S$-Inflation can occur if both $A_{1S} > 0$ ($\kappa_{12} < 0$) and $A_{2S} > 0$ ($\kappa_{21} < 0$).\footnote{ $\Phi_1\Phi_2$-Inflation is possible for the special case that $\gamma_{12} \simeq 0$ and $\lambda_{iS} > 0$, but any approximate $SO(2)$ symmetry of the Higgs potential realised when $\gamma_{12}$ is set to zero along with the inter-sector couplings is explicitly broken by the Yukawa terms. This means that $\gamma_{12} \simeq 0$ is not radiatively stable in the sense we define in Sec. \ref{sec:sec5} and so, unless $\xi_S \simeq 0$, requires a fine-tuning of the quartic parameters to become a valley.} The corresponding condition in brackets holds when the sectors are fully decoupled. As $\kappa_{12}$ and $\kappa_{21}$ cannot both be positive if $\xi_1, \xi_2 > 0$, there is exactly one candidate inflaton field for each of the three disjoint regions of $\{\kappa_{12},\kappa_{21}\}$ parameter space we identify.\footnote{We begin with $\sqrt{\lambda_1\lambda_2} - \lambda_{34} > 0$, which follows from $\gamma_{12} > 0$ and boundedness-from-below, and we assume that $\xi_1, \xi_2 > 0$. It follows that $(\sqrt{\lambda_1}\xi_1 - \sqrt{\lambda_1}\xi_2)^2 + 2\xi_1\xi_2(\sqrt{\lambda_1\lambda_2} - \lambda_{34}) > 0$, because we have multiplied by, and then added, positive real numbers. Rearrangement shows the left-hand quantity to be $- [\xi_1\kappa_{21} + \xi_2\kappa_{12}]$ which, accordingly, must be positive. This forbids the case where both $\kappa_{12} > 0$ and $\kappa_{21} > 0$.} 
Generalising our results, it is clear that a hidden-sector extension cannot constitute an inflaton sector if a hidden scalar develops a non-minimal gravitational coupling in addition to the SM Higgs. Symmetrically, scalars which are very weakly coupled to the Higgs boson inevitably play an inflationary role when both are non-minimally coupled to gravity. This is because directions aligned with the field axes constitute ridges, while hyper-valleys have a non-trivial orientation. For each of the three scenarios generically compatible with the VISH$\nu$ parameter space ($\Phi_1 \Phi_2 S$-Inflation and $\Phi_i S$-Inflation), masses associated to the orthogonal \textit{modulus} field directions exceed the Hubble scale of the inflation by several orders of magnitude (this is demonstrated in Appendix \ref{sec:AppendixA}), which, with the exception of fluctuations in the axion field discussed in Sec. \ref{sec:sec4}, suppresses isocurvature modes ($\beta_{\text{iso}} \sim 0$) even in the limit of sector decoupling.

However, there are special exceptions to these findings. To support either single component hyper-valleys (aligned with a field axis), or those which are misaligned in the Higgs plane but orthogonal to the $\sigma$ axis, we can specialise to a parameter space where only some fields develop sizeable non-minimal couplings, hence reversing the sign of the curvatures $\kappa_{ij}$ which otherwise forbid them, for the case where $\lambda_{iS} > 0$. Specifically, if $\xi_S \simeq 0$, then $\kappa_{iS},A_{12} > 0$ and so $\Phi_1\Phi_2$-Inflation, $\Phi_1$-Inflation and $\Phi_2$-Inflation become possible --- the Higgs sector can be an inflaton sector. While if $\xi_1, \xi_2 \simeq 0$ then $\kappa_{Si} > 0$, so $S$-Inflation becomes possible --- the hidden sector can be an inflaton sector.\footnote{Of course, the naturalness of this parameter space must be defended. Given the $\xi_k$ are coupled under renormalisation group evolution, their hierarchical organisation may not be radiatively stable. However, in the limit of sector decoupling, the running of $\xi_S$ should likewise decouple from $\xi_i$, so that $\xi_S \gg \xi_1, \xi_2$ and $\xi_S \ll \xi_1, \xi_2$ may not provoke a naturalness concern, while $\xi_S, \xi_1 \gg \xi_2$ may be considered unnatural. This decoupling is evinced in a two-scalar context by RG equations in Ref.~\cite{Clark:2009dc}.} For these special deformations of the potential, one finds that $\kappa_{ij} \simeq 0$, due to $|\lambda_{iS}| \ll 1$, approximating a regime of zero or extremely weak curvature considered in Refs.~\cite{greenwood2013,Kaiser:2013sna}.\footnote{It is worth noting that the resulting inflaton potential is, nonetheless, very different to the $O(3)$-symmetric case. It approaches flatness at large field values only along a single radial direction, or in the Higgs plane, but quartically diverges in a general direction. To study this scenario, we will conservatively assume the inflaton field is initialised sufficiently close to the hyper-valley, with negligible velocities in the orthogonal direction, to ensure the trajectory stabilises to the attractor before observationally relevant times. These initial data are more finely-tuned than the general case due to mass suppression of the orthogonal directions by $\lambda_{iS} \ll 1$ and may not be compulsory.} Within these limiting hyper-valleys, and preserving the hidden sector structure, the mass associated to at least one orthogonal direction in the modulus field-space can be very light compared to the Hubble scale, and the corresponding modulus field develops non-negligible isocurvature fluctuations during inflation. In general, these will be erased during reheating if PQ symmetry is restored (and since $\lambda_{iS} \ll 1$).\footnote{If non-restored, inflationary fluctuations in the axion field can lead to a significant isocurvature fraction (see Sec. \ref{sec:sec4}). A worst-case estimate for the preserved isocurvature fraction due only to modulus fluctuations is $\beta_{\text{iso}} \sim 10^{-5}$ (inferred from Ref.~\cite{Kaiser:2013sna}).}

\subsection{Inflation model at large non-minimal coupling}
\label{largexifit} 

The large $\xi_k$ inflation model is obtained in the limit that $\xi_k \gg 1$ whenever a hyper-valley has a non-trivial component in the $\varphi_k$ direction. We find that this regime, like Higgs inflation~\cite{HiggsInflation1} and the Starobinsky model~\cite{STAROBINSKY198099}, generally reproduces the potential (\ref{InflationPotential}), which remains among the best fits to CMB data~\cite{Planck:2018jri}, despite a multi-field scalar sector.

\subsubsection{Analytical approximation}

We have classified the orientations and existence conditions for the hyper-valleys using spherical coordinates. However, to analyse the inflationary dynamics in the large $\xi_k$ regime, it is most convenient to modify the spherical coordinates (\ref{sphericalxfm}) to an ellipsoidal form (with $\xi_1,\xi_2,\xi_S > 0$ and $c_{\psi}, s_{\psi}  \equiv \cos \psi, \sin \psi$):
\begin{equation}\label{ellipsoidalxfm}
\begin{split}
    \rho_1 &= \frac{M_P}{\sqrt{\xi_1}}\ e^{\frac{\chi_r}{M_P\sqrt{6}}} c_{\phi'} s_{\theta'} \simeq \frac{M_P}{\sqrt{\xi_1}}\ \Omega\  c_{\phi'} s_{\theta'}, \\
    \rho_2 &= \frac{M_P}{\sqrt{\xi_2}}\ e^{\frac{\chi_r}{M_P\sqrt{6}}} s_{\phi'} s_{\theta'} \simeq \frac{M_P}{\sqrt{\xi_2}}\ \Omega\  s_{\phi'} s_{\theta'}, \\
    \sigma &= \frac{M_P}{\sqrt{\xi_S}}\ e^{\frac{\chi_r}{M_P\sqrt{6}}} c_{\theta'} \simeq \frac{M_P}{\sqrt{\xi_S}}\ \Omega\ c_{\theta'};
\end{split}
\end{equation}
where $\chi_r \equiv \sqrt{\frac{3}{2}} M_P\log (\Omega^2 - 1)$ and $\Omega > 0$ parametrises large radial displacement:
\begin{equation}\label{radial}
    \Omega^2 - 1 \simeq \frac{\xi_1}{M_P^2} \rho_1^2 + \frac{\xi_2}{M_P^2} \rho_2^2 + \frac{\xi_S}{M_P^2} \sigma^2.
\end{equation}
This is because, up to $\mathcal{O}(\xi^{-1}_k)$ corrections, the kinetic mixing vanishes, $\chi_r$ is canonically normalised and the background angular fields are static (see Appendix \ref{sec:AppendixA}). In this parametrisation, the large-field Einstein frame VISH$\nu$ potential (\ref{EinsteinFramePot1}) becomes:
\begin{equation}
    V^{\mathcal{E}} \simeq \Lambda[\phi',\theta']\left[1 - e^{-\frac{2}{\sqrt{6}}\frac{\chi_r}{M_P}}\right]^2\ ;
\end{equation}
where, to leading order in the inter-sector couplings, the energy density:
\begin{equation}
\begin{split}
    &\Lambda[\phi',\theta'] \simeq \\
    &\ \frac{M_P^4}{4}\left( c^4_{\theta'} \frac{\lambda_S}{\xi_S^2} + s^4_{\theta'}\left[\frac{\lambda_1}{\xi_1^2}c_{\phi'}^4   + 
    \frac{2\lambda_{34}}{\xi_1 \xi_2}c_{\phi'}^2 s_{\phi'}^2 + \frac{\lambda_2}{\xi_2^2} s_{\phi'}^4  \right]\right),
\end{split}
\end{equation}
is minimised along a hyper-valley. While a reparameterisation of the field-space does not alter the topography of the potential, so that the existence conditions in Table \ref{tab:table2} as well as $\Lambda_{\text{min}}$ are invariant, the hyper-valley co-ordinates will transform, as $(\phi',\theta') \neq (\phi,\theta)$ in general. 

Recall that during the early stages of inflation, the trajectory rapidly settles into the hyper-valley before observationally relevant times and the corresponding background angular field values induce an effectively single-field potential for the inflaton field $\chi_r$:
\begin{equation}\label{InflationPotential}
    V^{\mathcal{E}} \simeq \frac{M_P^4}{8} \frac{\lambda_{\text{eff}}}{\xi^2_{\text{eff}}} \left[1 - e^{-\frac{2}{\sqrt{6}}\frac{\chi_r}{M_P}}\right]^2 ,
\end{equation}
which describes the slow-roll inflationary dynamics along the adiabatic direction. In $\Phi_1\Phi_2S$-Inflation and $\Phi_iS$-Inflation this is a general outcome and we respectively compute (defining $L \equiv \lambda_{1}\lambda_{2} - \lambda_{34}^2$):
\begin{equation}\label{leffone}
\begin{split}
    \frac{\lambda_{\text{eff}}}{\xi^2_{\text{eff}}} \simeq\ &\frac{\lambda_S L}{\lambda_S(\lambda_2\xi_1^2 - 2\lambda_{34}\xi_1\xi_2 + \lambda_1\xi_2^2)+ \xi_S^2L}\\
    &\text{and}\quad \frac{\lambda_S\lambda_i }{\lambda_S\xi_i^2 + \lambda_i\xi_S^2}.
\end{split}
\end{equation}
As explained in Sec. \ref{VISHnuInflation}, for $S$-Inflation ($\xi_1,\xi_2 \simeq 0$), $\Phi_1\Phi_2$-Inflation and $\Phi_i$-Inflation ($\xi_S \simeq 0$) the trajectory stabilises to the hyper-valley for more specific initial conditions due to sub-Hubble masses of the orthogonal directions. Respectively, we find that:
\begin{equation}\label{lefftwo}
    \frac{\lambda_{\text{eff}}}{\xi^2_{\text{eff}}} \simeq \frac{\lambda_S}{\xi_S^2},\ \ \frac{L}{\lambda_2\xi_1^2 - 2\lambda_{34}\xi_1\xi_2 + \lambda_1\xi_2^2},\ \ \text{and}\  \ \frac{\lambda_i}{\xi_i^2}.
\end{equation}
This information is summarised in Table \ref{tab:table3}. Observe that $\lambda_{\text{eff}}/\xi^2_{\text{eff}} > 0$ ensures positivity of the inflationary energy density in each scenario.

\begin{table*}
\caption{\label{tab:table3} We summarise some results from Sec. \ref{VISHnuInflation}, Sec. \ref{largexifit} and Sec. \ref{smallxifit}. The required parameter space is for the existence of a hyper-valley permitting each inflation scenario. Recall that $\kappa_{12} = \lambda_{34}\xi_1 - \lambda_1\xi_2$, $\kappa_{21}= \lambda_{34}\xi_2 - \lambda_2\xi_1$ and $L = \lambda_1\lambda_2 - \lambda_{34}^2$ (which is positive). The condition $\lambda_2\xi_1^2 - 2\lambda_{34}\xi_1\xi_2 + \lambda_1\xi_2^2 > 0$ ensures $V \sim M_P^4 \lambda_{\text{eff}} / 8 \xi^2_{\text{eff}} > 0$ for $\Phi_1\Phi_2S$-Inflation and $\Phi_1\Phi_2$-Inflation. Statements about initial conditions concern the neutral modulus field-space: we make more conservative assumptions in the final four rows due to the presence of at least one light direction, these assumptions are absent in the first three scenarios which we therefore designate as ``generic''. (Although we still assume, for example, a large displacement of the fields from the minimum of the potential at the onset of the inflation phase.) We use a $\checkmark$ in the final column to indicate when the small $\xi_k$ regime is viable and preserves naturalness requirements (we forbid small Higgs-sector parameters).} 
\centering
\def\arraystretch{2}
\begin{ruledtabular}
\begin{tabular}{ccccc}
    Inflation scenario & Hyper-valley conditions & $\lambda_{\text{eff}}/\xi^2_{\text{eff}}$ & Initial conditions & Small $\xi_k$ model 
    \\\hline
    $\Phi_1\Phi_2S$-Inflation & $\kappa_{12}, \kappa_{21} < 0 $ & $\dfrac{\lambda_S L}{\lambda_S (\lambda_2\xi_1^2 - 2\lambda_{34}\xi_1\xi_2 + \lambda_1\xi_2^2) + L \xi_S^2}$ & Generic & \checkmark 
    \\[1.5ex]
    $\Phi_1S$-Inflation & $\kappa_{12} > 0$ & $\dfrac{\lambda_S\lambda_1 }{\lambda_S\xi_1^2 + \lambda_1\xi_S^2}$ & Generic & \checkmark 
    \\[1.5ex]
    $\Phi_2S$-Inflation & $\kappa_{21} > 0$ & $\dfrac{\lambda_S\lambda_2 }{\lambda_S\xi_2^2 + \lambda_2\xi_S^2}$ & Generic & \checkmark 
    \\[1.5ex]
    $S$-Inflation & $\xi_i \simeq 0,\ \lambda_{iS} > 0$ 
    & $\dfrac{\lambda_S}{\xi_S^2}$ & $\rho_i, \Dot{\rho}_i \sim 0$ & \checkmark
    \\[1.5ex]
    $\Phi_1\Phi_2$-Inflation & $\xi_S \simeq 0,\ \lambda_{iS} > 0$ 
    & $\dfrac{L}{\lambda_2\xi_1^2 - 2\lambda_{34}\xi_1\xi_2 + \lambda_1\xi_2^2}$ & $\sigma, \Dot{\sigma} \sim 0$ & Unnatural
    \\[1.5ex]
    $\Phi_1$-Inflation & $\xi_S \simeq 0,\ \lambda_{iS},\kappa_{12} > 0$ 
    & $\dfrac{\lambda_1}{\xi_{1}^2}$ & $\sigma, \Dot{\sigma} \sim 0$ & Unnatural
    \\[1.5ex]
    $\Phi_2$-Inflation & $\xi_S \simeq 0,\ \lambda_{iS},\kappa_{21} > 0$ 
    & $\dfrac{\lambda_2}{\xi_{2}^2}$ & $\sigma, \Dot{\sigma} \sim 0$ & Unnatural \\[1.5ex]
\end{tabular}
\end{ruledtabular}
\end{table*}

\subsubsection{Fit to CMB data}

During the inflation, the inflaton field $\chi_r$ develops quantum fluctuations which source almost scale-invariant spectra of primordial scalar and tensor perturbations. Respectively, these give rise to large-scale structure formation and relic gravitational waves which, through observations of the CMB, constitute the observational basis by which the inflation model is constrained. One expresses the primordial spectra as a power law about a pivot scale $k_*$. For the case of the scalar power spectrum, we define:
\begin{equation}
    \Delta^2_s(k) = A_s(k_*)\left(k/k_*\right)^{-1 + n_s(k_*) + \frac{1}{2}\alpha_s(k_*) \log(k/k_*)},
\end{equation}
where $A_s$ is the amplitude, $n_s$ is the scalar spectral index and $\alpha_s$ describes first-order running in the spectral index. Analogous definitions follow for the tensor power spectrum, and we define the tensor-to-scalar ratio: $r \equiv A_t/A_s$.

The Planck 2018 release~\cite{Planck:2018jri}
of CMB temperature and polarisation data confirmed a small deviation from scale invariance in the scalar power spectrum, with the scalar spectral index measured as:
\begin{equation}\label{PLANCKns}
\begin{split}
    n_s =\ &0.9649 \pm 0.0042\\ &\text{(68 \% CL, Planck TT,TE,EE+lowE+lensing)}.
\end{split}
\end{equation}
If running in the scalar spectral index is also considered, it is constrained to be~\cite{Planck:2018jri}:
\begin{equation}\label{constraintonalpha}
\begin{split}
 \alpha_s \equiv &\frac{\rd n_s}{\rd \log k} =\ -0.0045 \pm 0.0067
 \\ &\text{(68 \% CL, Planck TT,TE,EE+lowE+lensing)}.
\end{split}
\end{equation}
The upper bound on the tensor-to-scalar ratio has recently been made more stringent by BICEP/Keck Array 2018 (BK18) data~\cite{BICEP:2021xfz}, given to 95\% confidence level as:
\begin{equation}\label{BICEPr}
    r < 0.036.
\end{equation}
These constraints (and $A_s$ below) are evaluated at the CMB pivot scale $k_* = 0.05$ Mpc$^{-1}$.

\begin{figure*}[t]
\includegraphics[width = 0.47\textwidth]{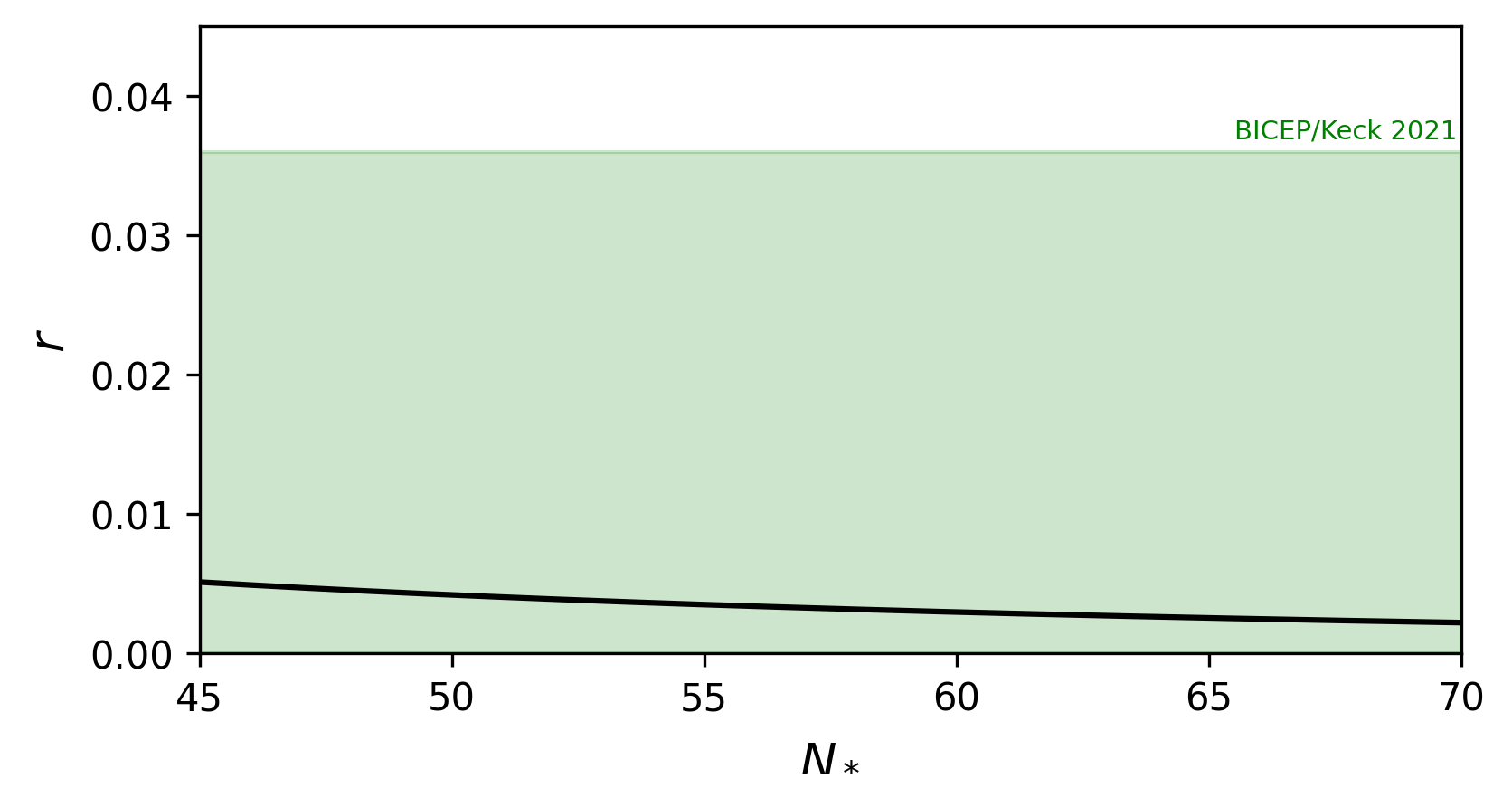}
\includegraphics[width = 0.47\textwidth]{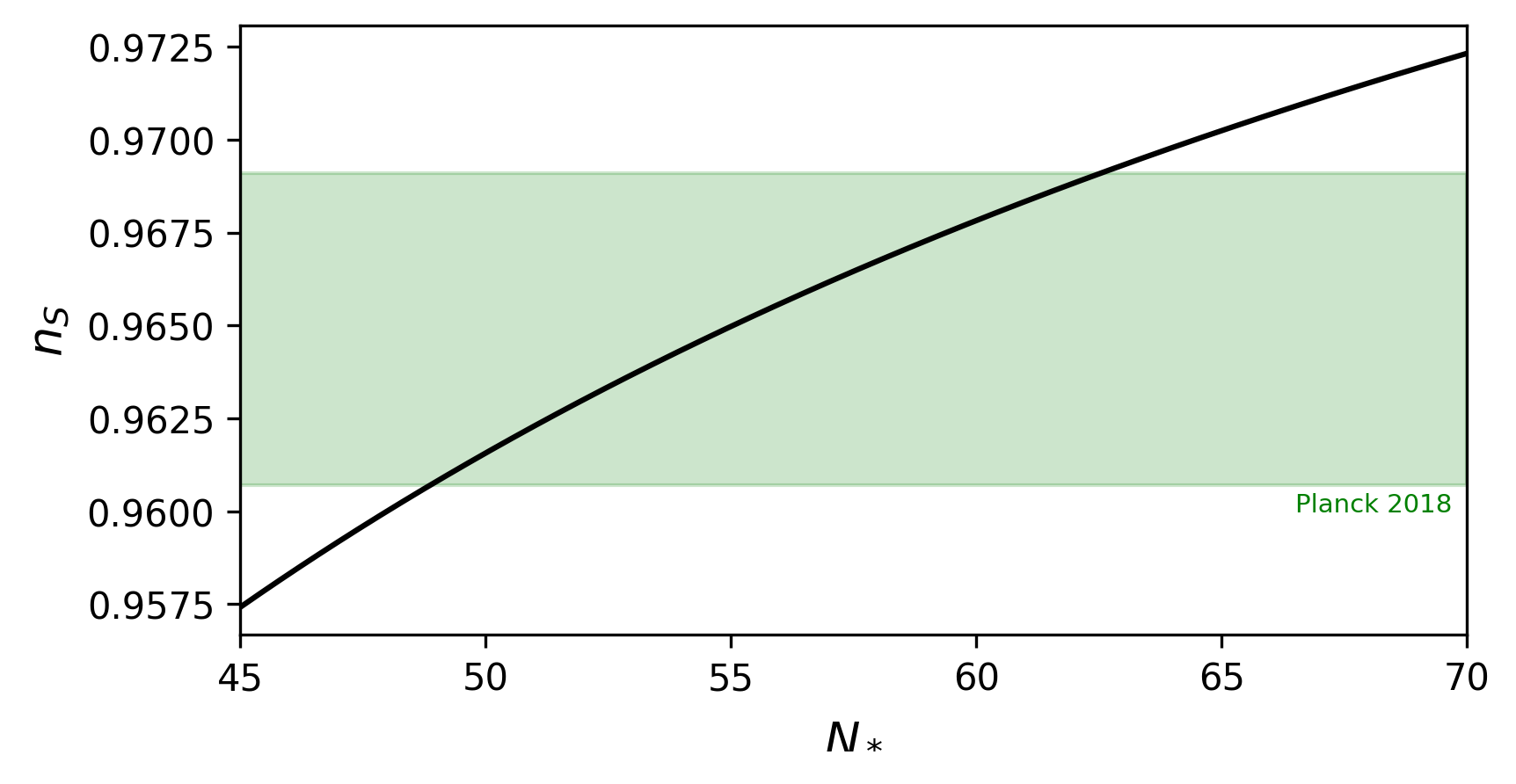}
\caption{\label{Figure1} Plots of the tensor to scalar ratio $r_*$ and spectral index $(n_s)_*$ at horizon-exit of the CMB pivot scale, exhibiting dependence on e-folds of inflation ($N_*$) in the large $\xi_k$ limit. The green regions correspond to those favoured by CMB data, leading to an acceptable range for $N_*$ which is ultimately determined by the reheating dynamics.}
\end{figure*}

Our predictions for $r,\ n_s$ and $\alpha_s$ may be computed directly from the Einstein frame model as follows. In the slow-roll approximation, we have that
\begin{equation}\label{randnsandalpha}
        r \simeq 16\tilde{\epsilon},\quad n_s \simeq 1 - 6\tilde{\epsilon} + 2\eta,\quad \alpha_s \simeq - 24\tilde{\epsilon}^2 + 6\tilde{\epsilon}\eta -2\zeta,
\end{equation}
where the first\footnote{Let us distinguish $\epsilon$, the $\nu$DFSZ model parameter, from $\tilde{\epsilon}$, the first slow-roll parameter, using a tilde.}, second and third slow-roll parameters are
\begin{equation}\label{SlowRollParameters}
    \tilde{\epsilon} \equiv \frac{M_P^2}{2}\left(\frac{V'}{V}\right)^2,\quad
    \eta \equiv M_P^2 \frac{V''}{V},\quad
    \zeta \equiv M_P^4 \frac{V'V'''}{V^2},
\end{equation}
Here $V \equiv V^\mathcal{E}$ is the inflaton potential (\ref{InflationPotential}) and $'$ denotes a derivative with respect to $\chi_r$. Hence, defining $x \equiv \exp[2\chi_r/(\sqrt{6}M_P)]$, we can re-express (\ref{randnsandalpha}) as~\cite{Ballesteros:2016xej}:
\begin{equation}\label{newrandnsandalpha}
        r \simeq \frac{64}{3(x - 1)^2},\ 
        n_s \simeq 1 - \frac{8(x+1)}{3(x - 1)^2},\ 
        \alpha_{s} \simeq \frac{-32x(x+3)}{9(x - 1)^4}.
\end{equation}
Let us define $N_*$ as the interval in e-folds between the end of the inflation phase and the earlier horizon exit of the CMB pivot scale $k_*$:
\begin{equation}
    N_* \simeq \frac{1}{M_P} \int^{\chi_*}_{\chi_f} \frac{\rd\chi}{\sqrt{2\tilde{\epsilon}}} \simeq \frac{3}{4} \left[x_* - x_f - \frac{2(\chi_* - \chi_{f})}{\sqrt{6}M_P} \right],
\end{equation}
where $\chi_f$ corresponds to the field value at the end of the inflation phase, where $\tilde{\epsilon}(\chi_f) \simeq 1$ and the slow-roll approximation breaks down. Then, the central value of $(n_s)_*$ is obtained for $N_* \simeq 55$ e-folds of inflation, corresponding to $\chi_* \simeq 5.4 M_P$. This means that (at least one) of the modulus fields takes a value $\sim \mathcal{O}(10) M_P/\sqrt{\xi}$ and we find that
\begin{equation}
    (n_s)_* \simeq 0.965,\ r_* \simeq 0.00351, \ (\alpha_{s})_* \simeq -0.000623,
\end{equation}
which are all in perfect agreement with available data. We plot the dependence of our $r$ and $n_s$ predictions on the e-folds of inflation in Figure \ref{Figure1}, which show a range of acceptable values in the large $\xi_k$ regime. In fact, at leading order, we can express (\ref{newrandnsandalpha}) in terms of $N_*$: 
\begin{equation}
    r_* \simeq \frac{12}{N_*^2} \quad (n_s)_* \simeq 1 - \frac{2}{N_*},
\end{equation}
if running in the scalar spectral index is ignored. 

Finally, we must also match the measured scalar power spectrum amplitude at horizon exit, $(A_s)_* \sim 2.1 \times 10^{-9}$~\cite{PlanckCosParam}. Since in the slow-roll regime this is
\begin{equation}
    (A_s)_* \simeq \frac{1}{24\pi^2\epsilon_*}\frac{V_*}{M_P^4} \simeq \frac{N_*}{72\pi^2}\frac{\lambda_{\text{eff}}}{\xi_{\text{eff}}^2},
\end{equation}
we obtain a relation between the effective quartic and non-minimal couplings (for acceptable values of $N_*$):
\begin{equation}\label{AsFit}
    \frac{\lambda_{\text{eff}}}{2\xi^2_{\text{eff}}} \sim 4.4 \times 10^{-10},
\end{equation}
which serves finally to constrain the parameters of the inflaton potential --- compare (\ref{newrandnsandalpha}), which depended only on $\chi_*$ (or $N_*$). In particular, (\ref{AsFit}) determines the height of the inflation plateau and, hence, the Hubble scale of the expansion, $H_{\text{inf}} = V/3M_P^2 \sim \text{few}\times 10^{13} $~GeV. The large relative size of the effective couplings implied by (\ref{AsFit}), poses a \emph{prima facie} challenge to the naturalness of the model, which we address in Sec. \ref{sec:sec5}. 

\subsection{Inflation model at small non-minimal coupling}
\label{smallxifit} 

In Sec. \ref{largexifit}, we analysed the case where all modulus fields which take a large value during the inflation phase have a large non-minimal coupling to gravity, so that $\max_k \{\xi_k\} \gg \mathcal{O}(1)$. In addition to this regime, we now show that a range of values $\max_k \{\xi_I\} \lesssim O(1)$ are also viable. As explained in Sec. \ref{sec:sec5}, this regime avoids possible issues associated to a low unitarity cutoff and can only be naturally realised for inflation scenarios where $\xi_S\sigma^2 > M^2_P$. This condition is satisfied in $\Phi_1\Phi_2S$-Inflation and  $\Phi_iS$-Inflation, which are compatible with a generic parameter space and initial data, as well as $S$-Inflation, which can result for $\xi_1,\xi_2 \simeq 0$.

\begin{figure}[t]
\centering
\includegraphics[width=0.48\textwidth]{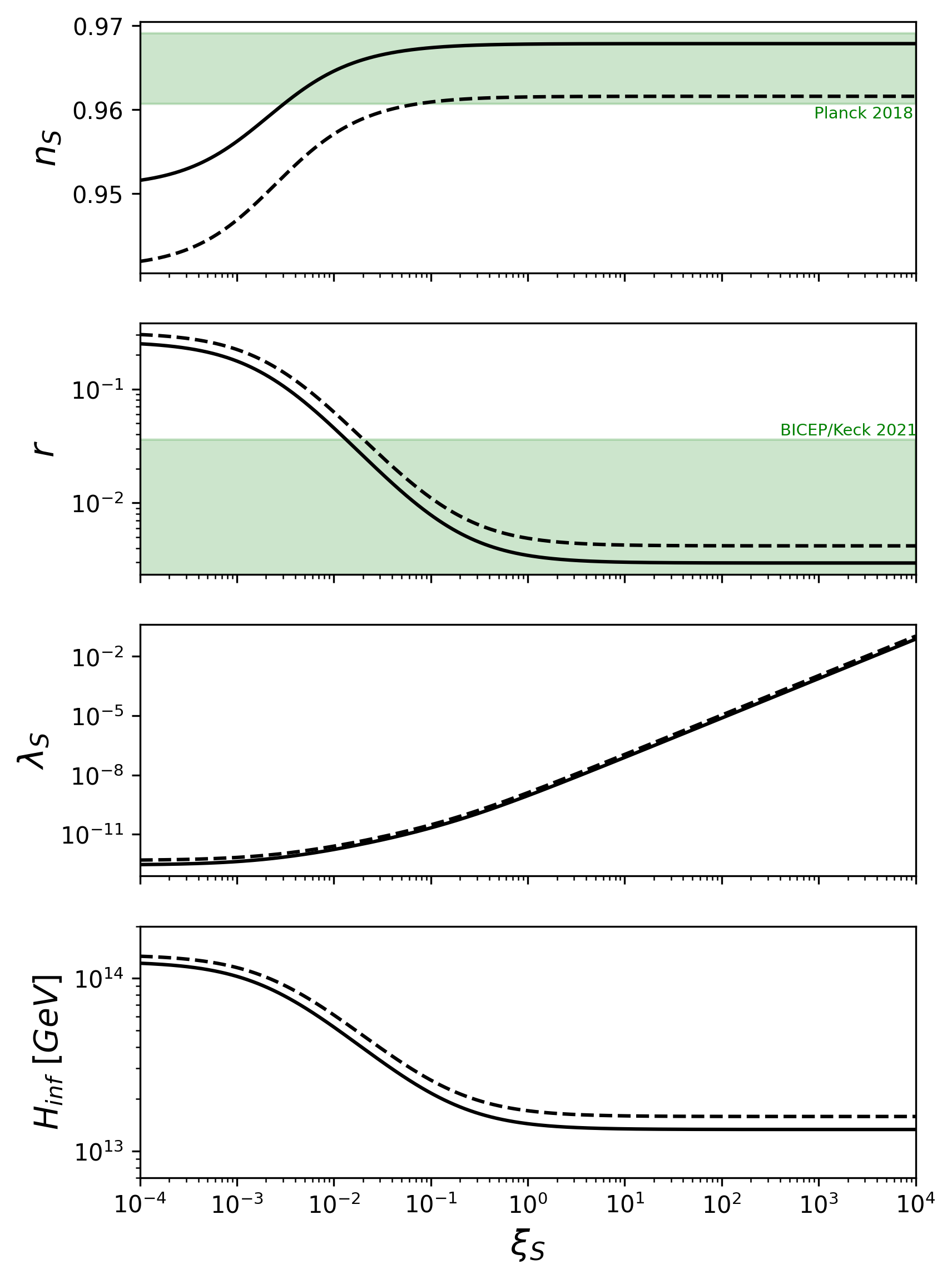}
\caption{\label{Figure2}
Plots exhibiting the $\xi_S$-dependence (common horizontal axis) of the tensor to scalar ratio $r_*$ and spectral index $(n_s)_*$ at the horizon exit of the CMB pivot scale in the parameter regimes $\xi_i \simeq \xi_S$ and $\xi_i \ll \xi_S$, as well as the quartic-coupling $\lambda_S$ at the inflation scale and the corresponding $H_\text{inf}$. The solid contours correspond to inflation scenarios with $N_*=60$, while $N_*=50$ is represented by the dashed contours and delimit the favoured values for $N_*$ \cite{Planck:2018jri}. The green regions correspond to (\ref{PLANCKns}) and (\ref{BICEPr}), and are favoured by CMB data, placing a lower limit on $\xi_S$.}
\end{figure}

While, outside of the large $\xi_k$ regime, the kinetic mixing of the modulus fields is non-vanishing for non-trivial hyper-valley orientations, this is not the case for hyper-valleys which are aligned with the field axis. To illustrate this, let us consider the relevant case of $S$-Inflation ($\overline{\rho_i} = 0$), for which the induced field metric has the non-zero components, where $\Omega^2 = 1 + \xi_S \sigma^2/M_P^2$:
\begin{equation}\label{metriccomponents}
    \overline{\mathcal{G}}_{11} = \overline{\mathcal{G}}_{22} = \frac{1}{\Omega^{2}},\quad \overline{\mathcal{G}}_{SS} = \frac{1}{\Omega^{4}} \left[1 + \xi_S(1 + 6\xi_S) \frac{\sigma^2}{M_P^2}\right].  
\end{equation}
In fact, since we require $\xi_i/\xi_S < \lambda_{iS}/\lambda_S \lesssim 10^{-8}$ for the $\sigma$ axis to form a valley, $\Phi_1\Phi_2S$-Inflation and $\Phi_iS$-Inflation essentially degenerate to $S$-Inflation for this parameter space. The field metric components for these scenarios are (\ref{metriccomponents}) up to $\mathcal{O}(\xi_i/\xi_S)$ corrections and the trajectories are, likewise, close to the $\sigma$ axis. Hence, (\ref{metriccomponents}) and the inflaton potential:
\begin{equation}\label{smallxipot}
    V(\sigma) \simeq \frac{1}{8} \frac{\lambda_S\sigma^4}{(1 + \xi_S\sigma^2/M_P^2)^2},
\end{equation}
describe the effectively single field dynamics appropriate to each hyper-valley, up to small $\mathcal{O}(\xi_i/\xi_S)$ and $\mathcal{O}(\lambda_{iS})$ corrections. In all four cases, we define the canonically normalised scalar inflaton field $\chi$:
\begin{equation}\label{canonicallynormalised}
    \frac{\rd \chi}{\rd \sigma} \simeq \sqrt{\mathcal{G}_{SS}} \simeq  \frac{\sqrt{1 + \xi_S(1+6\xi_S)(\sigma/M_P)^2}}{{1 + \xi_S(\sigma/M_P)^2}},
\end{equation}
where the equation is exact for the special case of $S$-Inflation and can be solved analytically~\cite{Garcia-Bellido:2008ycs}. The first and second slow-roll parameters (\ref{SlowRollParameters}) then become~\cite{Fairbairn:2014zta}:
\begin{equation}\label{smallxislowrollparameters}
    \tilde{\epsilon} = \frac{M_P^2}{2}\left(\frac{V'}{V\chi'}\right)^2, \ \eta = M_P^2 \left(\frac{V''}{V[\chi']^2} - \frac{V'\chi''}{V[\chi']^3}\right),
\end{equation}
where $'$ denotes a derivative with respect to $\sigma$. We summarise the resulting inflationary predictions in Figure \ref{Figure2}, where we have used $A_s$ to fix the relation between $\lambda_S$ and $\xi_S$, exhibiting the $\xi_S$-dependence of $r$ and $n_s$ (ignoring running in the latter), which promptly asymptote to values discussed above in the large $\xi_k$ regime. Observe that values of $\xi_S > \text{few} \times 10^{-2}$ abide by confidence limits on $n_s$ and the exclusion limit on $r$ for a range of standard values for $N_*$.

We now consider the general case, for which the $\sigma$ axis now forms a ridge and there are three possible inflation scenarios: $\Phi_1\Phi_2S$-Inflation and  $\Phi_iS$-Inflation. Although in the $\{\rho_1, \rho_2,\sigma \}$ field-basis there is non-trivial kinetic mixing,  it is possible to make a gauge transformation of the field-space coordinates which diagonalises the kinetic terms during the inflation phase, once the trajectory settles into a hyper-valley. For concreteness, consider $\Phi_1S$-Inflation ($\kappa_{12} > 0$) and the limiting case that $\xi_1 = \xi_S$. A transformation which accomplishes diagonal kinetic terms is merely a rotation of the field basis about the $\rho_2$ axis by the non-zero background angle which defines the hyper-valley orientation, $\overline{\theta}$:
\begin{equation}\label{rotation}
    \rho'_1 = \cos \overline{\theta}\ \rho_1 - \sin \overline{\theta}\ \sigma,\ \ \rho'_2 = \rho_2,\ \sigma' = \sin \overline{\theta}\ \rho_1 + \cos \overline{\theta}\  \sigma.
\end{equation}
The outcome of this transformation is that the $\sigma'$ direction is aligned with the hyper-valley (for which $\rho_1 =  \tan \overline{\theta}\ \sigma$). We obtain analogous transformations for the case of $\Phi_1\Phi_2S$-Inflation and $\Phi_2S$-Inflation. The reason we consider the limiting case $\xi_i = \xi_S$ is that this suppresses to zero any residual kinetic mixing of the primed fields after the transformation (\ref{rotation}). An analysis of the limiting case $\xi_i = \xi_S$ is, of course, the leading-order analysis of the more general situation where the sizes of non-minimal couplings are comparable ($\xi_1 \sim \xi_S$). Eqns. (\ref{metriccomponents}), (\ref{smallxipot}) and (\ref{canonicallynormalised}) are reproduced in the rotated field-basis for the substitution $\sigma \rightarrow \sigma'$ and with:
\begin{equation}
    \lambda_S \rightarrow \lambda_{\text{eff}} = \lambda_S \left[1 - \mathcal{O}\left(\frac{\lambda_S}{\lambda_1}\right)\right] \left[1 + \mathcal{O}\left(\frac{|\xi_S - \xi_1|}{\xi_S}\right)\right],
\end{equation}
provided corrections due to relative sizes of the $\xi_k$ are kept small (setting inter-sector couplings to zero). As a consequence of the assumed degeneracy in $\xi_k$, we plot the dependence of the inflationary observables on a single non-minimal coupling, $\xi_S$, and we re-obtain the plots in Figure \ref{Figure2} (assuming $\lambda_1, \lambda_2, \lambda_{34} \sim \mathcal{O}(1)$ for the plot of required $\lambda_S$). For the case of $\Phi_2S$-Inflation and $\Phi_1\Phi_2S$-Inflation, we also obtain expressions for $\lambda_{\text{eff}}$ which reproduce (\ref{leffone}) and (\ref{lefftwo}), and the resulting plots are degenerate with those in \ref{Figure2}, where $\lambda_S$ is responsible for the suppression of the scalar amplitude. 

Finally, we remark that the small values of $\lambda_S \lesssim 10^{-8}$ required for successful inflation in a regime of small $\xi_k$ results in a suppression of the Higgs component in the inflaton fields for $\Phi_1\Phi_2S$-Inflation and $\Phi_iS$-Inflation (which is to say, the orientation of the hyper-valley aligns more closely with the $\sigma$-axis). This is in contrast to the large non-minimal coupling regime, where the magnitude of the Higgs component can be up to a comparable size to the $S$ component. This is because, in $\Phi_iS$-Inflation, $\cos^2 \overline{\theta} \sim 1 - \mathcal{O}(\lambda_S/\lambda_i)$, while in $\Phi_1\Phi_2S$-Inflation, $\cos^2 \overline{\theta} \sim 1 - \mathcal{O}(\lambda_S(\kappa_{12}+\kappa_{21})/\xi_S L)$, using the results of Sec. \ref{valleyorientations}. So, it is also sensible to use the approximation $\overline{\rho}_i \simeq 0$, in these scenarios, if $\mathcal{O}(\lambda_S/\lambda_i)$ corrections can be safely ignored.

Beyond the two representative regimes for the $\xi_k$ (hierarchical and degenerate), the analysis is more involved due to \textit{a priori} large kinetic mixing but no important differences are anticipated: there will be an effective non-minimal coupling which, when large, reproduces in the findings of Sec. \ref{largexifit}, and, when small, remains perfectly compatible with confidence limits on the $r$-$n_s$ plane, up to lower bound $\sim \mathcal{O}(0.01)$.

\section{Dark matter}
\label{sec:sec4}

The solution offered by VISH$\nu$ to the origin and nature of the observed cold dark matter abundance depends on the existence of a range of $f_A$ for which the variant DFSZ axion, which arises as part of a solution to the strong $CP$ problem, also becomes a viable dark matter candidate. Hence, the predicted axion abundance becomes sensitive to the particular cosmological history and the possibility that the PQ symmetry was restored during the inflation or reheating phases. 

\subsection{Dependence on the reheating phase}
\label{reheatingdarkmatter}
Importantly, in the inflation scenarios we have analysed, the PQ symmetry is generally \emph{broken} during the inflation phase by the large instantaneous values of the modulus fields $\varphi^I$ driving the expansion, since both $S$ and $\Phi_{1,2}$ carry PQ charges, corresponding to a large displacement from the minimum of the VISH$\nu$ potential\footnote{\label{footnote11} The exception to the general situation that PQ is broken obtains when only \emph{one} of $\rho_1$ and $\rho_2$ takes large values, while $\sigma$ does not, since in those cases a linear combination of weak hypercharge and the original PQ charge remains unbroken, which is a redefined PQ charge. In each of those cases the associated redefined PQ symmetry can be restored during inflation because $f_A$ is less than $T_{\text{GH}} = H_{\text{inf}}/2\pi$, the inflationary Gibbons-Hawking temperature~\cite{Gibbons1977,Hertzberg:2008wr}, realising a post-inflationary axion.} (\ref{potential}). This can gives rise to an effective value for the axion decay constant~\cite{LINDE199138} ($f_{A,\text{inf}}$) --- either $\sigma_*$~\cite{Fairbairn:2014zta} or induced by large Higgs vevs~\cite{Nakayama2015} depending on the inflaton field --- and establishes the axion field during the inflation phase. Unless the axion field develops a large inflationary mass (see Ref.~\cite{Nakayama2015}), it will develop unsuppressed fluctuations.\footnote{Note that, because we can neglect the tri-linear term in the inflation, the Higgs pseudoscalar is also massless and develops fluctuations. As in Ref.~\cite{Nakayama2015}, we assume that these do not contribute to observable isocurvature perturbations.} If these axion fluctuations are not erased by a restoration of the PQ symmetry after inflation, these give rise to non-negligible (CDM) isocurvature fluctuations in the CMB temperature power spectrum which are severely constrained~\cite{Planck:2018jri}, imposing a bound on the inflationary Hubble scale which is \textit{prima facie} incompatible with the inflation model analysed in Sec. \ref{sec:sec3}.

It is possible for $U(1)_{PQ}$ to be non-thermally restored during the early stages of reheating in inflation scenarios where the scalar partner of the axion is embedded into the inflaton, which is the general case. Once inflation comes to an end (i.e. when $\tilde{\epsilon}(\chi_f) \simeq 1$), the identity of the inflaton field and the effectively single-field background field evolution are generically preserved by the non-minimal gravitational couplings~\cite{DeCross:2015uza}, and the inflaton condensate begins to oscillate about the minimum of the potential with a large initial amplitude. Non-perturbative effects will be responsible for the resonant amplification of scalar field fluctuations in a phase of pre-heating~\cite{Kofman:1994rk,Kofman:1997yn}, and the efficiency of this effect grows with non-minimal coupling size~\cite{DeCross:2016fdz}. In Ref.~\cite{Tkachev:1998dc}, it is shown that fluctuations in the axion field ($\sim \vartheta_S$) will grow to effectively restore the $U(1)_{\text{PQ}}$ symmetry for $f_A \lesssim 10^{16}$ GeV, and this result is replicated by more recent simulations in the context of the SMASH model~\cite{Ballesteros:2016xej,Ballesteros:2021bee} (see also~\cite{Kawasaki:2013iha,Kearney:2016vqw} for related discussions), for which the inflaton has a small Higgs component. This finding is based on three assumptions. The first is that dimensionful parameters in the potential can be ignored (this is what applies the upper-bound to $f_A$), the second is that non-minimal coupling sizes are at most $\sim 1$ and the third is that the Higgs fields do not affect the growth of the $S$ fluctuations~\cite{Ballesteros:2016xej}. The first and second conditions can be replicated in VISH$\nu$ models, as the tri-linear coupling and Higgs mass parameters can be ignored, while $m^2_{SS}$ induces the same bound, and Sec. \ref{smallxifit} outlines viable predictions for $\xi_k \sim 1$. VISH$\nu$ can also naturally realise the third condition by the effective decoupling of the inflaton field, which is mostly aligned with $\sigma$ for small $\xi_k$, from the Higgs fields. Hence, we can expect that $U(1)_{PQ}$ is non-thermally restored, in general, for the small $\xi_k$ inflation model. However, this conclusion should be supported by dedicated simulations, such as Ref.~\cite{Ballesteros:2021bee}, which we do not perform here.\footnote{Ref.~\cite{Ballesteros:2021bee} does not scan the hidden-sector parameter space, but uses inputs which are standard for the SMASH model, which requires significantly larger inter-sector couplings to successfully preserve the stability of the Higgs potential (see Sec. \ref{sec:sec5}).} We speculate that a regime of larger $\xi_k$ will increase the efficiency of the non-thermal restoration, rather than hampering it, but (to our knowledge) this has not been examined before. It has also been suggested that Higgs-sector oscillations following, for example, $\Phi_1\Phi_2$-Inflation and $\Phi_i$-Inflation, can induce fluctuations in $S$ that restore the symmetry~\cite{Nakayama2015}, and this was demonstrated in toy model simulations by Ref.~\cite{Ballesteros:2021bee}. We do not expect that this conclusion holds in VISH$\nu$, since the hidden-sector parameter space will suppress the effect in general.

An additional possibility is that $U(1)_{PQ}$ is thermally restored during the later stages of reheating, if the maximum temperature after inflation (which may be larger than the reheating temperature $T_{\text{reh}}$~\cite{chung1999}) exceeds the critical temperature for the PQ phase transition (which may be lower than $f_A$~\cite{Ballesteros:2016xej}). Non-minimally coupled models can predict a high reheating temperature when large amplitude oscillations of the background inflaton condensate lead to efficient SM gauge boson production. In VISH$\nu$, such a phase of reheating will occur through the Higgs fields, which either form a sizeable inflaton component ($\xi_k \gtrsim 1$), are effectively coupled to the inflaton ($\xi_k \ll 1$) or effectively decoupled ($\xi_1,\xi_2 \simeq 0$). In the large $\xi_k$ regime, Higgs inflation has been shown to result in $T_{\text{reh}} \sim 10^{13}$ GeV~\cite{Bezrukov:2008utReheat,Garcia-Bellido:2008ycs,HiggsInflation1}, while reheating through a tree-level coupling of the inflaton to the Higgs can vary in efficiency.\footnote{The variation is due primarily to the size of the portal coupling(s) compared with the effective inflaton self-coupling. However, comparable reheating temperatures are achievable~\cite{Lerner2011}.} A comprehensive study of the small $\xi_k$ regime has been completed for the SMASH model~\cite{Ballesteros:2016xej}, where reheating is less efficient ($T_{\text{reh}} \sim 10^{10}$ GeV) due to the largeness of the Higgs portal coupling compared to the effective inflaton self-coupling. As this is the opposite regime to VISH$\nu$, our results will differ and we can expect greater efficiency (like SMASH, the production of $\nu_R$ will not affect this result due to suppressed decays). Each of these temperatures are sufficient to realise a viable post-inflationary scenario, even if non-thermal restoration of the symmetry fails, and should result in axion thermalisation with the SM sector~\cite{Masso:2002np}, avoiding restrictive dark radiation constraints discussed in Ref.~\cite{Ballesteros:2016xej}.\footnote{A high reheating temperature is also consistent with our realisation of thermal leptogenesis.} Some efforts to achieve viable reheating without a tree-level inflaton coupling to the Higgs, for example Ref.~\cite{Figueroa:2016dsc}, cannot be successfully realised in the $S$-Inflation scenario due to light neutral Higgs modulus masses during the inflation.\footnote{We do not expect efficient reheating for the $S$-Inflation scenario in the absence of such a mechanism, due to $\lambda_{iS} \ll 1$.}

Let us summarise how dark matter predictions come to depend on reheating outcomes. For a given inflation scenario, and non-minimal coupling regime, there exists a particular value for the axion decay constant in VISH$\nu$ ($f^{res}_{A}$) below which the PQ symmetry is restored during reheating. For the region of parameter space $f_A < f^{res}_{A}$, there is only one value ($f^*_{A}$) which entirely reproduces the observed CDM density, provided $f^*_{A} < f^{res}_{A}$. We can expect VISH$\nu$ to realise the final condition, in general, due to thermal and/or non-thermal processes once inflation ends.\footnote{The generality of this statement applies to the non-minimal coupling sizes which, with the exception of limiting cases, lead to $\Phi_iS$-Inflation or $\Phi_1\Phi_2S$-Inflation, so guaranteeing both a singlet component in the inflaton (typically responsible for the efficient growth of fluctuations in the axion field during a phase of preheating) and a Higgs component (typically responsible for a high reheating temperature).} However, a complete analysis of the VISH$\nu$ reheating phase is beyond the scope of this paper. In the opposite regime ($f_A > f^{res}_{A}$), the PQ symmetry remains broken and there can be a range of $f_A$ values which respect the CDM constraint, up to a maximal value ($f^{iso}_A$) where isocurvature bounds are violated, provided $f^{iso}_A > f^{res}_{A}$. This requires a mechanism to suppress axion fluctuations, which can be fulfilled by a generic subset of the inflation scenarios, assuming the symmetry is not restored. We now consider both scenarios (restored and non-restored) in the context of VISH$\nu$ predictions for dark matter. 

\subsection{Restored $U(1)_{PQ}$ scenario}

In the scenario where the PQ symmetry is restored and PQ breaking occurs for the last time \textit{after} inflation, the predicted axion abundance depends only on a calculable axion mass~\cite{GorghettoAxions} which should not overproduce the measured CDM density and ideally reproduces it.  
In combination with astrophysical results, namely stellar cooling fits which constrain axion interactions, we can determine a representative range, see (\ref{vishnuaxionrange}) below, of viable VISH$\nu$ \emph{post-inflationary} axion masses.

One contribution to axion-production arises from the ``vacuum misalignment'' mechanism~\cite{MisAlign1,MisAlign2,MisAlign3}. The random selection of initial ``misalignment angle'' field-values in many distinct domains populating the universe leads to an averaged axion CDM contribution~\cite{RingwaldPostInf}:
\begin{equation}\label{RealignmentMechanism}
    \Omega^{\text{VR}}_A h^2 \simeq (3.8  \pm 0.6) \times 10^{-3} \left(\frac{f_A}{10^{10}\ \text{GeV}} \right)^{1.165}.
\end{equation}
Na\"ively, this gives rise to the totality of dark matter when $m_A \sim 28(2)\ \mu$eV~\cite{Borsnyi2016CalculationOT}, or equivalently $f_A \sim 2 \times 10^{11}$ GeV. However, there are additional sources of axion density, even besides thermal processes, which owe to the decay of one and two-dimensional topological defects that are not inflated away. The spontaneous breakdown of $U(1)_{\text{PQ}}$ results in the formation of cosmic strings which radiate axions as they evolve~\cite{Davis}. While this can furnish a second contribution to the CDM abundance, it is afflicted by large theoretical uncertainties which lead to a range of lower limits for $m_A$ below. Later, for temperatures around the QCD phase transition, instanton effects not only give the axion a mass, which at zero-temperature is~\cite{GorghettoTopSus}:
\begin{equation}\label{AxionMass}
    m_A \simeq 5.691(51)\left(\frac{10^{12}\ \text{GeV}}{f_A}\right)\ \mu
    \text{eV},
\end{equation} 
but also subsequently induce the formation of a network of domain walls bounded by axionic strings~\cite{Kawasaki2013}. As the VISH$\nu$ model has \emph{unit} domain wall number, the network is unstable~\cite{Vilenkin:1982ks} --- thus avoiding a domain wall problem --- and its ensuing decay into axions also generically contributes to the CDM density~\cite{Hiramatsu2012}. 

Assembling the three contributions, the \emph{full} observed CDM density is explained by the abundance of VISH$\nu$ axions of mass\footnote{The upper value is favoured by Refs.~\cite{HoofDomWalls,GorghettoAxions} and the lower value encloses a range favoured in Ref.~\cite{Buschmann2021}.} $m_A \sim (40$--$500)\, \mu \text{eV} $~\cite{HoofDomWalls,GorghettoAxions,Buschmann2021}, corresponding to values for the axion decay constant $f_A \sim (1.1 \times 10^{10}$--$1.4 \times 10^{11})\, \text{GeV}$.
To ensure that CDM is not overproduced beyond observational limits, we should understand this as the range of upper (lower) bounds for $f_A$ ($m_A$), allowing for theoretical uncertainties. Note that both bounds assume $N_{\text{DW}} = 1$, and the lower bound on the mass of the VISH$\nu$ axion is the same as that for minimal KSVZ models. 

Since interacting axions can accelerate energy transport out of stars, observed stellar cooling rates and lifetimes serve to constrain axion couplings to photons, electrons and nucleons. In the top-specific VISH$\nu$ model, the axion-photon coupling is the same as in the DFSZ-I model (see Appendix C), but may be enhanced in alternative flavour structures~\cite{Saikawa:2019lng,DiLuzio:2021ysg}. However, the coefficient of the axion-electron coupling ($C_{ae}$) is a factor of three larger~\cite{Saikawa:2019lng}, leading to rescaled lower bounds on $f_A$ from the white-dwarf luminosity function~\cite{MillerBertolami}, which becomes $f_A > (2.4 \times 10^9\ \text{GeV})\sin^2\beta$ and the tip of the red-giant branch constraint~\cite{Straniero:2020iyi,Capozzi2020}, which becomes $f_A > (\sim 3 \times 10^9\ \text{GeV})\sin^2\beta$, both slightly more stringent in the top-specific VISH$\nu$ model than for ordinary DFSZ models. 

The collection of anomalous stellar cooling hints discussed in Ref.~\cite{Giannotti:2017hny} may also be interpreted by a VISH$\nu$ model. An analysis of variant axion models in Ref.~\cite{Saikawa:2019lng}, suggests that the tree-level top-specific structure both accommodates WD, RGB and HB hints, and abides by the SN 1987A bound~\cite{Giannotti:2017hny} for the $2\sigma$ range of axion masses $m_A \in (0.46, 29)\, \text{meV}$ compatible with perturbativity.\footnote{We expect that the $R_2$ bounds discussed in Ref.~\cite{Dolan:2022kul} may further constrain this parameter space.} Interestingly, there is an overlap  around $0.5\, \textrm{meV}$ with the mass range, identified above, for which top-specific VISH$\nu$ axions explain the totality of dark matter. In comparison, the hints are not well-interpreted by KSVZ models, but may be in extensions by right-handed neutrinos due to loop-induced contributions to the axion couplings at the cost of a tension with perturbativity~\cite{Giannotti:2017hny,Saikawa:2019lng}. 

Taken together, cosmological and astrophysical constraints point to the following representative mass range for the post-inflationary VISH$\nu$ axion:
\begin{equation}\label{vishnuaxionrange}
     m_A\ \in\ ( 40\ \mu\text{eV},\ \sim 2\ \text{meV}),
\end{equation}
where, for the upper limit, we take $\tan \beta \gtrsim 8$, required by the natural explanation for leptogenesis~\cite{nu2HDM,nuDFSZ}, in the red giant bound mentioned above. Note that the lower bound\footnote{We recall that the lower bound is na\"ively as low as $28(2)\mu$eV if topological defects do not significantly contribute to axion CDM.} is comparable to SMASH~\cite{Ballesteros:2019tvf} (when updated using recent simulations of the topological defect contributions) while the generally $\beta$-dependent upper bound is DFSZ-like.

Of the direct detection experiments which probe the $\mu$eV ballpark for the axion mass in the QCD band, only ADMX has reached the sensitivity of the DFSZ axion~\cite{ADMX:2018gho,ADMX:2019uok,ADMX:2021nhd}, but for a mass range lighter than the  VISH$\nu$ prediction. In the years to come, our representative mass window (\ref{vishnuaxionrange}) may eventually be accessible to searches including, in addition to ADMX, MADMAX~\cite{Caldwell2017}, CAPP~\cite{Semertzidis:2019gkj}, HAYSTAC~\cite{HAYSTAC:2020kwv}, ALPHA~\cite{Lawson2019}, ORGAN~\cite{Quiskamp:2022pks} and others. Particularly if the Vissani-like bound on $\tan \beta$ can be somewhat relaxed, as suggested in Ref.~\cite{nu2HDM}, a larger VISH$\nu$ axion mass may also be in the projected reach of the helioscopes IAXO~\cite{IAXO:2019mpb} and IAXO+.

\subsection{\label{nonrestored}Non-restored $U(1)_{PQ}$ scenario}

We now consider the scenario where the PQ symmetry, broken during inflation, fails to be restored and discuss how the \textit{pre-inflationary} VISH$\nu$ (and $\nu$DFSZ) axion may still abide by CMB isocurvature constraints while reproducing the measured CDM density. 

In order for the non-restored case to be compatible with the high-scale inflation model we have analysed in Sec. \ref{sec:sec3}, the axion fluctuations during inflation must be suppressed and there are two ways this can be achieved in the model parameter space, distinguished by the composition of the inflaton field. The first case, which is more natural, consists in scenarios where there is a Planckian displacement of $\sigma$ during the inflation (as in $\Phi_1\Phi_2S$-Inflation, $\Phi_iS$-Inflation and $S$-Inflation). This induces $f_{A,\text{inf}} \gg f_A$ and there is a signficant suppression of axion isocurvature absent in the usual case~\cite{Fairbairn:2014zta,Ballesteros:2016xej,Boucenna:2017fna}. Such a scenario respects constraints for $f_A \lesssim 10^{14}$ GeV~\cite{Ballesteros:2016xej}, compatible with cosmological and astrophysical bounds. However, the growth of axion fluctuations during pre-heating should also be properly accounted for and this tends to restore the PQ symmetry unless $f_A$ is considerably larger~\cite{Ballesteros:2021bee}. 

The second case to consider is $\Phi_1\Phi_2$-Inflation (recall that $\Phi_i$-Inflation can satisfy the PQ-restored scenario). The required suppression of axion fluctuations can occur if the axion obtains a large inflationary mass which exceeds the Hubble scale. Remarkably, Ref.~\cite{Nakayama2015} have shown that this situation can arise in the DFSZ model through the $\epsilon$-coupled term, when the inflaton field is supplied only by the Higgs sector. Then, during the inflation, the axion mass exceeds the Hubble scale to suppress fluctuations, while the Higgs pseudoscalar is almost massless but develops harmless fluctuations. This mass relationship is reversed after inflation, because the Higgs are charged under $U(1)_{\text{PQ}}$, and the background Higgs field values subside. However, this explanation requires that inter-sector couplings exceed the tiny values, required to suppress tree-level corrections to the electro-weak scale, by many orders of magnitude, and so fails the strict naturalness criteria we adopt in Sec. \ref{sec:sec5}. This is not very serious for our model, as $\Phi_1\Phi_2$-Inflation only occurs for a small subset of the parameter space and non-thermal restoration is difficult to avoid~\cite{Ballesteros:2021bee}.

As axion contributions from the density of axion strings, formed when PQ symmetry is broken for the last time, are diluted away during the inflation, the only contribution to the axionic CDM abundance we need to consider is from the vacuum misalignment mechanism. In particular, there is no longer a domain wall problem to be overcome, and the pre-inflationary scenario is compatible with the ($N_{DW} \neq 1$) DFSZ axion of the $\nu$DFSZ. The explanation, however, differs to the post-inflationary case considered above, since the initial randomly-selected misalignment angle $\theta_i \in [-\pi,\pi]$ is now the same in our Hubble volume (and not averaged over, as other values are inflated away). Hence, the resulting CDM axion abundance~\cite{Borsnyi2016CalculationOT,ParticleDataGroup:2020ssz} depends also on $\theta_i$ (and a factor $F$ defined in Ref.~\cite{ParticleDataGroup:2020ssz}):
\begin{equation}\label{PreInfRealignmentMechanism}
    \Omega^{\text{VR}}_A h^2 \simeq 0.12 \left(\frac{f_A}{9 \times 10^{11}\ \text{GeV}} \right)^{1.165} F\theta_i^2.
\end{equation}
It follows that, for $F\theta_i^2 \sim 1$, the pre-inflationary axion can account for the totality of the measured CDM density for $f_A \sim 10^{12}$ GeV. As discussed in Ref.~\cite{nuDFSZ}, re-interpreting this as an upper bound preventing CDM over-production, smaller values of $f_A$ are permitted in so far as they are astrophysically viable --- including values which are relevant to stellar evolution considered above. Note that larger $f_A$ are possible under the assumption of smaller initial values $\theta^2_i$, provided they respect $f^{\text{res}}_A < f_A < f^{\text{iso}}_A$.

\section{Naturalness}
\label{sec:sec5}
In this section, we first review our naturalness philosophy and how our model obviates any new naturalness problems posed by the non-gravitational interactions of its hidden sector, and remains fully consistent with vacuum stability. We also address the particular naturalness challenges posed by the inflation, which arise for general values of non-minimal couplings, $\xi_k$. This is a non-trivial consideration, since some of the inflation scenarios we have considered could provoke a ``unitarity problem'' in the large $\xi_k$ limit of VISH$\nu$.

\subsection{Our naturalness philosophy}
\label{sec:hiddensectorphil}

A famous naturalness criterion, originally due to 't Hooft~\cite{tHooft:1979rat}, is that of \emph{technical naturalness}. It says that a parameter $\lambda$, \emph{or} a relation among model parameters $f(\lambda_I)$, can take a value\footnote{For parameters, $\mu_I$, of general mass dimension $d$, we define: $\mu_I \equiv \lambda_I \Lambda^d$, where $\Lambda$ is the fundamental scale and $\lambda_I$ are the dimensionless arguments of a (possibly trivial) relation $f$.} $|\lambda| \ll 1$ provided that $\lambda \rightarrow 0$ results in a new symmetry of the quantum theory. The upshot of such a postulate is that a small value for $\lambda$ is \emph{radiatively stable}, that is, quantum corrections do not drive $|\lambda|$ to orders-of-magnitude larger values. This is because when $\lambda = 0$ is set at tree level, radiative corrections cannot generate a nonzero value for it, because those corrections must obey the symmetry of the quantum theory. So, corrections to $|\lambda| \ll 1$ will only ever be \emph{proportional} to that small symmetry-violating parameter relation and thus themselves small.

The smallness of $|\lambda|$, which may be required for phenomenological reasons at a particular scale, thus becomes a consistent feature of the model across the entire regime of validity, as it is stable under renormalisation group evolution. We imagine that the model is replaced in the far UV by a more fundamental one --- as it surely must be to include gravity --- which ascribes a dynamical origin to the hierarchy of parameter sizes. Nevertheless, the particular origin of $|\lambda| \ll 1$ need not concern us in the IR of such a model, since the enhanced symmetry structure already furnishes \emph{a priori} justification.

However, this is merely a sufficient condition for naturalness in the sense of radiative stability. We now introduce an extension of this principle, which we call \emph{second-order technical naturalness}. We argue that a parameter $\lambda$, or a relation among model parameters $f(\lambda_I)$, can take a natural value $|\lambda| \ll 1$ provided that (i) $|g(\lambda_I)| \ll 1$ guarantees the radiative stability of $\lambda$ and (ii) $g(\lambda_I) \rightarrow 0$ results in a new symmetry of the quantum theory, where $g$ is another relation. Or, what is the same, $|g(\lambda_I)| \ll 1$ is technically natural and this is inherited by $\lambda$.

As hidden-sector extensions of the SM, the VISH$\nu$ and $\nu$DFSZ models come with a set of technically natural, ``inter-sector'' (or ``sector-mixing'') parameters: $y_\nu$, $\lambda_{1S}$, $\lambda_{2S}$ and either $\kappa$ or $\epsilon$. This is because, in the limit of sector decoupling, in which every inter-sector coupling is set to zero, there is an enhanced Poincar\'e symmetry,\footnote{This is only true in the limit that gravity is neglected (see footnote~\ref{foonote1}), because this trivially forbids the possibility of any surviving sector-mixing gravitational interaction, which must be treated separately. We note that a variation of this remaining issue already afflicts the SM.} which is manifest at the level of the action~\cite{Foot:2013hna}:
\begin{equation}
    S = S_{\text{SM}} + S_{\text{HS}} = \int \rd^4 x\ \mathcal{L}_{\text{SM}}(x)+ \int \rd^4 x'\ \mathcal{L}_{\text{HS}}(x'),
\end{equation}
where $S_{\text{SM}}$ is the electroweak scale action, $S_{\text{HS}}$ is the $f_A \sim 10^{11}$ GeV hidden sector ($\{S,\nu_R\}$) action and the inter-sector Lagrangian has disappeared. Observe that we can now make two independent Poincar\'e transformations of the spacetime  which leave the sector actions, $S_{\text{SM}}$ and $S_{\text{HS}}$, separately conserved~\cite{Foot:2013hna}. So, there is a custodial symmetry which automatically protects weak inter-sector interactions. Recall that there is also a custodial $U(1)$ in the limits $y_\nu, y_N, \kappa \rightarrow 0$, which separately protects small, technically natural values of each parameter~\cite{nuDFSZ}. 

In the plain SM, excluding potentially dangerous gravity effects, there is no naturalness problem. This is because there is now no other high physical scale which induces destabilising contributions to the Higgs mass parameter, $m_h$. Although $m_h \rightarrow 0$ is not a symmetry of the quantum theory, it does realise \emph{classical} scale invariance, resulting in only harmless logarithmic running of $m_h$, with the $\beta$-function vanishing in the $m_h \to 0$ limit. On the other hand, the introduction of a new heavy fundamental scale to solve the strong $CP$ problem and furnish axionic dark matter \emph{does} generically provoke a concern over naturalness, beyond existing concerns over gravity. In this situation, $m_h \rightarrow 0$ no longer realises classical scale invariance, owing to the presence of the new scale. The failure of the Higgs mass to be technically natural means some other way must be found of suppressing the accumulation of large physical corrections induced by inter-sector couplings of the SM to the new high-scale physics, which could serve to drive $m_h$ unnaturally large values, well above the electroweak scale ($\delta m^2_{h} \gg m_{h}^2$).

In VISH$\nu$, and the $\nu$DFSZ model, $m_{ii}^2$ is \emph{prima facie} vulnerable to large $ \sim v_S^2$ corrections from $\Phi_i$ couplings to the heavy sector. However, since all inter-sector couplings take values which ensure that $\delta m^2_{ii} \lesssim m_{ii}^2$, the Higgs mass parameters \emph{are} radiatively stable. If these custodial parameter values were assumed by fiat at a particular scale, but took general values elsewhere, this would ordinarily require fine-tuning. However we have seen that precisely this set of values is technically natural in the $\nu$DFSZ and VISH$\nu$, owing to a custodial Poincar\'e symmetry. 

Hence, while the Higgs mass parameters are not technically natural at first-order, our solution \emph{in the presence} of high-scale physics, is that they are technically natural at second-order: inheriting the technical naturalness of the inter-sector parameters. We may therefore call the electroweak sector ``Poincar\'e protected''~\cite{Foot:2013hna}. Note the importance of $S$ being a gauge-singlet, since additional tree-level couplings of the hidden sector to the SM can reintroduce large inter-sector couplings at higher loop order. That is, the decoupling limit must \emph{fully} decouple the heavy sector, as it does in both VISH$\nu$ and the $\nu$DFSZ. Note too that provided there is no intrinsic naturalness-violating hierarchy in the decoupled sector actions, the mass parameters of each are again subject to a custodial classical scale invariance that results in harmless logarithmic running about each fundamental scale. 

Ref.~\cite{nuDFSZ} gives a detailed analysis and demonstration of the naturalness of the $\nu$DFSZ model under RG evolution, in the sense just described, which we will not reproduce here for VISH$\nu$ as there are no important differences. On the other hand, the radiative stability of some fundamental parameters in theory must be reconsidered in light of parameter choices which permit viable or predictive inflation; we do this below. 

\subsection{Naturalness of the inflation model}

Since our results in Sec. \ref{sec:sec3} did not depend on the inter-sector couplings at leading order, the radiative stability of the electroweak scale is a consistent feature of the model supplemented by a viable period of inflation, as laid out in Sec. \ref{sec:sec3}.

For each of the seven inflation scenarios, (\ref{AsFit}) furnishes hierarchical constraints on the quartic parameters $\lambda_J$ of each sector Lagrangian as compared to the sizes of the non-minimal couplings $\xi_I$. One way the model can become unnatural is if, in some viable region of parameter space, the $\lambda_J$, or some relation between them (e.g. $L = \lambda_1\lambda_2 - \lambda^2_{34}$), must take a very small value without \emph{a priori} justification. If this is true, our naturalness expectations mean that such a scenario is ruled out in VISH$\nu$.  Since large values of the non-minimal couplings, $\max_I \{ \xi_I \} \in [10^3,10^4]$ satisfy (\ref{AsFit}) for naturally-sized $|\lambda_J| \in (0.01,1)$, this represents a regime where there is no tree-level conflict between our inflation predictions and our expectations for naturalness. 

\subsubsection{Small non-minimal couplings}

As we have demonstrated in Sec. \ref{smallxifit} for $\xi_i \ll \xi_S$ and $\xi_i \simeq \xi_S$, smaller values of $\xi_k$ additionally abide by constraints on primordial fluctuations. However, the case where $\xi_S \ll \xi_i$ is unnatural: the viability of Higgs-sector $\Phi_i$-Inflation and $\Phi_1\Phi_2$-Inflation scenarios requires that $\lambda_i, L \lesssim 10^{-8}$. Since the limits $\lambda_i, L \rightarrow 0$ do not correspond separately to enhanced symmetries of the theory, they are not technically natural limits. As neither can be made technically natural at second-order, the radiative stability of their small values cannot be guaranteed in the model. Hence, these three inflation scenarios meet our naturalness expectations only for $\xi_i \gg 1$.  

The remaining four possibilities --- $S$-Inflation, $\Phi_i S$-Inflation $\Phi_1 \Phi_2 S$-Inflation --- can all meet our naturalness expectations, even in a limited regime of small $\xi_k$. This is because, in the limit $y_N \rightarrow 0$, small values of $\lambda_S$ become natural at second order, because they inherit the technical naturalness of the inter-sector couplings in the decoupling limit. Hence, a window of radiative stability is opened up for small values of $\lambda_S$, the range of which depends on the values of $y_N$, which can take small values protected by a custodial $U(1)$ symmetry. So, while small $\lambda_S$ meets the criteria for second-order technical naturalness \emph{a priori}, we must also note that the right-handed neutrino Yukawas must still be sufficiently large to permit leptogenesis.

This can be seen through the RGE for $\lambda_S$, which to 1-loop reads: 
\begin{equation}
    \frac{\rd \lambda_S}{\rd \log \mu} \simeq \frac{1}{16\pi^2}
\left(10 \lambda_{S}^{2}
+ 2 \sum_i \left[\lambda_S(y^i_N)^2
- (y^i_N)^4\right] \right),
\end{equation}
and we suppress the contributions $\sim \lambda_{iS}^2$ which are at most $O(10^{-32})$. To ensure the radiative stability of a parameter $X$, namely that it is not driven to very different values under running from, say, the electroweak scale to the Planck scale, it suffices that:
\begin{equation}
    \left| \frac{\rd \log_{10} X}{\rd \log_{10} \mu} \right| = \left|\frac{1}{X}\frac{\rd X}{\rd \log \mu}\right| < \mathcal{O}(0.1),
\end{equation}
since $[\log_{10} (M_P/m_Z)]^{-1} \sim \mathcal{O}(0.1)$. As much, for example, is guaranteed for technically natural parameters, since the loop factor is $\mathcal{O}(0.1)$ and corrections are only ever \emph{proportional} to the symmetry-violating parameters (a glance at the RGEs for the inter-sector parameters in Appendix B makes this clear). Hence, an estimate for the right-handed neutrino Yukawa sizes permitted by the radiative stablity of $\lambda_S>0$ is $\sum_i (y_N^i)^4 < \lambda_S$. To discover a range of naturally small $\lambda_S$, we should therefore identify the range of $y^i_N$ deemed viable in the natural solutions to the origin of neutrino masses and baryogenesis found in Ref.~\cite{nuDFSZ}.

Much like $v_2$, the lightest right-handed neutrino mass, $M_1$, comes to be constrained by its role in leptogenesis. It has a Vissani-like upper bound and a Davidson-Ibarra lower bound, the latter being adapted for the two-Higgs doublet scenario~\cite{nuDFSZ}, yielding $10^5$~GeV $\lesssim M_1\lesssim10^7$~GeV. Our reproduction of the BAU employs hierarchical right-handed neutrino masses $M_1 \ll M_2,M_3$, setting aside the quasi-degenerate case~\cite{Buchmuller:2002rq} which usually requires some resonant enhancement~\cite{Pilaftsis:2003gt}. However, a Vissani-like upper bound still applies to the second and third generation right-handed neutrino masses~\cite{nu2HDM}, thus $M_2,M_3~\lesssim~10^7$~GeV. So, since $y^{2,3}_N \sim M_{2,3}/f_A \sim 10^{-4}$--$10^{-3}$, we expect that values $\lambda_S \gtrsim 10^{-13}$ are radiatively stable using our estimated neutrino Yukawa bound. This was confirmed under RG running for standard values of input parameters. The most stringent naturalness bound on $\lambda_S$, therefore, still permits the full range of viable values for $\xi_S$.

\subsubsection{Stability of the singlet direction}

As a direct application of the inherited radiative stability of $\lambda_S$ we show that the bounded-from-below condition ($\lambda_S>0$) in the $S$-direction is also preserved under radiative corrections away from the minimum of the potential, to accommodate small positive $\lambda_S$ at large inflationary field values. This can be seen using the effective Coleman-Weinberg~\cite{Coleman:1973jx} quartic potential in the $S$ direction, for which the largest radiative contribution to the effective quartic coupling, denoted $\Hat{\lambda}_S$, owes to the largest right-handed neutrino Yukawa, $y^{\text{max}}_N$, at first loop order. Hence, with $k = 1/16\pi^2$,
\begin{equation}
    \Hat{\lambda}_S(\sigma_{*}) \sim \lambda_S(\mu) - \mathcal{O}(k) \left(y^{\text{max}}_N\right)^4 \log\left(\sigma_*/\mu\right),
\end{equation}
where, as in Ref.~\cite{Ballesteros:2016xej},
we can set $\mu \sim m_\sigma$, the mass of the PQ scalar modulus (a renormalisation point relevant to the potential minimum), and $\sigma_*$ is the large inflationary field-value. Additionally, we use the fact that the $y_N$ evolve slowly, which is guaranteed by our custodial $U(1)$ symmetry, so that our neutrino mass bounds are not greatly modified under running. Even with a background field value $\sigma_* \sim \mathcal{O}(10)M_P$, $y^{\text{max}}_N < 10^{-3}$ is sufficient to keep $\Hat{\lambda}_S > 0$, and the correction smaller in magnitude than $\lambda_S \gtrsim 10^{-11}$ relevant to viable small $\xi_k$ inflation. Unlike Ref.~\cite{Ballesteros:2016xej}, we are therefore able to achieve the ensuing stability in the $S$-direction, along with a natural explanation of leptogenesis, without admitting degeneracy in the masses of the right-handed neutrinos.

\subsubsection{Stability of the Higgs sector}

The extended scalar sector of VISH$\nu$ evades other stability concerns addressed, for example, in Refs.~\cite{Ballesteros:2016euj,Ballesteros:2016xej,Ballesteros:2019tvf}, without compromising the hidden-sector structure. In particular, having demonstrated the stability of the S direction due to the second-order technical naturalness of $\lambda_S$, we must also consider the vacuum stability of the Higgs sector.

For current measured values of the Higgs and top masses, it is highly probable that the electroweak vacuum of the SM Higgs potential is metastable~\cite{Degrassi:2012ry}. This can pose a threat to the validity of the Higgs inflation models, since it is possible that the instability scale, at which the Higgs quartic coupling runs negative, may not be sufficiently large to accommodate the high-scale inflation model with Planckian field values. 

Let us understand how this concern may be ameliorated by the addition of an extra scalar state, as in Ref.~\cite{Ballesteros:2016xej}. The scalar potential of the ($N_{DW} = 1$) minimal KSVZ model~\cite{Kim:1979if,Shifman:1979if} can be written as:
\begin{equation}\label{KSVZpot}
\begin{split}
    V =\ &\lambda_H \left(H^\dagger H - v^2/2\right)^2 + \lambda_S \left( S^* S- v_S^2/2 \right)^2\\ 
    &+ 2\lambda_{HS}\left(H^\dagger H - v^2/2\right)
    \left(S^* S- v_S^2/2 \right),
\end{split}
\end{equation}
where $H$ is the Higgs field and $S$ is the gauge singlet complex scalar whose phase is the axion, which obtain the vevs $v \simeq 246$ GeV and $v_S = f_A$, giving rise to the Higgs mass parameter $M_H^2 = - \lambda_H v^2 - \lambda_{HS}v_S^2$. Furthermore, we will assume that $f_A \sim 10^{10}$--$10^{11}$ GeV, as in Ref.~\cite{Ballesteros:2016xej}. Once $|S|$ is integrated out to realise the SM Higgs potential, one requires~\cite{ParticleDataGroup:2020ssz}:
\begin{equation}\label{ksvzmass}
    m_H^2 = M^2_H + \lambda_{HS} v_S^2 \simeq -(88\ \text{GeV})^2,
\end{equation}
and for tree-level naturalness, we argue that the inter-sector coupling satisfies $|\lambda_{HS}|~\lesssim~10^{-18}$ to prevent a tuning.

In order to forbid unstable regions of the potential during the cosmological evolution, one can require that the bounded-from-below conditions are valid at high scales, and (\ref{KSVZpot}) is positive away from the absolute minimum. The point of this criterion is to eliminate the possibility that unsuppressed fluctuations developed during inflation, or thermally-sourced for very high reheating temperatures~\cite{DelleRose:2015bpo}, will harmfully probe or induce tunnelling into the unstable regions of the potential, even if the inflation na\"ively proceeds along a ``stable'' direction to a long-lived vacuum~\cite{Ballesteros:2016xej}. However, the inter-sector coupling $\lambda_{HS}$ --- which leads to positive contributions to the $\beta$ function of $\lambda_H$ that could help to counteract the destabilising effect of the top Yukawa --- can often be quite small, making the loop-level stabilisation difficult to achieve (and certainly unnatural). A remaining avenue to preserve $\lambda_H > 0$ is so-called threshold stabilisation~\cite{Lebedev:2012zw,Elias-Miro:2012eoi} (see also Ref.~\cite{Ballesteros:2016xej}) which can outsize the previous effect by leveraging the inter-sector tree-level contribution to induce a matching condition for $\lambda_H$, below the instability scale, that keeps it positive at high-scales. In Ref.~\cite{Ballesteros:2016xej}, it is found that this typically requires $|\lambda_{HS}| \gtrsim 10^{-7}$ at best, for the smallest values of $\lambda_S$ consistent with viable inflation in the $S$ direction. In comparison to the tree-level naturalness bound, this is sufficient to spoil the electroweak scale protection \emph{a priori} achievable in the model if an SM-like vacuum stability is accepted.

Let us turn our attention now to the top-specific VISH$\nu$ model. Setting aside the possibility of threshold stabilisation by the heavy sector, the need for additional positive contributions from the electroweak sector to the running of the Higgs quartic couplings $\lambda_{1,2}$ can be met by (and motivates) the extension of the Higgs sector to include an additional Higgs doublet. In general, this ensures the $\beta$-functions for $\lambda_{1,2}$ more easily become positive at high scales, due to the radiative accumulation of additional scalar couplings, whose sizes must nevertheless respect perturbativity bounds. In particular, Ref.~\cite{Oda:2019njo} demonstrates that (i) the absence of Landau poles, (ii) vacuum stability and (iii) the high-scale validity of the bounded-from-below conditions, can be met for a Poincar\'e-protected DFSZ Higgs sector\footnote{Since, as remarked in Sec. \ref{sec:sec2}, the decoupled Higgs sector is essentially the Type II 2HDM model with $U(1)_{PQ}$-limited charges, Refs.~\cite{Chakrabarty:2016smc,Chakrabarty:2017qkh,Branchina:2018qlf,Krauss:2018thf,Basler:2017nzu} are additionally relevant.} up to the Planck scale. Although with rather restrictive implications for the parameter space, as explained in Ref.~\cite{Oda:2019njo}, these predictions can still, quite remarkably, remain compatible with results of collider searches and --- importantly ---  be met naturally~\cite{nu2HDM} by $|\cos(\beta - \alpha)| \ll 1$, with $\approx m_{22}$ masses of the heavy scalar states~\cite{nuDFSZ} and for a range of $\tan \beta$ permitting the natural explanation of leptogenesis discussed in Refs.~\cite{nuDFSZ,nu2HDM}, for which the Vissani-like bound protects the Higgs sector from sizeable right-handed neutrino Yukawa contributions. Additionally, the top-specific Yukawa structure we consider here should not give rise to important differences; nor the replacement of the $\epsilon$-coupled quartic term with the cubic $\kappa$-coupled tri-linear term required by a ($N_{DW} = 1$) VISH$\nu$ model, since both induce, and therefore realise, the same bound on $m^2_{12}$ by construction. Hence, our analysis of an early period of inflation driven by the Poincar\'e-protected top-specific VISH$\nu$ model is well-motivated.

\subsubsection{Unitarity}
\label{524}

The small $\xi_k$ regime of VISH$\nu$ (Sec. \ref{smallxifit}) not only leads to successful inflationary predictions but is, in principle, a valid description of physics up to a Planckian cutoff. However, large non-minimal couplings ($\max_I \{\xi_I\} \gg 1$) could introduce new sub-Planckian unitarity-violating scale, $\Lambda_{UV}$, and this should be interpreted as a cutoff for the validity of the model. This was pointed out in the context of SM Higgs inflation, e.g. by Refs.~\cite{Barbon:2009ya,Burgess:2009ea}, for which the measured Higgs quartic coupling $\lambda \sim \mathcal{O}(0.1)$ requires $\xi \sim \mathcal{O}(10^4)$, though the running of $\lambda$ to small (positive) values may tolerate smaller $\xi$. We recall that this is not the case for VISH$\nu$, where the quartic couplings are not as tightly constrained and we expect smaller values of the $\xi_I$ to be likewise viable ($\lambda_S \lesssim 10^{-8}$ realise the Planckian cutoff for small $\xi_k$). Nevertheless, given that our model additionally provides viable predictions throughout a regime of large $\xi_k$, we find it necessary to understand what consequences to predictivity, if any, follow from a new scale $\Lambda_{UV}$ in VISH$\nu$, but will not add anything new to the debate over unitarity problems for non-minimally coupled inflation models. 

There are additional operators in the Einstein frame Lagrangian of the Higgs inflation model which induce the cutoff $\Lambda^{\text{Higgs}}_{UV} = M_P/\xi$ for $\xi \gg 1$, in the present vacuum.\footnote{The same result can be obtained in the Jordan frame by expanding the metric around the flat background~\cite{Bezrukov_2013}.} In VISH$\nu$, the corresponding cutoff is
$\Lambda_{UV} = \min_{IJ}\{M_P/\sqrt{\xi_I\xi_J}\}$. This is comparable to $H_{\text{inf}}$, and below both the energy density scale $V^{1/4}$ and the non-negligible background modulus field values during the inflation. This could lead to legitimate concern over the predictivity of the large $\xi_k$ inflation model, following the arguments of Ref.~\cite{Barbon:2009ya,Burgess:2009ea}, but does not afflict $\max_k \{ \xi_k \} \lesssim 1$ as the cutoff breaches the Planck scale.

An alternative approach takes into account the classical inflationary background, which renders the cutoff field-dependent. In this interpretation, the cutoff in the Einstein frame has a larger value during the inflation which is $\sim M_P$~\cite{Bezrukov:2010jz,Bezrukov_2013} (and this falls to $\sim M_P/\sqrt{\xi}$ during reheating). This is above the inflationary Hubble scale, and greater than (or comparable to) the inflationary modulus field-values and energy density scale. It is commonly claimed, then, that unitarity concerns over the tree-level \emph{inflation} are alleviated in non-minimally coupled models, such as the one we have analysed in Sec. \ref{sec:sec3}, although this is not a universal view. 

Unitarity violation may also occur during the production of longitudinal gauge bosons in the early stages of reheating~\cite{Ema_2017,DeCross:2016cbs,Sfakianakis2019}, but the scope of this problem has been limited by Ref.~\cite{Hamada:2020kuy}. The subset of large $\xi_k$ regime of VISH$\nu$ inflationary scenarios involving a large-valued neutral Higgs component may also confront this issue. Introducing $R^2$ gravitational corrections to the gravi-scalar action can avoid this problem and reinstate the Planck-scale cutoff~\cite{EMA2017403}, which may be a well-motivated UV completion to the large $\xi_k$ model if the unitarity issue is shown to be serious.\footnote{Ref.~\cite{Lee:2021rzy} has analysed inflation with an $R^2$-corrected 2HDM and we expect analogous tree-level results for a mixed $R^2$ and VISH$\nu$ $\Phi_1\Phi_2$-Inflation scenario, as $U(1)_{\text{PQ}}$ merely limits possible terms. We also note that Palatini Higgs inflation avoids this problem~\cite{Rubio:2019ypq} (we work in the metric formulation).} The problem is avoided without further modification in the small $\xi_k$ model.

\section{Summary and Conclusion}
\label{sec:con}

The top-specific VISH$\nu$ model, and its flavour-variants, serve as an existence proof that \textit{weakly-coupled high-scale physics}, which does not introduce new destabilising contributions to an electroweak-scale Higgs mass, can nonetheless address important phenomenological and theoretical problems of the Standard Model (SM) in a unified and minimal way. This is achieved without eschewing technical naturalness of non-gravitational interactions and with an experimentally accessible dark matter candidate. 

The model inherits the explanatory framework of the $\nu$DFSZ, predicting: small active neutrino masses, BAU through thermal hierarchial leptogenesis, axionic dark matter and $CP$ invariant strong interactions; while removing a domain wall problem through an $N_{DW} = 1$ flavoured Yukawa structure. The present work extends (and refines) these predictions in the following ways. 

A period of cosmic inflation is invoked by the standard cosmological paradigm, but lacks a unique particle physics explanation. A simple model is the slow-roll evolution of a scalar inflaton field initially displaced from a potential minimum and this is successfully realised in the Einstein frame of a theory with direct coupling of a single scalar field, such as the SM Higgs, to Ricci scalar curvature. Realistic extensions of the SM, such as invisible axion models, necessitate a multi-scalar explanation, which can be fulfilled by non-minimal gravitational couplings ($\xi_k$) of the additional scalars.

While the two-scalar case has been extensively analysed elsewhere, we classify the possible inflaton configurations for a \textit{generic} non-minimally coupled three-scalar case for the first time.\footnote{This analysis is made possible by arguments in Refs.~\cite{Kaiser:2013sna,DeCross:2015uza} that the attractors of generic inflaton trajectories in this class of models can often be identified with valleys of the Einstein frame potential, which realise effectively single-field inflation. Our findings are more general than a three-scalar discussion in Ref.~\cite{Lebedev:2011aq}.} We find that restriction to the VISH$\nu$ parameter space places a meaningful limitation on the seven inflationary scenarios, enforcing that a large-field valley of the conformally stretched Einstein frame potential is only realised \textit{in general} when there is both an electroweak sector ($\Phi_i$) and hidden-sector ($S$) component. The three possibilities are $\Phi_1S$-Inflation, $\Phi_2S$-Inflation and $\Phi_1\Phi_2S$-Inflation, corresponding to the disjoint regions of parameter space: $\kappa_{12} > 0$, $\kappa_{21} > 0$ and $\kappa_{12},\kappa_{21} < 0$, respectively.\footnote{These are parameters we introduce in Sec. \ref{sec:sec3} and have the definitions: $\kappa_{12}= \lambda_{34}\xi_1 - \lambda_1\xi_2$ and $\kappa_{21}= \lambda_{34}\xi_2 - \lambda_2\xi_1$.} It is also possible for $S$ ($\xi_1, \xi_2 \simeq 0$) or a Higgs direction ($\xi_S \simeq 0$) to play the role of the inflaton at the expense of one or more light neutral modulus field directions (absent in the general case for $\xi_S \gtrsim 0.2$). 

We show that, in addition to a Higgs sector which can be valid at high-scales~\cite{Oda:2019njo}, the model non-trivially preserves the stability of the singlet direction for $\lambda_S > 10^{-13}$. This fact means that the non-minimal coupling sizes can be naturally extended to smaller magnitudes for $\Phi_1\Phi_2S$-Inflation, $\Phi_iS$-Inflation and $S$-Inflation, while matching the amplitude of CMB temperature anisotropies. Accordingly, we distinguish two regimes for the non-minimal couplings (large and small), obtaining predictions of the tensor-to-scalar ratio, scalar spectral index and its running, which are in excellent agreement with present exclusion limits from CMB observations in both regimes (and for intermediate values). In particular, large $\xi_k \gg 1$ reproduce the inflationary potential of the Starobinsky and Higgs inflation models. While for small $\xi_S \lesssim \mathcal{O}(1)$, we cure the non-minimal inflation model of (actively debated) unitarity concerns and remain within excellent agreement with CMB data for $\xi_S > 0.01$. The suppression of $\lambda_S$ means the magnitude of the $S$ inflaton component will generally dominate the Higgs component, while in the large $\xi_k$ regime the ratio is democratic. Our analysis of the small $\xi_k$ regime is performed for two limiting cases that suppress kinetic mixing of the inflaton and orthogonal directions: $\xi_i \ll \xi_S$, which results in $S$-Inflation only, as well as $\xi_i \simeq \xi_S$, resulting in $\Phi_iS$-Inflation or $\Phi_1\Phi_2S$-Inflation.

Finally, as a remarkable bonus, we find that the top-specific structure unites natural explanations for a large top quark mass and hierarchical thermal leptogenesis due to a small protected hierarchy of the Higgs vevs. We also show that the mixing of the PQ and L symmetries does not alter the variant axion-photon and axion-electron couplings beyond top-specific DFSZ predictions.

As the PQ symmetry is generally broken during the high-scale inflation, there are significant implications for dark matter predictions. While the full details of the post-inflation reheating phase remain a subject for further research, a high reheating temperature and restoration of the PQ symmetry are reasonable expectations following $\Phi_iS$-Inflation and $\Phi_1\Phi_2S$-Inflation, due to the generic admixture of PQ scalar and Higgs-sector components preserved in the inflaton background. This results in a post-inflationary axion scenario with a tightly constrained axion mass range $(40\mu\text{eV},\sim 2 \text{meV})$ within the projected reach of many searches. The lower limit, subject to theoretical uncertainties, is KSVZ-like and forbids CDM over-production, while the upper limit is DFSZ-like and is imposed by astrophysical constraints. The mass-range also permits an interpretation of stellar cooling hints. As the inflaton generally carries an $S$ component, the axion isocurvature fluctuations during the inflation will be sufficiently suppressed to rule-in the pre-inflationary axion for any values $f_A 
\lesssim 10^{14}$~GeV~\cite{Ballesteros:2016xej} which do not lead to restoration. This scenario leads to a viable mass-range which is less predictively constrained. 

The DFSZ-based VISH$\nu$ model, like its KSVZ cousin SMASH, provides a remarkably simple and economical way to solve the currently most important phenomenological problems of particle physics within a cosmology that implements successful inflation using exclusively the scalar fields required to meet particle physics objectives. 

\acknowledgments

This work was supported in part by the Australian Research Council through the ARC Centre of Excellence for Dark Matter Particle Physics, CE200100008. AHS is supported by
the Australian Government Research Training Program
Scholarship initiative. We thank Jackson Clarke for helpful correspondence.

\appendix

\section{Discussion of the inflationary kinetic-mixing}
\label{sec:AppendixA}

In this appendix, we study the relationship between the masses of fluctuations in directions orthogonal to the inflaton direction in the neutral modulus field-space, and the inflationary Hubble scale. This will require the suppression of kinetic-mixing between the fields parameterising each direction, so that they may be canonically normalised. Our ultimate objective is to identify which scenarios have only one light modulus direction (for which slow-roll is induced generically) and distinguish these from scenarios where there is an additional light direction (which are made possible by the approximate decoupling of the Higgs and hidden sectors). We recall that light modes have sub-Hubble masses. Our analysis of the two-scalar case follows discussions in Refs.~\cite{Gong:2012ri,Modak:2020fij}.

Let us first consider the large $\xi_k$ regime and use an ellipsoidal parametrisation (with $\xi_1,\xi_2,\xi_3 > 0$):
\begin{widetext}
\begin{equation}
\begin{split}
    \chi_1 = \frac{M_P}{\sqrt{\xi_1}}\ e^{\frac{\chi_r}{M_P\sqrt{6}}} \cos \phi' \sin \theta',\quad
    \chi_2 = \frac{M_P}{\sqrt{\xi_2}}\ e^{\frac{\chi_r}{M_P\sqrt{6}}} \sin \phi' \sin \theta', \quad 
    \chi_3 = \frac{M_P}{\sqrt{\xi_3}}\ e^{\frac{\chi_r}{M_P\sqrt{6}}} \cos \theta';
\end{split}
\end{equation}
\end{widetext}
where $\chi_r \equiv \sqrt{\frac{3}{2}} M_P\log (\frac{\xi_1}{M_P^2} \chi_1^2 + \frac{\xi_2}{M_P^2} \chi_2^2 + \frac{\xi_3}{M_P^2})$. 

This brings the kinetic terms, $G'_{IJ} \partial \chi'^I \partial \chi'^J$ where $\chi'^I = (\chi_r,\phi', \theta')$, into an appealing form:

\begin{widetext}
\begin{equation}
    G'_{IJ} = 
    \pmat{1 + \frac{1}{6}\left[\frac{1}{\xi_1} c^2_{\phi'} + \frac{1}{\xi_2} s^2_{\phi'} \right]s^2_{\theta'} + \frac{1}{6\xi_3}c^2_{\theta'}\   & \frac{1}{\sqrt{6}}\left[\frac{1}{\xi_2} - \frac{1}{\xi_1}\right] c_{\phi'}s_{\phi'}s^2_{\theta'}  & \frac{1}{\sqrt{6}}\left[\frac{1}{\xi_1} c^2_{\phi'} + \frac{1}{\xi_2} s^2_{\phi'} - \frac{1}{\xi_3} \right] c_{\theta'}s_{\theta'} \\[1em]
    \frac{1}{\sqrt{6}}\left[\frac{1}{\xi_2} - \frac{1}{\xi_1}\right] c_{\phi'}s_{\phi'}s^2_{\theta'} &  \left[\frac{1}{\xi_2} c^2_{\phi'} + \frac{1}{\xi_1} s^2_{\phi'} \right]s^2_{\theta'}& \left[\frac{1}{\xi_2} - \frac{1}{\xi_1} \right] c_{\phi'} s_{\phi'} c_{\theta'}s_{\theta'} \\[1em]
    \frac{1}{\sqrt{6}} \left[\frac{1}{\xi_1} c^2_{\phi'} + \frac{1}{\xi_2} s^2_{\phi'} - \frac{1}{\xi_3} \right] c_{\theta'}s_{\theta'} & \left[\frac{1}{\xi_2} - \frac{1}{\xi_1} \right] c_{\phi'} s_{\phi'} c_{\theta'}s_{\theta'} &  \left[\frac{1}{\xi_1} c^2_{\phi'} + \frac{1}{\xi_2} s^2_{\phi'} \right]c^2_{\theta'} + \frac{1}{\xi_3} s^2_{\theta'}}.
\end{equation}
\end{widetext}
If we consider the limit $\xi_k \rightarrow \infty$, the radial field ($\chi_r$) is canonically normalised and all other components are exactly zero. Retaining $\mathcal{O}(\xi_k^{-1})$ terms, a field transformation must be made to bring $G'_{IJ}$ into a canonical form.
As we are interested in the effective mass of perturbations about the false vacua of the leading order potential (\ref{largesurfacesphere}), which are orthogonal to the radial time-evolution parametrised by $\chi_r$, we define the shifted angular fields $\phi' = \overline{\phi}' + \delta \phi' /M_P $ and $\theta' = \overline{\theta}' + \delta \theta' /M_P $. We denote by $(\overline{\phi}', \overline{\theta}')$, the background field values attained at a given critical point of (\ref{largesurfacesphere}), and by $(\delta \phi',\delta \theta')$, the massive first-order perturbations. 

Let us consider the simplest case where a hyper-valley is aligned with an axis in the $\{\chi_k\}$ basis, so the candidate inflaton field is proportional to the single modulus field parametrising the large-field radial displacement along the axis. We choose the $\chi_1$ axis without loss of generality, as results generalise under a permutation of indices. The background angular field-values are thus $\phi' = 0$ and $\theta' = \pi/2$, bringing the kinetic terms into a diagonal form:
\begin{equation}
    \mathcal{L} \supset \frac{1}{2} (\partial {\chi_r})^2 + \frac{1}{2\xi_2} [\partial ( \delta \phi' )]^2 + \frac{1}{2\xi_3} [\partial ( \delta \theta' ) ]^2
\end{equation}
where we retain only the leading order term for $G_{\chi_r\chi_r}$ (we only need to assume $\xi_1 \gg 1$) and we rescale the angular perturbations ($\delta \phi'' = \delta \phi' / \sqrt{\xi_2}$, $\delta \theta'' = \delta \theta' / \sqrt{\xi_3}$) to achieve the desired canonical normalisation. The effective mass terms for the perturbations arise from the quadratic part of the perturbative expansion of the leading-order potential around the background angular fields:
\begin{equation}
    \Lambda^{(2)}[\delta \phi'', \delta \theta''] \supset \frac{\kappa_{12} }{2\xi_1^2} M_P^2 (\delta \phi'')^2 + \frac{\kappa_{13} }{2\xi_1^2} M_P^2 (\delta \theta'')^2
\end{equation}
As the inflationary Hubble scale predicted for this regime is $H_{\text{inf}} \sim 10^{13}\  \text{GeV}$, the masses of the rescaled, orthogonal modes exceed $H_{\text{inf}}$ when both $\kappa_{12}, \kappa_{13} \gtrsim (\xi_1/10^5)^2$.

In the context of the VISH$\nu$ model, $\Phi_i$-Inflation ($\xi_S \simeq 0$) it is possible to find a generic region of parameter space where the orthogonal Higgs direction is heavy, however the orthogonal direction associated to the modulus of the PQ scalar is generically light (and positive for $\lambda_{iS} > 0$), since $\lambda_{iS} \ll 10^{-10}\xi_i$. In $S$-Inflation, both the orthogonal Higgs directions have sub-Hubble masses. To ensure the motion is stabilised in the hyper-valley it suffices to assume that initial field velocities in the orthogonal directions are negligible, and the fields are initialised close to the attractor, which is a more finely-tuned scenario than the general case, where these requirements are absent. For these cases, the dynamics in the adabiatic direction follows the discussion in Sec. \ref{largexifit} and Sec. \ref{smallxifit}, while the isocurvature direction experiences non-negligible fluctuations which must be separately analysed (we do not expect these to survive reheating). 

Consider, now, a hyper-valley misaligned in the $\chi_1\chi_2$-plane (requiring $\kappa_{12}, \kappa_{21}, A_{12} < 0$), the results for the trajectories in the $\chi_1\chi_3$- and $\chi_2\chi_3$-planes ultimately being identical under the interchanges of indices: $2 \leftrightarrow 3$ and $1 \leftrightarrow 2$, respectively. The kinetic terms read, to leading order in $\xi_1^{-1}$ and $\xi_2^{-1}$:
\begin{equation}
\begin{split}
    &\overline{G}'_{IJ} \simeq \pmat{1  & \alpha  & 0 \\ \alpha & \beta & 0 \\ 0 & 0 & \xi_3^{-1}}\quad \text{where:} \\
    &\alpha = \left[\frac{1}{\xi_2} - \frac{1}{\xi_1}\right] c_{ \overline{\phi}'}s_{\overline{\phi}'},\ \beta = \frac{1}{\xi_2} c^2_{\overline{\phi}'} + \frac{1}{\xi_1} s^2_{\overline{\phi}'}.
\end{split}
\end{equation}
To analyse the upper-left submatrix we may closely follow the discussions for the two-scalar case in Refs.~\cite{Gong:2012ri,Modak:2020fij}. We make the transformation:
$\chi_r' = \sqrt{\lambda_+}(c_\psi \chi_r + s_\psi \delta \phi'$), $\delta \phi'' = \sqrt{\lambda_-}( - s_\psi \chi_r + c_\psi \delta \phi')$ and $\delta \theta'' = \sqrt{\lambda_2} \delta \theta'$, where to diagonalise the kinetic terms:
\begin{equation}
    \tan \psi = \frac{2\alpha}{1 - \beta + \sqrt{(1 - \beta)^2 + 4\alpha^2}},
\end{equation}
and we have rescaled by the eigenvalues: $\lambda_{\pm} = \frac{1}{2}(1+\beta \pm \sqrt{(1- \beta)^2 + 4\alpha^2})$ and $\lambda_2 = \xi^{-1}_3$ to achieve a canonical normalisation. 
Hence, the effective mass terms give rise to the mass eigenvalues:
\begin{equation}
\begin{split}
    &m_{-}^2 = 0,\quad  m_{+}^2 = \frac{2M_P^2}{\xi_1\xi_2} \frac{- \kappa_{12}\kappa_{21}}{\kappa_{21}\xi_{1}  +  \kappa_{12}\xi_{2}} \left[ \frac{s_\psi^2}{\lambda_+} + \frac{c^2_\psi}{\lambda_-}\right],\\
    &m^2_2 = M_P^2 \cdot \frac{A_{12}}{\kappa_{21}\xi_{1}  +  \kappa_{12}\xi_{2}}.
\end{split}
\end{equation}
Because we assume $\xi_1, \xi_2 \gg 1$, we know that $\alpha, \beta \ll 1$, from which it follows that $\lambda_+ \sim 1$, $\theta \sim \mathcal{O}(\xi^{-1}_k)$ and $ \lambda_- \sim \beta$. So, the mode which is approximately massless is $\simeq \chi_r$, and the orthogonal motion is generically stabilised for:
\begin{equation}\label{twofieldcase}
    \frac{- \kappa_{12}\kappa_{21}}{\kappa_{21}\xi_1^2 + \kappa_{12}\xi_2^2},\  \frac{A_{12}}{
    \kappa_{21}\xi_1  +  \kappa_{12}\xi_2} \gtrsim 10^{-10}.
\end{equation}
Let us now consider the generic VISH$\nu$ scenario of $\Phi_iS$-Inflation. Setting $\lambda_{iS} \simeq 0$ in the analogue of (\ref{twofieldcase}), assuming $\lambda_i, \lambda_{34} \sim \mathcal{O}(0.1)$ and $\xi_i \sim \xi_S$, the orthogonal direction in the $\rho_i\sigma$-plane is heavy since $- \kappa_{iS}\kappa_{Si}/(\kappa_{Si}\xi_i^2 + \kappa_{iS}\xi_S^2) \sim \lambda_S/\xi_S^2 \sim \text{few} \times 10^{-10}$ (fitting to $A_s$), while the Higgs modulus direction is heavy due to $A_{iS}/(\kappa_{Si}\xi_i  +  \kappa_{iS}\xi_S) \sim \lambda_S/\xi_S \gtrsim 10^{-10}$. Hence, we can integrate out the orthogonal degrees of freedom~\cite{Achucarro:20101,Achucarro:20102,Cespedes:2012hu,Achucarro:20123} to obtain the effective field theory in general. In $\Phi_1\Phi_2$-Inflation, we find that while the orthogonal direction aligned with the Higgs plane is generically heavy, the $\sigma$ direction is always light due to suppressed inter-sector couplings (but positive for $\lambda_{iS} > 0$).

The final situation to analyse concerns the hyper-valley misaligned with respect to all three axes. We make statements only for the special case of the hidden-sector parameter space where the results are more simplified and pertinent. The more general calculation is, of course, the direct generalisation of the two-scalar cases discussed above. We set $\lambda_{13}, \lambda_{23} \simeq 0$ to simplify the mass matrix, and assume $\xi_1 \simeq \xi_2$, which suppresses kinetic mixing between the angular directions to replicate the block-diagonal form just discussed, and we also set $\lambda_i, \lambda_{34} \sim \mathcal{O}(0.1)$. Hence, we find that the masses of the orthogonal directions are $\sim (\lambda_S/\xi_S)M_P^2$, which are heavy when compared with $H_{\text{inf}}$. Taken together, this demonstrates that a tilt of the hyper-valley towards both the $S$ direction and a Higgs direction is sufficient to steepen the orthogonal direction of the valley in spite of an approximate decoupling of the Higgs and singlet sectors, in the large $\xi_k$ regime.

The above conclusions hold for $\max_{k} \{ \xi_k \} \gtrsim \mathcal{O}(10)$, but can be suitably extended into a the $\max_k \{ \xi_k \} \lesssim \mathcal{O}(1)$ regime as follows. The parametrisation used above no longer canonically normalises the kinetic terms and we resort to the regimes discussed in Sec. \ref{smallxifit}. Let us assume that $\xi_S \gg \xi_i$ which diagonalises the kinetic terms and enforces that $\rho_i \simeq 0$. As remarked in Sec. \ref{smallxifit},  $\Phi_1\Phi_2S$-Inflation, $\Phi_i S$-Inflation degenerate to $S$-Inflation for this extremal parameter space. This leads to the mass eigenvalues for canonically normalised orthogonal modes: $m^2_{\rho'_i} \simeq \lambda_{iS} \sigma^2 \Omega^{-2}(\sigma)$. The non-negligible dependence of the masses on the inflaton field-value is a generic consequence of the small $\xi_k$ regime absent from the large $\xi_k$ discussion. In particular, this allows for the masses of the orthogonal directions to become light as the inflaton field evolves. To prevent this, we consider when the masses remain heavy up until the inflaton field-value ($\sigma_f$) at the end of the inflation, $\widetilde{\epsilon}(\sigma_f) \simeq 1$. For the case that $\xi \simeq 0.01$, we find that $\sigma_f \simeq 2.7M_P$ while for $\xi \simeq 1$, we find that $\sigma \simeq M_P$. In either case (and for the range of $\xi_k$ so delimited) we find that the orthogonal directions are light, but positive for $\lambda_{iS} > 0$, during the entire inflation and are aligned with the Higgs modulus directions. For $\xi_S \simeq \xi_i$, the orthogonal directions in $\Phi_iS$-Inflation and $\Phi_1\Phi_2S$-Inflation are heavy throughout the entire phase for $\xi_S \gtrsim 0.2$, but become slightly sub-Hubble as the inflation phase ends for smaller values of the non-minimal coupling. This is because, for $\lambda_i, \lambda_{34} \sim \mathcal{O}(0.1)$, the masses of the orthogonal directions are $\sim \lambda_S M_P^2$, and $\lambda_S$ can be expressed as a function of $\xi_S$ through the fit to $A_s$. Hence, for $\xi \gtrsim 0.2$, the large $\xi_k$ findings are replicated: the inflation model is compatible with generic initial conditions and negligible isocurvature fluctuations develop in the neutral modulus directions. For $\xi_S < 0.2$, the initial conditions which ensure the motion is stabilised in the hyper-valley are potentially less general (the orthogonal masses will be $\sim \mathcal{O}(0.1) H_{\text{inf}}$), and this could be investigated by numerical simulations.

To summarise: for $\max\{\xi_k\} \gtrsim 0.2$, the orthogonal directions in the neutral modulus field-space for the generic $\Phi_iS$-Inflation and $\Phi_1\Phi_2S$-Inflation scenarios obtain an effective mass which remains heavy in comparison to the Hubble scale throughout the inflation phase, while in all other inflation scenarios, and for smaller values of the non-minimal couplings, there is at least one additional light neutral modulus direction (with sub-Hubble but positive mass for $\lambda_{iS}>0$). 

\section{$\beta$-functions}

To illustrate features of the running in both the top-specific VISH$\nu$ and $\nu$DFSZ models, we give the $\beta$ functions to 1-loop order, which we have calculated using PyR@TE~\cite{Sartore:2020gou}. In what follows, we adopt the convention $\mu \frac{d X}{d \mu}\equiv \frac{1}{\left(4 \pi\right)^{2}}\beta^{(1)}[X]$. 

\subsection{Top-specific VISH$\nu$ model}

The $\beta$ functions for the fundamental top-specific VISH$\nu$ couplings are:
\begingroup
\allowdisplaybreaks
\begin{widetext}
\begin{align}
    \beta^{(1)}[g_{\{1,2,3\}}] &= \{7, -3, -7\} g_{\{1,2,3\}}^{3},
    \\
    \beta^{(1)}[M^2_{SS}] &=  M^2_{SS}\left[4 \lambda_{S} + \tr\left(y_N y_N^{\dagger} \right)\right] + 4 \lambda_{1S} M^2_{11}  + 4 \lambda_{2S} M^2_{22} + 4\kappa^2,
    \\
    \beta^{(1)}[M^2_{11}] &= M^2_{11}\left[6 \lambda_1 - \frac{3}{2} g_1^{2} - \frac{9}{2} g_2^{2} + 6\tr\left(y_{u1} y_{u1}^{\dagger} \right)\right]  + M^2_{22}[4 \lambda_3 + 2 \lambda_4] + 2 \lambda_{1S} M^2_{SS} + 2\kappa^2,
    \\
    \beta^{(1)}[M^2_{22}] &= M^2_{22}\bigg[6 \lambda_2 - \frac{3}{2} g_1^{2} - \frac{9}{2} g_2^{2}  + 2\tr\left(y_e y_e^{\dagger} \right) + 2 \tr\left(y_\nu y_\nu^{\dagger} \right) + 6 \tr\left(y_d y_d^{\dagger} \right) + 6 \tr\left(y_{u2} y_{u2}^{\dagger}\right)\bigg]  \\
    &\quad \ \ + M^2_{11}\left[4 \lambda_3  + 2 \lambda_4 \right]  + 2 \lambda_{2S} M^2_{SS} + 2\kappa^2, \nonumber
    \\
    \beta^{(1)}[\kappa] &= \kappa \Bigg[\frac{3}{2} g_1^{2} -  \frac{9}{2} g_2^{2} + 2  \lambda_3 + 4  \lambda_4 + 2  \lambda_{1S} + 2  \lambda_{2S} + 3  \tr\left(y_{u1} y_{u1}^{\dagger}\right) + 3  \tr\left(y_{u2} y_{u2}^{\dagger} \right) + 3  \tr\left(y_d y_d^{\dagger}\right) \\
    &\quad \ \ \ \ \ + \tr\left(y_e y_e^{\dagger} \right) +  \tr\left(y_\nu y_\nu^{\dagger} \right) + \frac{1}{2}\tr\left(y_N y_N^{\dagger} \right)\Bigg], \nonumber
    \\
    \beta^{(1)}[\lambda_1] &=
    \lambda_1\left[12\lambda_1 - 3 g_1^{2} - 9 g_2^{2} + 12 \tr\left(y_{u1} y_{u1}^{\dagger} \right) \right] 
    + 4 \lambda_3^{2} 
    + 4 \lambda_3 \lambda_4 
    + 2 \lambda_4^{2} 
    + 2 \lambda_{1S}^{2}
    + \frac{3}{4} g_1^{4}
    + \frac{3}{2} g_1^{2} g_2^{2}
    + \frac{9}{4} g_2^{4}
    \\
    &\quad \
    - 12 \tr\left(y_{u1} y_{u1}^{\dagger} y_{u1} y_{u1}^{\dagger} \right),
    \nonumber
    \\
    \beta^{(1)}[\lambda_2] &=
    \lambda_2\bigg[12 \lambda_2
    - 3 g_1^{2}
    - 9 g_2^{2}
    + 4  \tr\left(y_e y_e^{\dagger} \right)
    + 4 \tr\left(y_\nu y_\nu^{\dagger} \right)
    + 12\tr\left(y_d y_d^{\dagger} \right) + 12\tr\left(y_{u2} y_{u2}^{\dagger} \right)
    \bigg] 
    \\
    &\quad \ + 4 \lambda_3^{2} 
    + 4 \lambda_3 \lambda_4 
    + 2 \lambda_4^{2} + 
    2 \lambda_{2S}^{2}
    + \frac{3}{4} g_1^{4}
    + \frac{3}{2} g_1^{2} g_2^{2}
    + \frac{9}{4} g_2^{4} 
    - 12 \tr\left(y_d y_d^{\dagger} y_d y_d^{\dagger} \right) - 12 \tr\left(y_{u2} y_{u2}^{\dagger} y_{u2} y_{u2}^{\dagger} \right) \nonumber \\
    &\quad \ - 4 \tr\left(y_e y_e^{\dagger} y_e y_e^{\dagger} \right)
    - 4 \tr\left(y_\nu y_\nu^{\dagger} y_\nu y_\nu^{\dagger} \right), \nonumber
    \\
    \beta^{(1)}[\lambda_3] &=
    \lambda_3\bigg[4 \lambda_3
    + 6 \lambda_1
    + 6 \lambda_2
    - 3 g_1^{2}
    - 9 g_2^{2}
    + 6 \tr\left(y_{u1} y_{u1}^{\dagger} \right)
    + 6 \tr\left(y_{u2} y_{u2}^{\dagger} \right)
    + 6 \tr\left(y_d y_d^{\dagger} \right)  + 2 \tr\left(y_e y_e^{\dagger} \right) \\ 
    &\quad \ \ \ \ \ + 2 \tr\left(y_\nu y_\nu^{\dagger} \right) \bigg]
    + 2 \lambda_1 \lambda_4
    + 2 \lambda_2 \lambda_4
    + 2 \lambda_4^{2}
    + 2 \lambda_{1S} \lambda_{2S}  + \frac{3}{4} g_1^{4}
    -  \frac{3}{2} g_1^{2} g_2^{2}
    + \frac{9}{4} g_2^{4} 
    - 12 \tr\left(y_{u1} y_{u1}^{\dagger} y_d y_d^{\dagger} \right), \nonumber
    \\
    \beta^{(1)}[\lambda_4]&=
    \lambda_4\bigg[4 \lambda_4
    + 2 \lambda_1 
    + 2 \lambda_2 
    + 8 \lambda_3 
    - 3 g_1^{2} 
    - 9 g_2^{2} 
    + 6  \tr\left(y_{u1} y_{u1}^{\dagger} \right)
    + 6  \tr\left(y_{u2} y_{u2}^{\dagger} \right) 
    + 6  \tr\left(y_d y_d^{\dagger} \right) \\ 
    &\quad \ \ \ \ \ 
    + 2  \tr\left(y_e y_e^{\dagger} \right) 
    + 2  \tr\left(y_\nu y_\nu^{\dagger} \right)
    \bigg]
    + 4 \epsilon^{2}
    + 3 g_1^{2} g_2^{2} 
    - 12 \tr\left(y_{u1} y_{u1}^{\dagger} y_{u2} y_{u2}^{\dagger} \right)
    + 12 \tr\left(y_{u1} y_{u1}^{\dagger} y_d y_d^{\dagger} \right), \nonumber
    \\
    \beta^{(1)}[\lambda_{S}] &=
    \lambda_S\left[10 \lambda_{S} + 2 \tr\left(y_N y_N^{\dagger} \right)\right]
    + 4 \lambda_{1S}^{2}
    + 4 \lambda_{2S}^{2}
    - 2 \tr\left(y_N y_N^{\dagger} y_N y_N^{\dagger} \right),
    \\
    \beta^{(1)}[\lambda_{1S}] &=
    \lambda_{1S}\left[4 \lambda_{1S} 
    + 6 \lambda_1
    + 4 \lambda_{S}
    -  \frac{3}{2} g_1^{2} 
    -  \frac{9}{2} g_2^{2} 
    + 6 \tr\left(y_{u1} y_{u1}^{\dagger} \right)
    + \tr\left(y_N y_N^{\dagger} \right)
    \right]
    + \lambda_{2S}\left[4 \lambda_3
    + 2 \lambda_4 \right], 
    \\
    \beta^{(1)}[\lambda_{2S}] &= 
    \lambda_{2S} \bigg[4 \lambda_{2S}
    + 6 \lambda_2
    + 4 \lambda_{S}
    -  \frac{3}{2} g_1^{2}
    -  \frac{9}{2} g_2^{2} 
    + 6 \tr\left(y_d y_d^{\dagger} \right)
    + 6 \tr\left(y_{u2} y_{u2}^{\dagger} \right)
    + 2 \tr\left(y_e y_e^{\dagger} \right) 
    + 2 \tr\left(y_\nu y_\nu^{\dagger} \right) \\
    &\quad \ \ \ \ \ \
    + \tr\left(y_N y_N^{\dagger} \right)
    \bigg]
    + 2 \lambda_{1S} [2\lambda_3
    +\lambda_4] - 4 \tr\left(y_\nu^{\dagger} y_\nu y_N^{\dagger} y_N \right), \nonumber \\
    \beta^{(1)}[y_{u1}] &= 
    y_{u1}\Bigg[-  \frac{17}{12} g_1^{2}   -  \frac{9}{4} g_2^{2} 
    - 8 g_3^{2} + 3 \tr\left(y_{u1} y_{u1}^{\dagger} \right) 
    \Bigg]
    + \frac{3}{2} y_{u1} y_{u1}^{\dagger} y_{u1} 
    + \frac{1}{2} y_{u2} y_{u2}^{\dagger} y_{u1} 
    + \frac{1}{2} y_d y_d^{\dagger} y_{u1},
    \\
    \beta^{(1)}[y_{u2}] &=
    y_{u2}\Bigg[-  \frac{17}{12} g_1^{2} 
    -  \frac{9}{4} g_2^{2} 
    - 8 g_3^{2} + 3 \tr\left(y_{u2} y_{u2}^{\dagger} \right) 
    + 3 \tr\left(y_d y_d^{\dagger} \right) 
    + \tr\left(y_e y_e^{\dagger} \right) 
    + \tr\left(y_\nu y_\nu^{\dagger} \right) 
    \Bigg] 
    \\
    & \quad \
    + \frac{1}{2} y_{u1} y_{u1}^{\dagger} y_{u2}
    + \frac{3}{2} y_{u2} y_{u2}^{\dagger} y_{u2}
    -  \frac{3}{2} y_d y_d^{\dagger} y_{u2},
    \nonumber \\
    \beta^{(1)}[y_d] &=
    y_d \Bigg[ -  \frac{5}{12} g_1^{2} 
    -  \frac{9}{4} g_2^{2} 
    - 8 g_3^{2} 
    + 3 \tr\left(y_{u2} y_{u2}^{\dagger} \right) 
    + 3 \tr\left(y_d y_d^{\dagger} \right) 
    + \tr\left(y_e y_e^{\dagger} \right)
    + \tr\left(y_\nu y_\nu^{\dagger} \right) 
    \Bigg]
    + \frac{1}{2} y_{u1} y_{u1}^{\dagger} y_d 
    \\
    & \quad \
    -  \frac{3}{2} y_{u2} y_{u2}^{\dagger} y_d
    + \frac{3}{2} y_d y_d^{\dagger} y_d,
    \nonumber \\
    \beta^{(1)}[y_e] &=
    y_e \Bigg[ -  \frac{15}{4} g_1^{2} 
    -  \frac{9}{4} g_2^{2} 
    + 3 \tr\left(y_{u2} y_{u2}^{\dagger} \right) 
    + 3 \tr\left(y_d y_d^{\dagger} \right)  
    + \tr\left(y_e y_e^{\dagger} \right)
    + \tr\left(y_\nu y_\nu^{\dagger} \right) 
    \Bigg]
    + \frac{3}{2} y_e y_e^{\dagger} y_e
    -  \frac{3}{2} y_\nu y_\nu^{\dagger} y_e,
    \\
    \beta^{(1)}[y_\nu] &=
    y_\nu \Bigg[ -  \frac{3}{4} g_1^{2} 
    -  \frac{9}{4} g_2^{2} 
    + 3 \tr\left(y_{u2} y_{u2}^{\dagger} \right) 
    + 3 \tr\left(y_d y_d^{\dagger} \right)  
    + \tr\left(y_e y_e^{\dagger} \right)
    + \tr\left(y_\nu y_\nu^{\dagger} \right) 
    \Bigg]
    -  \frac{3}{2} y_e y_e^{\dagger} y_\nu
    \\
    & \quad \
    + \frac{3}{2} y_\nu y_\nu^{\dagger} y_\nu 
    + \frac{1}{2} y_\nu y_N y_N^{\dagger},
    \nonumber \\
    \beta^{(1)}[y_N] &= \frac{1}{2} \tr\left(y_N y_N^{\dagger} \right) y_N + y_N y_N^{\dagger} y_N 
    + y_\nu^{\text{T}} y_\nu^{*} y_N
    + y_N y_\nu^{\dagger} y_\nu.
\end{align}
\end{widetext}
\endgroup

\subsection{$\nu$DFSZ model}
For comparison, we give the $\beta$ functions for the $\nu$DFSZ model (underlying Type-II 2HDM version), see Ref.~\cite{nuDFSZ} for details. In the $\nu$DFSZ, the PQ charge assignment is generation universal and $\mathcal{L}^u \supset y_u \Bar{q_L}\widetilde{\Phi}_1u_R$, where we implicitly define $y_u$ as it occurs in the following:
\begingroup
\allowdisplaybreaks
\begin{widetext}
\begin{align}
    \beta^{(1)}[g_{\{1,2,3\}}] &= \{7, -3, -7\} g_{\{1,2,3\}}^{3},
    \\
    \beta^{(1)}[M^2_{SS}] &=  M^2_{SS}\left[4 \lambda_{S} + \tr\left(y_N y_N^{\dagger} \right)\right] + 4 \lambda_{1S} M^2_{11} + 4 \lambda_{2S} M^2_{22},
    \\
    \beta^{(1)}[M^2_{11}] &= M^2_{11}\left[6 \lambda_1 - \frac{3}{2} g_1^{2} - \frac{9}{2} g_2^{2} + 6\tr\left(y_u y_u^{\dagger} \right)\right] + M^2_{22}[4 \lambda_3 + 2 \lambda_4] + 2 \lambda_{1S} M^2_{SS},
    \\
    \beta^{(1)}[M^2_{22}] &= M^2_{22}\left[6 \lambda_2 - \frac{3}{2} g_1^{2} - \frac{9}{2} g_2^{2}  + 2\tr\left(y_e y_e^{\dagger} \right) + 2 \tr\left(y_\nu y_\nu^{\dagger} \right) + 6 \tr\left(y_d y_d^{\dagger} \right) \right]  + M^2_{11}\left[4 \lambda_3  + 2 \lambda_4 \right]  + 2 \lambda_{2S} M^2_{SS},  
    \\
    \beta^{(1)}[\lambda_1] &=
    \lambda_1\left[12\lambda_1 - 3 g_1^{2} - 9 g_2^{2} + 12 \tr\left(y_u y_u^{\dagger} \right) \right] 
    + 4 \lambda_3^{2}
    + 4 \lambda_3 \lambda_4
    + 2 \lambda_4^{2}
    + 2 \lambda_{1S}^{2} + \frac{3}{4} g_1^{4}
    + \frac{3}{2} g_1^{2} g_2^{2}
    + \frac{9}{4} g_2^{4} \\
    & \quad \
    - 12 \tr\left(y_u y_u^{\dagger} y_u y_u^{\dagger} \right), \nonumber
    \\
    \beta^{(1)}[\lambda_2] &=
    \lambda_2\left[12 \lambda_2
    - 3 g_1^{2}
    - 9 g_2^{2}
    + 4  \tr\left(y_e y_e^{\dagger} \right)
    + 4 \tr\left(y_\nu y_\nu^{\dagger} \right)
    + 12\tr\left(y_d y_d^{\dagger} \right)
    \right] + 4 \lambda_3^{2} 
    + 4 \lambda_3 \lambda_4 
    + 2 \lambda_4^{2} + 2 \lambda_{2S}^{2} \\
    & \quad \
    + \frac{3}{4} g_1^{4}
    + \frac{3}{2} g_1^{2} g_2^{2}
    + \frac{9}{4} g_2^{4}
    - 12 \tr\left(y_d y_d^{\dagger} y_d y_d^{\dagger} \right)
    - 4 \tr\left(y_e y_e^{\dagger} y_e y_e^{\dagger} \right)
    - 4 \tr\left(y_\nu y_\nu^{\dagger} y_\nu y_\nu^{\dagger} \right), \nonumber
    \\
    \beta^{(1)}[\lambda_3] &=
    \lambda_3\bigg[4 \lambda_3
    + 6 \lambda_1
    + 6 \lambda_2
    - 3 g_1^{2}
    - 9 g_2^{2}
    + 6 \tr\left(y_u y_u^{\dagger} \right)
    + 6 \tr\left(y_d y_d^{\dagger} \right) + 2 \tr\left(y_e y_e^{\dagger} \right)  + 2 \tr\left(y_\nu y_\nu^{\dagger} \right) \bigg] \\
    &\quad \
    + 2 \lambda_1 \lambda_4
    + 2 \lambda_2 \lambda_4
    + 2 \lambda_4^{2}
    + 2 \lambda_{1S} \lambda_{2S} 
    + \frac{3}{4} g_1^{4}
    -  \frac{3}{2} g_1^{2} g_2^{2}
    + \frac{9}{4} g_2^{4} 
    - 12 \tr\left(y_u y_u^{\dagger} y_d y_d^{\dagger} \right), \nonumber
    \\
    \beta^{(1)}[\lambda_4]&=
    \lambda_4\bigg[4 \lambda_4
    + 2 \lambda_1 
    + 2 \lambda_2 
    + 8 \lambda_3 
    - 3 g_1^{2} 
    - 9 g_2^{2} 
    + 6  \tr\left(y_u y_u^{\dagger} \right)
    + 6  \tr\left(y_d y_d^{\dagger} \right) + 2  \tr\left(y_e y_e^{\dagger} \right)
    + 2  \tr\left(y_\nu y_\nu^{\dagger} \right)
    \bigg] \\
    & \quad \
    + 4 \epsilon^{2}
    + 3 g_1^{2} g_2^{2}
    + 12 \tr\left(y_u y_u^{\dagger} y_d y_d^{\dagger} \right), \nonumber
    \\
    \beta^{(1)}[\lambda_{S}] &=
    \lambda_S\left[10 \lambda_{S} + 2 \tr\left(y_N y_N^{\dagger} \right)\right]
    + 4 \lambda_{1S}^{2}
    + 4 \lambda_{2S}^{2}
    + 8 \epsilon^{2}
    - 2 \tr\left(y_N y_N^{\dagger} y_N y_N^{\dagger} \right),
    \\
    \beta^{(1)}[\lambda_{1S}] &=
    \lambda_{1S}\left[4 \lambda_{1S} 
    + 6 \lambda_1
    + 4 \lambda_{S}
    -  \frac{3}{2} g_1^{2} 
    -  \frac{9}{2} g_2^{2} 
    + 6 \tr\left(y_u y_u^{\dagger} \right)
    + \tr\left(y_N y_N^{\dagger} \right)
    \right]
    + \lambda_{2S}\left[4 \lambda_3
    + 2 \lambda_4 \right]
    + 8 \epsilon^{2}, 
    \\
    \beta^{(1)}[\lambda_{2S}] &= 
    \lambda_{2S} \bigg[4 \lambda_{2S}
    + 6 \lambda_2
    + 4 \lambda_{S}
    -  \frac{3}{2} g_1^{2}
    -  \frac{9}{2} g_2^{2} 
    + 6 \tr\left(y_d y_d^{\dagger} \right)
    + 2 \tr\left(y_e y_e^{\dagger} \right)  + 2 \tr\left(y_\nu y_\nu^{\dagger} \right)
    + \tr\left(y_N y_N^{\dagger} \right)
    \bigg] \\
    & \quad \
    + 2 \lambda_{1S} [2\lambda_3
    +\lambda_4]
    + 8 \epsilon^{2}  - 4 \tr\left(y_\nu^{\dagger} y_\nu y_N^{\dagger} y_N \right), \nonumber
    \\
    \beta^{(1)}[\epsilon] &=
    \epsilon  \bigg[
    2 \lambda_3
    + 4 \lambda_4
    + 2 \lambda_{S}
    + 4 \lambda_{1S}
    + 4 \lambda_{2S}
    - \frac{3}{2}  g_1^{2}
    - \frac{9}{2}  g_2^{2}
    + 3 \tr\left(y_u y_u^{\dagger} \right)
    + 3 \tr\left(y_d y_d^{\dagger} \right)
    + \tr\left(y_e y_e^{\dagger} \right) \\
    &\quad \ \ \ 
    + \tr\left(y_\nu y_\nu^{\dagger} \right)
    + \tr\left(y_N y_N^{\dagger} \right)
    \bigg],
    \nonumber \\
    \beta^{(1)}[y_u] &=
    y_u \Bigg[-  \frac{17}{12} g_1^{2} 
    -  \frac{9}{4} g_2^{2} 
    - 8 g_3^{2} 
    + 3 \tr\left(y_u y_u^{\dagger} \right) \Bigg]
     + \frac{3}{2} y_u y_u^{\dagger} y_u
    + \frac{1}{2} y_d y_d^{\dagger} y_u,
    \\
    \beta^{(1)}[y_d] &=
    y_d \Bigg[ -  \frac{5}{12} g_1^{2} 
    -  \frac{9}{4} g_2^{2} 
    - 8 g_3^{2} 
    + 3 \tr\left(y_d y_d^{\dagger} \right) 
    + \tr\left(y_e y_e^{\dagger} \right) 
    + \tr\left(y_\nu y_\nu^{\dagger} \right) 
    \Bigg] 
    + \frac{1}{2} y_{u} y_{u}^{\dagger} y_d
    + \frac{3}{2} y_d y_d^{\dagger} y_d,
    \\
    \beta^{(1)}[y_e] &=
    y_e \Bigg[ -  \frac{15}{4} g_1^{2} 
    -  \frac{9}{4} g_2^{2} 
    + 3 \tr\left(y_d y_d^{\dagger} \right)  
    + \tr\left(y_e y_e^{\dagger} \right) 
    + \tr\left(y_\nu y_\nu^{\dagger} \right) 
    \Bigg]
    + \frac{3}{2} y_e y_e^{\dagger} y_e
    -  \frac{3}{2} y_\nu y_\nu^{\dagger} y_e,
    \\
    \beta^{(1)}[y_\nu] &=
    y_\nu \Bigg[ -  \frac{3}{4} g_1^{2} 
    -  \frac{9}{4} g_2^{2} 
    + 3 \tr\left(y_d y_d^{\dagger} \right)  
    + \tr\left(y_e y_e^{\dagger} \right) 
    + \tr\left(y_\nu y_\nu^{\dagger} \right) 
    \Bigg]
    -  \frac{3}{2} y_e y_e^{\dagger} y_\nu
    + \frac{3}{2} y_\nu y_\nu^{\dagger} y_\nu
    + \frac{1}{2} y_\nu y_N y_N^{\dagger},
    \\
    \beta^{(1)}[y_N] &= \frac{1}{2} \tr\left(y_N y_N^{\dagger} \right) y_N + y_N y_N^{\dagger} y_N 
    + y_\nu^{\text{T}} y_\nu^{*} y_N
    + y_N y_\nu^{\dagger} y_\nu.
\end{align}
\end{widetext}
\endgroup

\section{Axion couplings}

In this appendix, we present calculations of the axion-photon ($g_{a\gamma}$) and axion-electron couplings ($g_{ae}$) in both the top-specific VISH$\nu$ and $\nu$DFSZ models, using the physical basis where the axion is orthogonal to the Goldstone eaten by the $Z$. In the first model, we simply expect the tree-level axion-nucleon couplings to be as discussed in Ref.~\cite{Saikawa:2019lng} while in the second they are DFSZ-like. However, the leptonic contributions in both models deserve careful consideration since, although generation-universal, they include shifts in their PQ charges beyond the standard DFSZ-I\footnote{In the Type I DFSZ model (DFSZ-I), the Higgs doublet ($\Phi_2$, say) which is responsible for giving mass to the down-type quarks has tree-level lepton Yukawa couplings, while $\Phi_1$ does not.} assignment to accommodate a Majorana term for the right-handed neutrinos after PQ breaking.

In both models, the PQ current is conserved up to colour and electromagnetic (EM) anomalies, given by:
\begin{equation}\label{anomalyfactors}
    \partial_\mu J^\mu_{PQ} = \mathcal{N}\frac{g_3^2}{16\pi^2}G\widetilde{G} + \mathcal{E}\frac{e^2}{16\pi^2}F\widetilde{F} ,
\end{equation}
and we analyse the following terms in the low energy axion Lagrangian~\cite{Saikawa:2019lng,DILUZIO20201}:
\begin{equation}\label{Lagrangian}
\begin{split}
    -\mathcal{L}_a \supset\ &\frac{g_3^2}{32\pi^2} \frac{a}{f_A}G\widetilde{G} + \frac{e^2}{32\pi^2}\frac{a}{f_A}\frac{\mathcal{E}}{\mathcal{N}}F\widetilde{F} \\
    &+ \frac{\partial_{\mu} a}{2f_A} \sum_{\text{leptons}} \frac{1}{\mathcal{N}} \left[ \overline{\psi}_{L} X_{\psi_{L}} \gamma^\mu \psi_{L} + \overline{\psi}_{R} X_{\psi_{R}} \gamma^\mu \psi_{iR}\right],
\end{split}
\end{equation}
where the PQ charges $X_{\psi}$ of the leptons $\psi_{R,L}$ are represented as diagonal\footnote{In both models, the leptonic PQ charge assignment is generation-universal and, hence, the coupling is the same in both weak interaction basis and the mass basis~\cite{Saikawa:2019lng}.} matrices, and:
\begin{equation}
    f_A \equiv v_{PQ}/2|\mathcal{N}|, \quad  v_{PQ} \simeq |X_S| v_S.
\end{equation} 
We can generically express the color and EM anomaly factors, respectively, as:
\begin{equation}\label{colouranomaly}
    \mathcal{N} = \frac{1}{2} \sum^{N_{g}}_{i = 1} \left[2X_q - X_u - X_d \right], \quad \mathcal{E} = \sum_{L-R} C_i Q_i^2 X_i,
\end{equation}
where $N_g$ is the number of generations, $X_i$ is the PQ charge of the field $i$ and we take the colour index of the quark representation to be $T(q) = \frac{1}{2}$~\cite{DFSZaxioncouplings}; and for the $i^{\text{th}}$ left- or right-handed representation, $C_i$ is the number of colours, $X_i$ is again its PQ charge and $Q_i$ is the electric charge. 
We also distinguish the \textit{domain wall number} of the model, which is~\cite{Geng:1990dv}:
\begin{equation}\label{domainwallnumber}
    N_{DW} = 2 \left| \frac{\mathcal{N}}{X_S} \right|,
\end{equation}
from the colour anomaly $\mathcal{N}$, where $X_S$ is the PQ charge of the complex scalar $S$, noting that $f_a \simeq v_{S}/N_{DW}$. We therefore compute directly for the VISH$\nu$ model (setting $s^2 \equiv \sin^2 \beta$ and $c^2 \equiv \cos^2 \beta$):
\begin{equation}
\begin{split}
    &\mathcal{N} = \frac{1}{2} \cdot \left[ 0 - (c^2 - 2s^2) - 3s^2 \right]  = - \frac{1}{2}, \\
    &\mathcal{E}  = 3(-1)^2X_l - 3 \left[ \frac{4}{9} \sum_{i} X^{(i)}_{u} - 3 \cdot \frac{1}{9}X_d\right] - 3 (-1)^2 X_e \\
    &\quad = 3\left(s^2 - \frac{1}{2} \right) - \frac{4}{3} \left(c^2 - 2s^2 \right) + s^2 - 3\left(2s^2 - \frac{1}{2} \right) \\
    &\quad = - \frac{4}{3},\\
    &N_{DW} = 2\left(\frac{1}{2}\right) = 1.
\end{split}
\end{equation}
Importantly, the top-specific structure of the VISH$\nu$ model means that $N_{DW} = 1$ as earlier claimed. For completeness, we also include the $\nu$DFSZ model values:
\begin{equation}
\begin{split}
    &\mathcal{N} = \frac{1}{2} \cdot 3 \cdot \left[ 0 - c^2 - s^2 \right]  = - \frac{3}{2}, \\
    &\mathcal{E} = 3(-1)^2X_l - 3\left[ 3 \left( \frac{4}{9}X_u + \frac{1}{9} X_d \right) + (-1)^2 X_e \right] \\
    & \quad = 3X_l - 4 X_u - X_d - 3 X_e\\
    &\quad = 3\left(s^2 - \frac{1}{4} \right)  - 4c^2 - s^2 - 3\left(2s^2 - \frac{1}{4}\right) \\
    &\quad = -4,\\
    &N_{DW} = 2\left(\frac{3}{2}\right)\left(\frac{1}{2}\right)^{-1} = 6.
\end{split}
\end{equation}

Now, to arrive at the axion-electron coupling we can eliminate the vector-part of the axion-lepton derivative couplings using the equations of motion. So that:
\begin{equation}
    \mathcal{L}_a \supset  \frac{\partial_{\mu} a}{2f_A} C_{ae} \overline{e} \gamma^\mu \gamma_5 e, \quad C_{ae} = \frac{X_{e_L} - X_{e_R}}{2\mathcal{N}},
\end{equation}
at tree-level (as in Ref.~\cite{DILUZIO20201}),
giving the axion-electron interaction\footnote{Notice that $f_A$ also contains a factor of $\mathcal{N}^{-1}$, which means that, in general, although the $C_{ae}$ of DFSZ-I is rescaled with the new charge assignments, $g_{ae}$ remains the same in the top-specific VISH$\nu$ model~\cite{Saikawa:2019lng}.} as~\cite{Saikawa:2019lng}:
\begin{align}
    \mathcal{L}_a \supset - ig_{ae}a\overline{e}\gamma_5 e, \quad g_{ae} \equiv \frac{C_{ae}m_e}{f_A}.
\end{align}
The axion-photon coupling is given by~\cite{ParticleDataGroup:2020ssz}:
\begin{align}
    g_{a\gamma} = \frac{\alpha}{2\pi f_A} \left[\frac{\mathcal{E}}{\mathcal{N}} - 1.92(4) \right],
\end{align}
where the first term is model-dependent and the second term comes from an axion-field-dependent transformation of the quark fields that forbids axion-pion mass-mixing~\cite{DILUZIO20201}. 

Hence, we compute for the top-specific VISH$\nu$ model:
\begin{equation}
\begin{split}
    C_{ae} &= \frac{s^2 - \frac{1}{2} - \left(2s^2 - \frac{1}{2}\right) }{- 1} = \sin^2 \beta, \\
    \frac{\mathcal{E}}{\mathcal{N}} &= \frac{-4/3}{-1/2} = \frac{8}{3};
\end{split}
\end{equation}
while for the $\nu$DFSZ:
\begin{equation}
\begin{split}
    C_{ae} &= \frac{s^2 - \frac{1}{4} - \left(2s^2 - \frac{1}{4}\right) }{- 3} = \frac{1}{3} \sin^2 \beta, \\ \frac{\mathcal{E}}{\mathcal{N}} &= \frac{-4}{-3/2} = \frac{8}{3}.
\end{split}
\end{equation}
Comparing these results to DFSZ-I~\cite{DILUZIO20201}:
\begin{equation}
\begin{split}
    C_{ae} &= \frac{-s^2}{-3}= \frac{1}{3}\sin^2 \beta, \\
    \frac{\mathcal{E}}{\mathcal{N}} &= \frac{-4}{-3/2} = \frac{8}{3},
\end{split}
\end{equation}
we see that the axion-electron coupling \textit{coefficient} is a factor of three larger in the top-specific VISH$\nu$ model, while the axion-photon coupling is the same. 

\bibliography{vishnu}% Produces the bibliography via BibTeX.

\providecommand{\noopsort}[1]{}\providecommand{\singleletter}[1]{#1}%
\begin{thebibliography}{149}%
\makeatletter
\providecommand \@ifxundefined [1]{%
 \@ifx{#1\undefined}
}%
\providecommand \@ifnum [1]{%
 \ifnum #1\expandafter \@firstoftwo
 \else \expandafter \@secondoftwo
 \fi
}%
\providecommand \@ifx [1]{%
 \ifx #1\expandafter \@firstoftwo
 \else \expandafter \@secondoftwo
 \fi
}%
\providecommand \natexlab [1]{#1}%
\providecommand \enquote  [1]{``#1''}%
\providecommand \bibnamefont  [1]{#1}%
\providecommand \bibfnamefont [1]{#1}%
\providecommand \citenamefont [1]{#1}%
\providecommand \href@noop [0]{\@secondoftwo}%
\providecommand \href [0]{\begingroup \@sanitize@url \@href}%
\providecommand \@href[1]{\@@startlink{#1}\@@href}%
\providecommand \@@href[1]{\endgroup#1\@@endlink}%
\providecommand \@sanitize@url [0]{\catcode `\\12\catcode `\$12\catcode
  `\&12\catcode `\#12\catcode `\^12\catcode `\_12\catcode `\%12\relax}%
\providecommand \@@startlink[1]{}%
\providecommand \@@endlink[0]{}%
\providecommand \url  [0]{\begingroup\@sanitize@url \@url }%
\providecommand \@url [1]{\endgroup\@href {#1}{\urlprefix }}%
\providecommand \urlprefix  [0]{URL }%
\providecommand \Eprint [0]{\href }%
\providecommand \doibase [0]{https://doi.org/}%
\providecommand \selectlanguage [0]{\@gobble}%
\providecommand \bibinfo  [0]{\@secondoftwo}%
\providecommand \bibfield  [0]{\@secondoftwo}%
\providecommand \translation [1]{[#1]}%
\providecommand \BibitemOpen [0]{}%
\providecommand \bibitemStop [0]{}%
\providecommand \bibitemNoStop [0]{.\EOS\space}%
\providecommand \EOS [0]{\spacefactor3000\relax}%
\providecommand \BibitemShut  [1]{\csname bibitem#1\endcsname}%
\let\auto@bib@innerbib\@empty
%</preamble>
\bibitem [{\citenamefont {Bezrukov}\ and\ \citenamefont
  {Shaposhnikov}(2008)}]{HiggsInflation1}%
  \BibitemOpen
  \bibfield  {author} {\bibinfo {author} {\bibfnamefont {F.~L.}\ \bibnamefont
  {Bezrukov}}\ and\ \bibinfo {author} {\bibfnamefont {M.}~\bibnamefont
  {Shaposhnikov}},\ }\bibfield  {title} {\bibinfo {title} {{The Standard Model
  Higgs boson as the inflaton}},\ }\href
  {https://doi.org/10.1016/j.physletb.2007.11.072} {\bibfield  {journal}
  {\bibinfo  {journal} {Phys. Lett. B}\ }\textbf {\bibinfo {volume} {659}},\
  \bibinfo {pages} {703} (\bibinfo {year} {2008})},\ \Eprint
  {https://arxiv.org/abs/0710.3755} {arXiv:0710.3755 [hep-th]} \BibitemShut
  {NoStop}%
\bibitem [{\citenamefont {Asaka}\ \emph {et~al.}(2005)\citenamefont {Asaka},
  \citenamefont {Blanchet},\ and\ \citenamefont {Shaposhnikov}}]{Asaka:2005an}%
  \BibitemOpen
  \bibfield  {author} {\bibinfo {author} {\bibfnamefont {T.}~\bibnamefont
  {Asaka}}, \bibinfo {author} {\bibfnamefont {S.}~\bibnamefont {Blanchet}},\
  and\ \bibinfo {author} {\bibfnamefont {M.}~\bibnamefont {Shaposhnikov}},\
  }\bibfield  {title} {\bibinfo {title} {{The $\nu$MSM, dark matter and
  neutrino masses}},\ }\href {https://doi.org/10.1016/j.physletb.2005.09.070}
  {\bibfield  {journal} {\bibinfo  {journal} {Phys. Lett. B}\ }\textbf
  {\bibinfo {volume} {631}},\ \bibinfo {pages} {151} (\bibinfo {year}
  {2005})},\ \Eprint {https://arxiv.org/abs/hep-ph/0503065}
  {arXiv:hep-ph/0503065} \BibitemShut {NoStop}%
\bibitem [{\citenamefont {Asaka}\ and\ \citenamefont
  {Shaposhnikov}(2005)}]{Asaka:2005pn}%
  \BibitemOpen
  \bibfield  {author} {\bibinfo {author} {\bibfnamefont {T.}~\bibnamefont
  {Asaka}}\ and\ \bibinfo {author} {\bibfnamefont {M.}~\bibnamefont
  {Shaposhnikov}},\ }\bibfield  {title} {\bibinfo {title} {{The $\nu$MSM, dark
  matter and baryon asymmetry of the universe}},\ }\href
  {https://doi.org/10.1016/j.physletb.2005.06.020} {\bibfield  {journal}
  {\bibinfo  {journal} {Phys. Lett. B}\ }\textbf {\bibinfo {volume} {620}},\
  \bibinfo {pages} {17} (\bibinfo {year} {2005})},\ \Eprint
  {https://arxiv.org/abs/hep-ph/0505013} {arXiv:hep-ph/0505013} \BibitemShut
  {NoStop}%
\bibitem [{\citenamefont {Minkowski}(1977)}]{Minkowski:1977sc}%
  \BibitemOpen
  \bibfield  {author} {\bibinfo {author} {\bibfnamefont {P.}~\bibnamefont
  {Minkowski}},\ }\bibfield  {title} {\bibinfo {title} {{$\mu \to e\gamma$ at a
  Rate of One Out of $10^{9}$ Muon Decays?}},\ }\href
  {https://doi.org/10.1016/0370-2693(77)90435-X} {\bibfield  {journal}
  {\bibinfo  {journal} {Phys. Lett.}\ }\textbf {\bibinfo {volume} {B67}},\
  \bibinfo {pages} {421} (\bibinfo {year} {1977})}\BibitemShut {NoStop}%
%%CITATION = PHLTA,B67,421;%%
\bibitem [{\citenamefont {Yanagida}(1979)}]{Yanagida:1979as}%
  \BibitemOpen
  \bibfield  {author} {\bibinfo {author} {\bibfnamefont {T.}~\bibnamefont
  {Yanagida}},\ }\bibfield  {title} {\bibinfo {title} {{Horizontal symmetry and
  masses of neutrinos}},\ }\bibfield  {booktitle} {\emph {\bibinfo {booktitle}
  {{Proceedings: Workshop on the Unified Theories and the Baryon Number in the
  Universe: Tsukuba, Japan, February 13-14, 1979}}},\ }\href@noop {} {\bibfield
   {journal} {\bibinfo  {journal} {Conf. Proc.}\ }\textbf {\bibinfo {volume}
  {C7902131}},\ \bibinfo {pages} {95} (\bibinfo {year} {1979})}\BibitemShut
  {NoStop}%
%%CITATION = CONFP,C7902131,95;%%
\bibitem [{\citenamefont {Gell-Mann}\ \emph {et~al.}(1979)\citenamefont
  {Gell-Mann}, \citenamefont {Ramond},\ and\ \citenamefont
  {Slansky}}]{GellMann:1980vs}%
  \BibitemOpen
  \bibfield  {author} {\bibinfo {author} {\bibfnamefont {M.}~\bibnamefont
  {Gell-Mann}}, \bibinfo {author} {\bibfnamefont {P.}~\bibnamefont {Ramond}},\
  and\ \bibinfo {author} {\bibfnamefont {R.}~\bibnamefont {Slansky}},\
  }\bibfield  {title} {\bibinfo {title} {{Complex Spinors and Unified
  Theories}},\ }\bibfield  {booktitle} {\emph {\bibinfo {booktitle}
  {{Supergravity Workshop Stony Brook, New York, September 27-28, 1979}}},\
  }\href@noop {} {\bibfield  {journal} {\bibinfo  {journal} {Conf. Proc.}\
  }\textbf {\bibinfo {volume} {C790927}},\ \bibinfo {pages} {315} (\bibinfo
  {year} {1979})}\BibitemShut {NoStop}%
%%CITATION = ARXIV:1306.4669;%%
\bibitem [{\citenamefont {Mohapatra}\ and\ \citenamefont
  {Senjanovic}(1980)}]{Mohapatra:1979ia}%
  \BibitemOpen
  \bibfield  {author} {\bibinfo {author} {\bibfnamefont {R.~N.}\ \bibnamefont
  {Mohapatra}}\ and\ \bibinfo {author} {\bibfnamefont {G.}~\bibnamefont
  {Senjanovic}},\ }\bibfield  {title} {\bibinfo {title} {{Neutrino Mass and
  Spontaneous Parity Violation}},\ }\href
  {https://doi.org/10.1103/PhysRevLett.44.912} {\bibfield  {journal} {\bibinfo
  {journal} {Phys. Rev. Lett.}\ }\textbf {\bibinfo {volume} {44}},\ \bibinfo
  {pages} {912} (\bibinfo {year} {1980})}\BibitemShut {NoStop}%
%%CITATION = PRLTA,44,912;%%
\bibitem [{\citenamefont {Akhmedov}\ \emph {et~al.}(1998)\citenamefont
  {Akhmedov}, \citenamefont {Rubakov},\ and\ \citenamefont {Smirnov}}]{ARS}%
  \BibitemOpen
  \bibfield  {author} {\bibinfo {author} {\bibfnamefont {E.~K.}\ \bibnamefont
  {Akhmedov}}, \bibinfo {author} {\bibfnamefont {V.~A.}\ \bibnamefont
  {Rubakov}},\ and\ \bibinfo {author} {\bibfnamefont {A.~Y.}\ \bibnamefont
  {Smirnov}},\ }\bibfield  {title} {\bibinfo {title} {{Baryogenesis via
  neutrino oscillations}},\ }\href
  {https://doi.org/10.1103/PhysRevLett.81.1359} {\bibfield  {journal} {\bibinfo
   {journal} {Phys. Rev. Lett.}\ }\textbf {\bibinfo {volume} {81}},\ \bibinfo
  {pages} {1359} (\bibinfo {year} {1998})},\ \Eprint
  {https://arxiv.org/abs/hep-ph/9803255} {arXiv:hep-ph/9803255} \BibitemShut
  {NoStop}%
\bibitem [{\citenamefont {Peccei}\ and\ \citenamefont
  {Quinn}(1977)}]{Peccei:1977hh}%
  \BibitemOpen
  \bibfield  {author} {\bibinfo {author} {\bibfnamefont {R.~D.}\ \bibnamefont
  {Peccei}}\ and\ \bibinfo {author} {\bibfnamefont {H.~R.}\ \bibnamefont
  {Quinn}},\ }\bibfield  {title} {\bibinfo {title} {{CP Conservation in the
  Presence of Instantons}},\ }\href
  {https://doi.org/10.1103/PhysRevLett.38.1440} {\bibfield  {journal} {\bibinfo
   {journal} {Phys. Rev. Lett.}\ }\textbf {\bibinfo {volume} {38}},\ \bibinfo
  {pages} {1440} (\bibinfo {year} {1977})}\BibitemShut {NoStop}%
\bibitem [{\citenamefont {Foot}\ \emph {et~al.}(2014)\citenamefont {Foot},
  \citenamefont {Kobakhidze}, \citenamefont {McDonald},\ and\ \citenamefont
  {Volkas}}]{Foot:2013hna}%
  \BibitemOpen
  \bibfield  {author} {\bibinfo {author} {\bibfnamefont {R.}~\bibnamefont
  {Foot}}, \bibinfo {author} {\bibfnamefont {A.}~\bibnamefont {Kobakhidze}},
  \bibinfo {author} {\bibfnamefont {K.~L.}\ \bibnamefont {McDonald}},\ and\
  \bibinfo {author} {\bibfnamefont {R.~R.}\ \bibnamefont {Volkas}},\ }\bibfield
   {title} {\bibinfo {title} {{Poincar\'e protection for a natural electroweak
  scale}},\ }\href {https://doi.org/10.1103/PhysRevD.89.115018} {\bibfield
  {journal} {\bibinfo  {journal} {Phys. Rev. D}\ }\textbf {\bibinfo {volume}
  {89}},\ \bibinfo {pages} {115018} (\bibinfo {year} {2014})},\ \Eprint
  {https://arxiv.org/abs/1310.0223} {arXiv:1310.0223 [hep-ph]} \BibitemShut
  {NoStop}%
\bibitem [{\citenamefont {Volkas}\ \emph {et~al.}(1988)\citenamefont {Volkas},
  \citenamefont {Davies},\ and\ \citenamefont {Joshi}}]{Volkas1988}%
  \BibitemOpen
  \bibfield  {author} {\bibinfo {author} {\bibfnamefont {R.~R.}\ \bibnamefont
  {Volkas}}, \bibinfo {author} {\bibfnamefont {A.~J.}\ \bibnamefont {Davies}},\
  and\ \bibinfo {author} {\bibfnamefont {G.~C.}\ \bibnamefont {Joshi}},\
  }\bibfield  {title} {\bibinfo {title} {{Naturalness of the invisible axion
  model}},\ }\href {https://doi.org/10.1016/0370-2693(88)91084-2} {\bibfield
  {journal} {\bibinfo  {journal} {Phys. Lett. B}\ }\textbf {\bibinfo {volume}
  {215}},\ \bibinfo {pages} {133} (\bibinfo {year} {1988})}\BibitemShut
  {NoStop}%
\bibitem [{\citenamefont {Clarke}\ and\ \citenamefont {Volkas}(2016)}]{nuDFSZ}%
  \BibitemOpen
  \bibfield  {author} {\bibinfo {author} {\bibfnamefont {J.~D.}\ \bibnamefont
  {Clarke}}\ and\ \bibinfo {author} {\bibfnamefont {R.~R.}\ \bibnamefont
  {Volkas}},\ }\bibfield  {title} {\bibinfo {title} {{Technically natural
  nonsupersymmetric model of neutrino masses, baryogenesis, the strong CP
  problem, and dark matter}},\ }\href
  {https://doi.org/10.1103/PhysRevD.93.035001} {\bibfield  {journal} {\bibinfo
  {journal} {Phys. Rev. D}\ }\textbf {\bibinfo {volume} {93}},\ \bibinfo
  {pages} {035001} (\bibinfo {year} {2016})},\ \Eprint
  {https://arxiv.org/abs/1509.07243} {arXiv:1509.07243 [hep-ph]} \BibitemShut
  {NoStop}%
\bibitem [{\citenamefont {Dine}\ \emph {et~al.}(1981)\citenamefont {Dine},
  \citenamefont {Fischler},\ and\ \citenamefont {Srednicki}}]{DFSZ}%
  \BibitemOpen
  \bibfield  {author} {\bibinfo {author} {\bibfnamefont {M.}~\bibnamefont
  {Dine}}, \bibinfo {author} {\bibfnamefont {W.}~\bibnamefont {Fischler}},\
  and\ \bibinfo {author} {\bibfnamefont {M.}~\bibnamefont {Srednicki}},\
  }\bibfield  {title} {\bibinfo {title} {A simple solution to the strong cp
  problem with a harmless axion},\ }\href
  {https://doi.org/https://doi.org/10.1016/0370-2693(81)90590-6} {\bibfield
  {journal} {\bibinfo  {journal} {Physics Letters B}\ }\textbf {\bibinfo
  {volume} {104}},\ \bibinfo {pages} {199} (\bibinfo {year}
  {1981})}\BibitemShut {NoStop}%
\bibitem [{\citenamefont {Zhitnitsky}(1980)}]{Zhitnitsky:1980tq}%
  \BibitemOpen
  \bibfield  {author} {\bibinfo {author} {\bibfnamefont {A.~R.}\ \bibnamefont
  {Zhitnitsky}},\ }\bibfield  {title} {\bibinfo {title} {{On Possible
  Suppression of the Axion Hadron Interactions.}},\ }\href@noop {} {\bibfield
  {journal} {\bibinfo  {journal} {Sov. J. Nucl. Phys.}\ }\textbf {\bibinfo
  {volume} {31}},\ \bibinfo {pages} {260} (\bibinfo {year} {1980})}\BibitemShut
  {NoStop}%
\bibitem [{\citenamefont {Langacker}\ \emph {et~al.}(1986)\citenamefont
  {Langacker}, \citenamefont {Peccei},\ and\ \citenamefont
  {Yanagida}}]{Langacker1986}%
  \BibitemOpen
  \bibfield  {author} {\bibinfo {author} {\bibfnamefont {P.}~\bibnamefont
  {Langacker}}, \bibinfo {author} {\bibfnamefont {R.}~\bibnamefont {Peccei}},\
  and\ \bibinfo {author} {\bibfnamefont {T.}~\bibnamefont {Yanagida}},\
  }\bibfield  {title} {\bibinfo {title} {Invisible axions and light neutrinos
  — are they connected?},\ }\href {https://doi.org/10.1142/S0217732386000683}
  {\bibfield  {journal} {\bibinfo  {journal} {Modern Physics Letters A}\
  }\textbf {\bibinfo {volume} {01}},\ \bibinfo {pages} {541} (\bibinfo {year}
  {1986})},\ \Eprint
  {https://arxiv.org/abs/https://doi.org/10.1142/S0217732386000683}
  {https://doi.org/10.1142/S0217732386000683} \BibitemShut {NoStop}%
\bibitem [{\citenamefont {Shin}(1987)}]{Shin1987}%
  \BibitemOpen
  \bibfield  {author} {\bibinfo {author} {\bibfnamefont {M.}~\bibnamefont
  {Shin}},\ }\bibfield  {title} {\bibinfo {title} {Light-neutrino masses and
  the strong $\mathrm{CP}$ problem},\ }\href
  {https://doi.org/10.1103/PhysRevLett.59.2515} {\bibfield  {journal} {\bibinfo
   {journal} {Phys. Rev. Lett.}\ }\textbf {\bibinfo {volume} {59}},\ \bibinfo
  {pages} {2515} (\bibinfo {year} {1987})}\BibitemShut {NoStop}%
\bibitem [{\citenamefont {Clarke}\ \emph {et~al.}(2015)\citenamefont {Clarke},
  \citenamefont {Foot},\ and\ \citenamefont {Volkas}}]{nu2HDM}%
  \BibitemOpen
  \bibfield  {author} {\bibinfo {author} {\bibfnamefont {J.~D.}\ \bibnamefont
  {Clarke}}, \bibinfo {author} {\bibfnamefont {R.}~\bibnamefont {Foot}},\ and\
  \bibinfo {author} {\bibfnamefont {R.~R.}\ \bibnamefont {Volkas}},\ }\bibfield
   {title} {\bibinfo {title} {{Natural leptogenesis and neutrino masses with
  two Higgs doublets}},\ }\href {https://doi.org/10.1103/PhysRevD.92.033006}
  {\bibfield  {journal} {\bibinfo  {journal} {Phys. Rev. D}\ }\textbf {\bibinfo
  {volume} {92}},\ \bibinfo {pages} {033006} (\bibinfo {year} {2015})},\
  \Eprint {https://arxiv.org/abs/1505.05744} {arXiv:1505.05744 [hep-ph]}
  \BibitemShut {NoStop}%
\bibitem [{\citenamefont {Fukugita}\ and\ \citenamefont
  {Yanagida}(1986)}]{Fukugita:1986hr}%
  \BibitemOpen
  \bibfield  {author} {\bibinfo {author} {\bibfnamefont {M.}~\bibnamefont
  {Fukugita}}\ and\ \bibinfo {author} {\bibfnamefont {T.}~\bibnamefont
  {Yanagida}},\ }\bibfield  {title} {\bibinfo {title} {{Baryogenesis Without
  Grand Unification}},\ }\href {https://doi.org/10.1016/0370-2693(86)91126-3}
  {\bibfield  {journal} {\bibinfo  {journal} {Phys. Lett. B}\ }\textbf
  {\bibinfo {volume} {174}},\ \bibinfo {pages} {45} (\bibinfo {year}
  {1986})}\BibitemShut {NoStop}%
\bibitem [{\citenamefont {Salvio}(2015)}]{Salvio:2015cja}%
  \BibitemOpen
  \bibfield  {author} {\bibinfo {author} {\bibfnamefont {A.}~\bibnamefont
  {Salvio}},\ }\bibfield  {title} {\bibinfo {title} {{A Simple Motivated
  Completion of the Standard Model below the Planck Scale: Axions and
  Right-Handed Neutrinos}},\ }\href
  {https://doi.org/10.1016/j.physletb.2015.03.015} {\bibfield  {journal}
  {\bibinfo  {journal} {Phys. Lett. B}\ }\textbf {\bibinfo {volume} {743}},\
  \bibinfo {pages} {428} (\bibinfo {year} {2015})},\ \Eprint
  {https://arxiv.org/abs/1501.03781} {arXiv:1501.03781 [hep-ph]} \BibitemShut
  {NoStop}%
\bibitem [{\citenamefont {Ballesteros}\ \emph
  {et~al.}(2017{\natexlab{a}})\citenamefont {Ballesteros}, \citenamefont
  {Redondo}, \citenamefont {Ringwald},\ and\ \citenamefont
  {Tamarit}}]{Ballesteros:2016euj}%
  \BibitemOpen
  \bibfield  {author} {\bibinfo {author} {\bibfnamefont {G.}~\bibnamefont
  {Ballesteros}}, \bibinfo {author} {\bibfnamefont {J.}~\bibnamefont
  {Redondo}}, \bibinfo {author} {\bibfnamefont {A.}~\bibnamefont {Ringwald}},\
  and\ \bibinfo {author} {\bibfnamefont {C.}~\bibnamefont {Tamarit}},\
  }\bibfield  {title} {\bibinfo {title} {{Unifying inflation with the axion,
  dark matter, baryogenesis and the seesaw mechanism}},\ }\href
  {https://doi.org/10.1103/PhysRevLett.118.071802} {\bibfield  {journal}
  {\bibinfo  {journal} {Phys. Rev. Lett.}\ }\textbf {\bibinfo {volume} {118}},\
  \bibinfo {pages} {071802} (\bibinfo {year} {2017}{\natexlab{a}})},\ \Eprint
  {https://arxiv.org/abs/1608.05414} {arXiv:1608.05414 [hep-ph]} \BibitemShut
  {NoStop}%
\bibitem [{\citenamefont {Ballesteros}\ \emph
  {et~al.}(2017{\natexlab{b}})\citenamefont {Ballesteros}, \citenamefont
  {Redondo}, \citenamefont {Ringwald},\ and\ \citenamefont
  {Tamarit}}]{Ballesteros:2016xej}%
  \BibitemOpen
  \bibfield  {author} {\bibinfo {author} {\bibfnamefont {G.}~\bibnamefont
  {Ballesteros}}, \bibinfo {author} {\bibfnamefont {J.}~\bibnamefont
  {Redondo}}, \bibinfo {author} {\bibfnamefont {A.}~\bibnamefont {Ringwald}},\
  and\ \bibinfo {author} {\bibfnamefont {C.}~\bibnamefont {Tamarit}},\
  }\bibfield  {title} {\bibinfo {title} {{Standard
  Model\textemdash{}axion\textemdash{}seesaw\textemdash{}Higgs portal
  inflation. Five problems of particle physics and cosmology solved in one
  stroke}},\ }\href {https://doi.org/10.1088/1475-7516/2017/08/001} {\bibfield
  {journal} {\bibinfo  {journal} {JCAP}\ }\textbf {\bibinfo {volume} {08}},\
  \bibinfo {pages} {001}},\ \Eprint {https://arxiv.org/abs/1610.01639}
  {arXiv:1610.01639 [hep-ph]} \BibitemShut {NoStop}%
\bibitem [{\citenamefont {Ballesteros}\ \emph {et~al.}(2019)\citenamefont
  {Ballesteros}, \citenamefont {Redondo}, \citenamefont {Ringwald},\ and\
  \citenamefont {Tamarit}}]{Ballesteros:2019tvf}%
  \BibitemOpen
  \bibfield  {author} {\bibinfo {author} {\bibfnamefont {G.}~\bibnamefont
  {Ballesteros}}, \bibinfo {author} {\bibfnamefont {J.}~\bibnamefont
  {Redondo}}, \bibinfo {author} {\bibfnamefont {A.}~\bibnamefont {Ringwald}},\
  and\ \bibinfo {author} {\bibfnamefont {C.}~\bibnamefont {Tamarit}},\
  }\bibfield  {title} {\bibinfo {title} {{Several Problems in Particle Physics
  and Cosmology Solved in One SMASH}},\ }\href
  {https://doi.org/10.3389/fspas.2019.00055} {\bibfield  {journal} {\bibinfo
  {journal} {Front. Astron. Space Sci.}\ }\textbf {\bibinfo {volume} {6}},\
  \bibinfo {pages} {55} (\bibinfo {year} {2019})},\ \Eprint
  {https://arxiv.org/abs/1904.05594} {arXiv:1904.05594 [hep-ph]} \BibitemShut
  {NoStop}%
\bibitem [{\citenamefont {Salvio}(2019)}]{Salvio:2018rv}%
  \BibitemOpen
  \bibfield  {author} {\bibinfo {author} {\bibfnamefont {A.}~\bibnamefont
  {Salvio}},\ }\bibfield  {title} {\bibinfo {title} {{Critical Higgs inflation
  in a Viable Motivated Model}},\ }\href
  {https://doi.org/10.1103/PhysRevD.99.015037} {\bibfield  {journal} {\bibinfo
  {journal} {Phys. Rev. D}\ }\textbf {\bibinfo {volume} {99}},\ \bibinfo
  {pages} {015037} (\bibinfo {year} {2019})},\ \Eprint
  {https://arxiv.org/abs/1810.00792} {arXiv:1810.00792 [hep-ph]} \BibitemShut
  {NoStop}%
\bibitem [{\citenamefont {Kim}(1979)}]{Kim:1979if}%
  \BibitemOpen
  \bibfield  {author} {\bibinfo {author} {\bibfnamefont {J.~E.}\ \bibnamefont
  {Kim}},\ }\bibfield  {title} {\bibinfo {title} {{Weak Interaction Singlet and
  Strong CP Invariance}},\ }\href {https://doi.org/10.1103/PhysRevLett.43.103}
  {\bibfield  {journal} {\bibinfo  {journal} {Phys. Rev. Lett.}\ }\textbf
  {\bibinfo {volume} {43}},\ \bibinfo {pages} {103} (\bibinfo {year}
  {1979})}\BibitemShut {NoStop}%
\bibitem [{\citenamefont {Shifman}\ \emph {et~al.}(1980)\citenamefont
  {Shifman}, \citenamefont {Vainshtein},\ and\ \citenamefont
  {Zakharov}}]{Shifman:1979if}%
  \BibitemOpen
  \bibfield  {author} {\bibinfo {author} {\bibfnamefont {M.~A.}\ \bibnamefont
  {Shifman}}, \bibinfo {author} {\bibfnamefont {A.~I.}\ \bibnamefont
  {Vainshtein}},\ and\ \bibinfo {author} {\bibfnamefont {V.~I.}\ \bibnamefont
  {Zakharov}},\ }\bibfield  {title} {\bibinfo {title} {{Can Confinement Ensure
  Natural CP Invariance of Strong Interactions?}},\ }\href
  {https://doi.org/10.1016/0550-3213(80)90209-6} {\bibfield  {journal}
  {\bibinfo  {journal} {Nucl. Phys. B}\ }\textbf {\bibinfo {volume} {166}},\
  \bibinfo {pages} {493} (\bibinfo {year} {1980})}\BibitemShut {NoStop}%
\bibitem [{\citenamefont {Oda}\ \emph {et~al.}(2020)\citenamefont {Oda},
  \citenamefont {Shoji},\ and\ \citenamefont {Takahashi}}]{Oda:2019njo}%
  \BibitemOpen
  \bibfield  {author} {\bibinfo {author} {\bibfnamefont {S.}~\bibnamefont
  {Oda}}, \bibinfo {author} {\bibfnamefont {Y.}~\bibnamefont {Shoji}},\ and\
  \bibinfo {author} {\bibfnamefont {D.-S.}\ \bibnamefont {Takahashi}},\
  }\bibfield  {title} {\bibinfo {title} {{High Scale Validity of the DFSZ Axion
  Model with Precision}},\ }\href {https://doi.org/10.1007/JHEP03(2020)011}
  {\bibfield  {journal} {\bibinfo  {journal} {JHEP}\ }\textbf {\bibinfo
  {volume} {03}},\ \bibinfo {pages} {011}},\ \Eprint
  {https://arxiv.org/abs/1912.01147} {arXiv:1912.01147 [hep-ph]} \BibitemShut
  {NoStop}%
\bibitem [{\citenamefont {Starobinsky}(1980)}]{STAROBINSKY198099}%
  \BibitemOpen
  \bibfield  {author} {\bibinfo {author} {\bibfnamefont {A.}~\bibnamefont
  {Starobinsky}},\ }\bibfield  {title} {\bibinfo {title} {A new type of
  isotropic cosmological models without singularity},\ }\href
  {https://doi.org/https://doi.org/10.1016/0370-2693(80)90670-X} {\bibfield
  {journal} {\bibinfo  {journal} {Physics Letters B}\ }\textbf {\bibinfo
  {volume} {91}},\ \bibinfo {pages} {99} (\bibinfo {year} {1980})}\BibitemShut
  {NoStop}%
\bibitem [{\citenamefont {Guth}(1981)}]{Guth1981}%
  \BibitemOpen
  \bibfield  {author} {\bibinfo {author} {\bibfnamefont {A.~H.}\ \bibnamefont
  {Guth}},\ }\bibfield  {title} {\bibinfo {title} {Inflationary universe: A
  possible solution to the horizon and flatness problems},\ }\href
  {https://doi.org/10.1103/PhysRevD.23.347} {\bibfield  {journal} {\bibinfo
  {journal} {Phys. Rev. D}\ }\textbf {\bibinfo {volume} {23}},\ \bibinfo
  {pages} {347} (\bibinfo {year} {1981})}\BibitemShut {NoStop}%
\bibitem [{\citenamefont {Kaiser}\ and\ \citenamefont
  {Sfakianakis}(2014)}]{Kaiser:2013sna}%
  \BibitemOpen
  \bibfield  {author} {\bibinfo {author} {\bibfnamefont {D.~I.}\ \bibnamefont
  {Kaiser}}\ and\ \bibinfo {author} {\bibfnamefont {E.~I.}\ \bibnamefont
  {Sfakianakis}},\ }\bibfield  {title} {\bibinfo {title} {{Multifield Inflation
  after Planck: The Case for Nonminimal Couplings}},\ }\href
  {https://doi.org/10.1103/PhysRevLett.112.011302} {\bibfield  {journal}
  {\bibinfo  {journal} {Phys. Rev. Lett.}\ }\textbf {\bibinfo {volume} {112}},\
  \bibinfo {pages} {011302} (\bibinfo {year} {2014})},\ \Eprint
  {https://arxiv.org/abs/1304.0363} {arXiv:1304.0363 [astro-ph.CO]}
  \BibitemShut {NoStop}%
\bibitem [{\citenamefont {DeCross}\ \emph
  {et~al.}(2018{\natexlab{a}})\citenamefont {DeCross}, \citenamefont {Kaiser},
  \citenamefont {Prabhu}, \citenamefont {Prescod-Weinstein},\ and\
  \citenamefont {Sfakianakis}}]{DeCross:2015uza}%
  \BibitemOpen
  \bibfield  {author} {\bibinfo {author} {\bibfnamefont {M.~P.}\ \bibnamefont
  {DeCross}}, \bibinfo {author} {\bibfnamefont {D.~I.}\ \bibnamefont {Kaiser}},
  \bibinfo {author} {\bibfnamefont {A.}~\bibnamefont {Prabhu}}, \bibinfo
  {author} {\bibfnamefont {C.}~\bibnamefont {Prescod-Weinstein}},\ and\
  \bibinfo {author} {\bibfnamefont {E.~I.}\ \bibnamefont {Sfakianakis}},\
  }\bibfield  {title} {\bibinfo {title} {{Preheating after Multifield Inflation
  with Nonminimal Couplings, I: Covariant Formalism and Attractor Behavior}},\
  }\href {https://doi.org/10.1103/PhysRevD.97.023526} {\bibfield  {journal}
  {\bibinfo  {journal} {Phys. Rev. D}\ }\textbf {\bibinfo {volume} {97}},\
  \bibinfo {pages} {023526} (\bibinfo {year} {2018}{\natexlab{a}})},\ \Eprint
  {https://arxiv.org/abs/1510.08553} {arXiv:1510.08553 [astro-ph.CO]}
  \BibitemShut {NoStop}%
\bibitem [{\citenamefont {Akrami}\ \emph {et~al.}(2020)\citenamefont {Akrami}
  \emph {et~al.}}]{Planck:2018jri}%
  \BibitemOpen
  \bibfield  {author} {\bibinfo {author} {\bibfnamefont {Y.}~\bibnamefont
  {Akrami}} \emph {et~al.} (\bibinfo {collaboration} {Planck}),\ }\bibfield
  {title} {\bibinfo {title} {{Planck 2018 results. X. Constraints on
  inflation}},\ }\href {https://doi.org/10.1051/0004-6361/201833887} {\bibfield
   {journal} {\bibinfo  {journal} {Astron. Astrophys.}\ }\textbf {\bibinfo
  {volume} {641}},\ \bibinfo {pages} {A10} (\bibinfo {year} {2020})},\ \Eprint
  {https://arxiv.org/abs/1807.06211} {arXiv:1807.06211 [astro-ph.CO]}
  \BibitemShut {NoStop}%
\bibitem [{\citenamefont {Sikivie}(1982)}]{Sikivie:1982qv}%
  \BibitemOpen
  \bibfield  {author} {\bibinfo {author} {\bibfnamefont {P.}~\bibnamefont
  {Sikivie}},\ }\bibfield  {title} {\bibinfo {title} {{Axions, Domain Walls and
  the Early Universe}},\ }\href {https://doi.org/10.1103/PhysRevLett.48.1156}
  {\bibfield  {journal} {\bibinfo  {journal} {Phys. Rev. Lett.}\ }\textbf
  {\bibinfo {volume} {48}},\ \bibinfo {pages} {1156} (\bibinfo {year}
  {1982})}\BibitemShut {NoStop}%
\bibitem [{\citenamefont {Peccei}\ \emph {et~al.}(1986)\citenamefont {Peccei},
  \citenamefont {Wu},\ and\ \citenamefont {Yanagida}}]{Peccei:1986pn}%
  \BibitemOpen
  \bibfield  {author} {\bibinfo {author} {\bibfnamefont {R.~D.}\ \bibnamefont
  {Peccei}}, \bibinfo {author} {\bibfnamefont {T.~T.}\ \bibnamefont {Wu}},\
  and\ \bibinfo {author} {\bibfnamefont {T.}~\bibnamefont {Yanagida}},\
  }\bibfield  {title} {\bibinfo {title} {{A Viable Axion Model}},\ }\href
  {https://doi.org/10.1016/0370-2693(86)90284-4} {\bibfield  {journal}
  {\bibinfo  {journal} {Phys. Lett. B}\ }\textbf {\bibinfo {volume} {172}},\
  \bibinfo {pages} {435} (\bibinfo {year} {1986})}\BibitemShut {NoStop}%
\bibitem [{\citenamefont {Krauss}\ and\ \citenamefont
  {Wilczek}(1986)}]{KRAUSS1986189}%
  \BibitemOpen
  \bibfield  {author} {\bibinfo {author} {\bibfnamefont {L.~M.}\ \bibnamefont
  {Krauss}}\ and\ \bibinfo {author} {\bibfnamefont {F.}~\bibnamefont
  {Wilczek}},\ }\bibfield  {title} {\bibinfo {title} {A short-lived axion
  variant},\ }\href
  {https://doi.org/https://doi.org/10.1016/0370-2693(86)90244-3} {\bibfield
  {journal} {\bibinfo  {journal} {Physics Letters B}\ }\textbf {\bibinfo
  {volume} {173}},\ \bibinfo {pages} {189} (\bibinfo {year}
  {1986})}\BibitemShut {NoStop}%
\bibitem [{\citenamefont {Davidson}\ and\ \citenamefont
  {Vozmediano}(1984{\natexlab{a}})}]{Davidson:1983tp}%
  \BibitemOpen
  \bibfield  {author} {\bibinfo {author} {\bibfnamefont {A.}~\bibnamefont
  {Davidson}}\ and\ \bibinfo {author} {\bibfnamefont {M.~A.~H.}\ \bibnamefont
  {Vozmediano}},\ }\bibfield  {title} {\bibinfo {title} {{Domain walls:
  horizontal epilogue}},\ }\href {https://doi.org/10.1016/0370-2693(84)90198-9}
  {\bibfield  {journal} {\bibinfo  {journal} {Phys. Lett. B}\ }\textbf
  {\bibinfo {volume} {141}},\ \bibinfo {pages} {177} (\bibinfo {year}
  {1984}{\natexlab{a}})}\BibitemShut {NoStop}%
\bibitem [{\citenamefont {Davidson}\ and\ \citenamefont
  {Vozmediano}(1984{\natexlab{b}})}]{Davidson:1984ik}%
  \BibitemOpen
  \bibfield  {author} {\bibinfo {author} {\bibfnamefont {A.}~\bibnamefont
  {Davidson}}\ and\ \bibinfo {author} {\bibfnamefont {M.~A.~H.}\ \bibnamefont
  {Vozmediano}},\ }\bibfield  {title} {\bibinfo {title} {{The Horizontal Axion
  Alternative: The Interplay of Vacuum Structure and Flavor Interactions}},\
  }\href {https://doi.org/10.1016/0550-3213(84)90616-3} {\bibfield  {journal}
  {\bibinfo  {journal} {Nucl. Phys. B}\ }\textbf {\bibinfo {volume} {248}},\
  \bibinfo {pages} {647} (\bibinfo {year} {1984}{\natexlab{b}})}\BibitemShut
  {NoStop}%
\bibitem [{\citenamefont {Geng}\ and\ \citenamefont {Ng}(1989)}]{Geng1989}%
  \BibitemOpen
  \bibfield  {author} {\bibinfo {author} {\bibfnamefont {C.~Q.}\ \bibnamefont
  {Geng}}\ and\ \bibinfo {author} {\bibfnamefont {J.~N.}\ \bibnamefont {Ng}},\
  }\bibfield  {title} {\bibinfo {title} {Flavor connections and neutrino-mass
  hierarchy in variant invisible-axion models without domain-wall problem},\
  }\href {https://doi.org/10.1103/PhysRevD.39.1449} {\bibfield  {journal}
  {\bibinfo  {journal} {Phys. Rev. D}\ }\textbf {\bibinfo {volume} {39}},\
  \bibinfo {pages} {1449} (\bibinfo {year} {1989})}\BibitemShut {NoStop}%
\bibitem [{\citenamefont {Geng}\ and\ \citenamefont {Ng}(1990)}]{Geng:1990dv}%
  \BibitemOpen
  \bibfield  {author} {\bibinfo {author} {\bibfnamefont {C.~Q.}\ \bibnamefont
  {Geng}}\ and\ \bibinfo {author} {\bibfnamefont {J.~N.}\ \bibnamefont {Ng}},\
  }\bibfield  {title} {\bibinfo {title} {{The domain wall number in various
  invisible axion models}},\ }\href {https://doi.org/10.1103/PhysRevD.41.3848}
  {\bibfield  {journal} {\bibinfo  {journal} {Phys. Rev. D}\ }\textbf {\bibinfo
  {volume} {41}},\ \bibinfo {pages} {3848} (\bibinfo {year}
  {1990})}\BibitemShut {NoStop}%
\bibitem [{\citenamefont {Chiang}\ \emph {et~al.}(2015)\citenamefont {Chiang},
  \citenamefont {Fukuda}, \citenamefont {Takeuchi},\ and\ \citenamefont
  {Yanagida}}]{Chiang:2015cba}%
  \BibitemOpen
  \bibfield  {author} {\bibinfo {author} {\bibfnamefont {C.-W.}\ \bibnamefont
  {Chiang}}, \bibinfo {author} {\bibfnamefont {H.}~\bibnamefont {Fukuda}},
  \bibinfo {author} {\bibfnamefont {M.}~\bibnamefont {Takeuchi}},\ and\
  \bibinfo {author} {\bibfnamefont {T.~T.}\ \bibnamefont {Yanagida}},\
  }\bibfield  {title} {\bibinfo {title} {{Flavor-Changing Neutral-Current
  Decays in Top-Specific Variant Axion Model}},\ }\href
  {https://doi.org/10.1007/JHEP11(2015)057} {\bibfield  {journal} {\bibinfo
  {journal} {JHEP}\ }\textbf {\bibinfo {volume} {11}},\ \bibinfo {pages}
  {057}},\ \Eprint {https://arxiv.org/abs/1507.04354} {arXiv:1507.04354
  [hep-ph]} \BibitemShut {NoStop}%
\bibitem [{\citenamefont {Chiang}\ \emph {et~al.}(2018)\citenamefont {Chiang},
  \citenamefont {Fukuda}, \citenamefont {Takeuchi},\ and\ \citenamefont
  {Yanagida}}]{Chiang:2017fjr}%
  \BibitemOpen
  \bibfield  {author} {\bibinfo {author} {\bibfnamefont {C.-W.}\ \bibnamefont
  {Chiang}}, \bibinfo {author} {\bibfnamefont {H.}~\bibnamefont {Fukuda}},
  \bibinfo {author} {\bibfnamefont {M.}~\bibnamefont {Takeuchi}},\ and\
  \bibinfo {author} {\bibfnamefont {T.~T.}\ \bibnamefont {Yanagida}},\
  }\bibfield  {title} {\bibinfo {title} {{Current Status of Top-Specific
  Variant Axion Model}},\ }\href {https://doi.org/10.1103/PhysRevD.97.035015}
  {\bibfield  {journal} {\bibinfo  {journal} {Phys. Rev. D}\ }\textbf {\bibinfo
  {volume} {97}},\ \bibinfo {pages} {035015} (\bibinfo {year} {2018})},\
  \Eprint {https://arxiv.org/abs/1711.02993} {arXiv:1711.02993 [hep-ph]}
  \BibitemShut {NoStop}%
\bibitem [{\citenamefont {Klimenko}(1985)}]{Klimenko:1984qx}%
  \BibitemOpen
  \bibfield  {author} {\bibinfo {author} {\bibfnamefont {K.~G.}\ \bibnamefont
  {Klimenko}},\ }\bibfield  {title} {\bibinfo {title} {{On Necessary and
  Sufficient Conditions for Some Higgs Potentials to Be Bounded From Below}},\
  }\href {https://doi.org/10.1007/BF01034825} {\bibfield  {journal} {\bibinfo
  {journal} {Theor. Math. Phys.}\ }\textbf {\bibinfo {volume} {62}},\ \bibinfo
  {pages} {58} (\bibinfo {year} {1985})}\BibitemShut {NoStop}%
\bibitem [{\citenamefont {Deshpande}\ and\ \citenamefont
  {Ma}(1978)}]{Deshpande1978}%
  \BibitemOpen
  \bibfield  {author} {\bibinfo {author} {\bibfnamefont {N.~G.}\ \bibnamefont
  {Deshpande}}\ and\ \bibinfo {author} {\bibfnamefont {E.}~\bibnamefont {Ma}},\
  }\bibfield  {title} {\bibinfo {title} {Pattern of symmetry breaking with two
  higgs doublets},\ }\href {https://doi.org/10.1103/PhysRevD.18.2574}
  {\bibfield  {journal} {\bibinfo  {journal} {Phys. Rev. D}\ }\textbf {\bibinfo
  {volume} {18}},\ \bibinfo {pages} {2574} (\bibinfo {year}
  {1978})}\BibitemShut {NoStop}%
\bibitem [{\citenamefont {Maniatis}\ \emph {et~al.}(2006)\citenamefont
  {Maniatis}, \citenamefont {von Manteuffel}, \citenamefont {Nachtmann},\ and\
  \citenamefont {Nagel}}]{Maniatis:2006fs}%
  \BibitemOpen
  \bibfield  {author} {\bibinfo {author} {\bibfnamefont {M.}~\bibnamefont
  {Maniatis}}, \bibinfo {author} {\bibfnamefont {A.}~\bibnamefont {von
  Manteuffel}}, \bibinfo {author} {\bibfnamefont {O.}~\bibnamefont
  {Nachtmann}},\ and\ \bibinfo {author} {\bibfnamefont {F.}~\bibnamefont
  {Nagel}},\ }\bibfield  {title} {\bibinfo {title} {{Stability and symmetry
  breaking in the general two-Higgs-doublet model}},\ }\href
  {https://doi.org/10.1140/epjc/s10052-006-0016-6} {\bibfield  {journal}
  {\bibinfo  {journal} {Eur. Phys. J. C}\ }\textbf {\bibinfo {volume} {48}},\
  \bibinfo {pages} {805} (\bibinfo {year} {2006})},\ \Eprint
  {https://arxiv.org/abs/hep-ph/0605184} {arXiv:hep-ph/0605184} \BibitemShut
  {NoStop}%
\bibitem [{\citenamefont {Hou}\ and\ \citenamefont
  {Modak}(2021)}]{Hou:2020chc}%
  \BibitemOpen
  \bibfield  {author} {\bibinfo {author} {\bibfnamefont {W.-S.}\ \bibnamefont
  {Hou}}\ and\ \bibinfo {author} {\bibfnamefont {T.}~\bibnamefont {Modak}},\
  }\bibfield  {title} {\bibinfo {title} {{Probing Top Changing Neutral Higgs
  Couplings at Colliders}},\ }\href {https://doi.org/10.1142/S0217732321300068}
  {\bibfield  {journal} {\bibinfo  {journal} {Mod. Phys. Lett. A}\ }\textbf
  {\bibinfo {volume} {36}},\ \bibinfo {pages} {2130006} (\bibinfo {year}
  {2021})},\ \Eprint {https://arxiv.org/abs/2012.05735} {arXiv:2012.05735
  [hep-ph]} \BibitemShut {NoStop}%
\bibitem [{\citenamefont {Tumasyan}\ \emph {et~al.}(2022)\citenamefont
  {Tumasyan} \emph {et~al.}}]{CMS:2021cqc}%
  \BibitemOpen
  \bibfield  {author} {\bibinfo {author} {\bibfnamefont {A.}~\bibnamefont
  {Tumasyan}} \emph {et~al.} (\bibinfo {collaboration} {CMS}),\ }\bibfield
  {title} {\bibinfo {title} {{Search for Flavor-Changing Neutral Current
  Interactions of the Top Quark and Higgs Boson in Final States with Two
  Photons in Proton-Proton Collisions at $\sqrt{s}=13\text{ }\text{
  }\mathrm{TeV}$}},\ }\href {https://doi.org/10.1103/PhysRevLett.129.032001}
  {\bibfield  {journal} {\bibinfo  {journal} {Phys. Rev. Lett.}\ }\textbf
  {\bibinfo {volume} {129}},\ \bibinfo {pages} {032001} (\bibinfo {year}
  {2022})},\ \Eprint {https://arxiv.org/abs/2111.02219} {arXiv:2111.02219
  [hep-ex]} \BibitemShut {NoStop}%
\bibitem [{\citenamefont {Kohda}\ \emph {et~al.}(2018)\citenamefont {Kohda},
  \citenamefont {Modak},\ and\ \citenamefont {Hou}}]{KOHDA2018379}%
  \BibitemOpen
  \bibfield  {author} {\bibinfo {author} {\bibfnamefont {M.}~\bibnamefont
  {Kohda}}, \bibinfo {author} {\bibfnamefont {T.}~\bibnamefont {Modak}},\ and\
  \bibinfo {author} {\bibfnamefont {W.-S.}\ \bibnamefont {Hou}},\ }\bibfield
  {title} {\bibinfo {title} {Searching for new scalar bosons via triple-top
  signature in $cg \rightarrow t \mathrm{S}^0 \rightarrow tt\overline{t}$},\
  }\href {https://doi.org/https://doi.org/10.1016/j.physletb.2017.11.056}
  {\bibfield  {journal} {\bibinfo  {journal} {Physics Letters B}\ }\textbf
  {\bibinfo {volume} {776}},\ \bibinfo {pages} {379} (\bibinfo {year}
  {2018})}\BibitemShut {NoStop}%
\bibitem [{\citenamefont {Ghosh}\ \emph {et~al.}(2020)\citenamefont {Ghosh},
  \citenamefont {Hou},\ and\ \citenamefont {Modak}}]{Ghosh2020}%
  \BibitemOpen
  \bibfield  {author} {\bibinfo {author} {\bibfnamefont {D.~K.}\ \bibnamefont
  {Ghosh}}, \bibinfo {author} {\bibfnamefont {W.-S.}\ \bibnamefont {Hou}},\
  and\ \bibinfo {author} {\bibfnamefont {T.}~\bibnamefont {Modak}},\ }\bibfield
   {title} {\bibinfo {title} {Sub-tev ${H}^{+}$ boson production as probe of
  extra top yukawa couplings},\ }\href
  {https://doi.org/10.1103/PhysRevLett.125.221801} {\bibfield  {journal}
  {\bibinfo  {journal} {Phys. Rev. Lett.}\ }\textbf {\bibinfo {volume} {125}},\
  \bibinfo {pages} {221801} (\bibinfo {year} {2020})}\BibitemShut {NoStop}%
\bibitem [{\citenamefont {Schutz}\ \emph {et~al.}(2014)\citenamefont {Schutz},
  \citenamefont {Sfakianakis},\ and\ \citenamefont {Kaiser}}]{Schutz:2013fua}%
  \BibitemOpen
  \bibfield  {author} {\bibinfo {author} {\bibfnamefont {K.}~\bibnamefont
  {Schutz}}, \bibinfo {author} {\bibfnamefont {E.~I.}\ \bibnamefont
  {Sfakianakis}},\ and\ \bibinfo {author} {\bibfnamefont {D.~I.}\ \bibnamefont
  {Kaiser}},\ }\bibfield  {title} {\bibinfo {title} {{Multifield Inflation
  after Planck: Isocurvature Modes from Nonminimal Couplings}},\ }\href
  {https://doi.org/10.1103/PhysRevD.89.064044} {\bibfield  {journal} {\bibinfo
  {journal} {Phys. Rev. D}\ }\textbf {\bibinfo {volume} {89}},\ \bibinfo
  {pages} {064044} (\bibinfo {year} {2014})},\ \Eprint
  {https://arxiv.org/abs/1310.8285} {arXiv:1310.8285 [astro-ph.CO]}
  \BibitemShut {NoStop}%
\bibitem [{\citenamefont {Gong}\ \emph {et~al.}(2012)\citenamefont {Gong},
  \citenamefont {Lee},\ and\ \citenamefont {Kang}}]{Gong:2012ri}%
  \BibitemOpen
  \bibfield  {author} {\bibinfo {author} {\bibfnamefont {J.-O.}\ \bibnamefont
  {Gong}}, \bibinfo {author} {\bibfnamefont {H.~M.}\ \bibnamefont {Lee}},\ and\
  \bibinfo {author} {\bibfnamefont {S.~K.}\ \bibnamefont {Kang}},\ }\bibfield
  {title} {\bibinfo {title} {{Inflation and dark matter in two Higgs doublet
  models}},\ }\href {https://doi.org/10.1007/JHEP04(2012)128} {\bibfield
  {journal} {\bibinfo  {journal} {JHEP}\ }\textbf {\bibinfo {volume} {04}},\
  \bibinfo {pages} {128}},\ \Eprint {https://arxiv.org/abs/1202.0288}
  {arXiv:1202.0288 [hep-ph]} \BibitemShut {NoStop}%
\bibitem [{\citenamefont {Modak}\ and\ \citenamefont
  {Oda}(2020)}]{Modak:2020fij}%
  \BibitemOpen
  \bibfield  {author} {\bibinfo {author} {\bibfnamefont {T.}~\bibnamefont
  {Modak}}\ and\ \bibinfo {author} {\bibfnamefont {K.-y.}\ \bibnamefont
  {Oda}},\ }\bibfield  {title} {\bibinfo {title} {{Echoes of 2HDM inflation at
  the collider experiments}},\ }\href
  {https://doi.org/10.1140/epjc/s10052-021-09295-2} {\bibfield  {journal}
  {\bibinfo  {journal} {Eur. Phys. J. C}\ }\textbf {\bibinfo {volume} {80}},\
  \bibinfo {pages} {863} (\bibinfo {year} {2020})},\ \bibinfo {note} {[Erratum:
  Eur.Phys.J.C 81, 518 (2021)]},\ \Eprint {https://arxiv.org/abs/2007.08141}
  {arXiv:2007.08141 [hep-ph]} \BibitemShut {NoStop}%
\bibitem [{\citenamefont {Nakayama}\ and\ \citenamefont
  {Takimoto}(2015)}]{Nakayama2015}%
  \BibitemOpen
  \bibfield  {author} {\bibinfo {author} {\bibfnamefont {K.}~\bibnamefont
  {Nakayama}}\ and\ \bibinfo {author} {\bibfnamefont {M.}~\bibnamefont
  {Takimoto}},\ }\bibfield  {title} {\bibinfo {title} {{Higgs inflation and
  suppression of axion isocurvature perturbation}},\ }\href
  {https://doi.org/10.1016/j.physletb.2015.07.001} {\bibfield  {journal}
  {\bibinfo  {journal} {Phys. Lett. B}\ }\textbf {\bibinfo {volume} {748}},\
  \bibinfo {pages} {108} (\bibinfo {year} {2015})},\ \Eprint
  {https://arxiv.org/abs/1505.02119} {arXiv:1505.02119 [hep-ph]} \BibitemShut
  {NoStop}%
\bibitem [{\citenamefont {Chernikov}\ and\ \citenamefont
  {Tagirov}(1968)}]{Chernikov:1968zm}%
  \BibitemOpen
  \bibfield  {author} {\bibinfo {author} {\bibfnamefont {N.~A.}\ \bibnamefont
  {Chernikov}}\ and\ \bibinfo {author} {\bibfnamefont {E.~A.}\ \bibnamefont
  {Tagirov}},\ }\bibfield  {title} {\bibinfo {title} {{Quantum theory of scalar
  fields in de Sitter space-time}},\ }\href@noop {} {\bibfield  {journal}
  {\bibinfo  {journal} {Ann. Inst. H. Poincare Phys. Theor. A}\ }\textbf
  {\bibinfo {volume} {9}},\ \bibinfo {pages} {109} (\bibinfo {year}
  {1968})}\BibitemShut {NoStop}%
\bibitem [{\citenamefont {Callan}\ \emph {et~al.}(1970)\citenamefont {Callan},
  \citenamefont {Coleman},\ and\ \citenamefont {Jackiw}}]{CALLAN197042}%
  \BibitemOpen
  \bibfield  {author} {\bibinfo {author} {\bibfnamefont {C.~G.}\ \bibnamefont
  {Callan}}, \bibinfo {author} {\bibfnamefont {S.}~\bibnamefont {Coleman}},\
  and\ \bibinfo {author} {\bibfnamefont {R.}~\bibnamefont {Jackiw}},\
  }\bibfield  {title} {\bibinfo {title} {A new improved energy-momentum
  tensor},\ }\href
  {https://doi.org/https://doi.org/10.1016/0003-4916(70)90394-5} {\bibfield
  {journal} {\bibinfo  {journal} {Annals of Physics}\ }\textbf {\bibinfo
  {volume} {59}},\ \bibinfo {pages} {42} (\bibinfo {year} {1970})}\BibitemShut
  {NoStop}%
\bibitem [{\citenamefont {Birrell}\ and\ \citenamefont
  {Davies}(1980)}]{Birrell1980}%
  \BibitemOpen
  \bibfield  {author} {\bibinfo {author} {\bibfnamefont {N.~D.}\ \bibnamefont
  {Birrell}}\ and\ \bibinfo {author} {\bibfnamefont {P.~C.~W.}\ \bibnamefont
  {Davies}},\ }\bibfield  {title} {\bibinfo {title} {Conformal-symmetry
  breaking and cosmological particle creation in
  $\ensuremath{\lambda}{\ensuremath{\varphi}}^{4}$ theory},\ }\href
  {https://doi.org/10.1103/PhysRevD.22.322} {\bibfield  {journal} {\bibinfo
  {journal} {Phys. Rev. D}\ }\textbf {\bibinfo {volume} {22}},\ \bibinfo
  {pages} {322} (\bibinfo {year} {1980})}\BibitemShut {NoStop}%
\bibitem [{\citenamefont {Bunch}\ \emph {et~al.}(1980)\citenamefont {Bunch},
  \citenamefont {Panangaden},\ and\ \citenamefont {Parker}}]{Bunch_1980}%
  \BibitemOpen
  \bibfield  {author} {\bibinfo {author} {\bibfnamefont {T.~S.}\ \bibnamefont
  {Bunch}}, \bibinfo {author} {\bibfnamefont {P.}~\bibnamefont {Panangaden}},\
  and\ \bibinfo {author} {\bibfnamefont {L.}~\bibnamefont {Parker}},\
  }\bibfield  {title} {\bibinfo {title} {On renormalisation of $\lambda\phi^4$
  field theory in curved space-time. i},\ }\href
  {https://doi.org/10.1088/0305-4470/13/3/022} {\bibfield  {journal} {\bibinfo
  {journal} {Journal of Physics A: Mathematical and General}\ }\textbf
  {\bibinfo {volume} {13}},\ \bibinfo {pages} {901} (\bibinfo {year}
  {1980})}\BibitemShut {NoStop}%
\bibitem [{\citenamefont {Bunch}\ and\ \citenamefont
  {Panangaden}(1980)}]{Bunch_1980_2}%
  \BibitemOpen
  \bibfield  {author} {\bibinfo {author} {\bibfnamefont {T.~S.}\ \bibnamefont
  {Bunch}}\ and\ \bibinfo {author} {\bibfnamefont {P.}~\bibnamefont
  {Panangaden}},\ }\bibfield  {title} {\bibinfo {title} {On renormalisation of
  $\lambda\phi^4$ field theory in curved space-time. ii},\ }\href
  {https://doi.org/10.1088/0305-4470/13/3/023} {\bibfield  {journal} {\bibinfo
  {journal} {Journal of Physics A: Mathematical and General}\ }\textbf
  {\bibinfo {volume} {13}},\ \bibinfo {pages} {919} (\bibinfo {year}
  {1980})}\BibitemShut {NoStop}%
\bibitem [{\citenamefont {Birrell}\ and\ \citenamefont
  {Davies}(1984)}]{Birrell:1982ix}%
  \BibitemOpen
  \bibfield  {author} {\bibinfo {author} {\bibfnamefont {N.~D.}\ \bibnamefont
  {Birrell}}\ and\ \bibinfo {author} {\bibfnamefont {P.~C.~W.}\ \bibnamefont
  {Davies}},\ }\href {https://doi.org/10.1017/CBO9780511622632} {\emph
  {\bibinfo {title} {{Quantum Fields in Curved Space}}}},\ Cambridge Monographs
  on Mathematical Physics\ (\bibinfo  {publisher} {Cambridge Univ. Press},\
  \bibinfo {address} {Cambridge, UK},\ \bibinfo {year} {1984})\BibitemShut
  {NoStop}%
\bibitem [{\citenamefont {Nelson}\ and\ \citenamefont
  {Panangaden}(1982)}]{Nelson1982}%
  \BibitemOpen
  \bibfield  {author} {\bibinfo {author} {\bibfnamefont {B.~L.}\ \bibnamefont
  {Nelson}}\ and\ \bibinfo {author} {\bibfnamefont {P.}~\bibnamefont
  {Panangaden}},\ }\bibfield  {title} {\bibinfo {title} {Scaling behavior of
  interacting quantum fields in curved spacetime},\ }\href
  {https://doi.org/10.1103/PhysRevD.25.1019} {\bibfield  {journal} {\bibinfo
  {journal} {Phys. Rev. D}\ }\textbf {\bibinfo {volume} {25}},\ \bibinfo
  {pages} {1019} (\bibinfo {year} {1982})}\BibitemShut {NoStop}%
\bibitem [{\citenamefont {Ford}\ and\ \citenamefont {Toms}(1982)}]{Ford1982}%
  \BibitemOpen
  \bibfield  {author} {\bibinfo {author} {\bibfnamefont {L.~H.}\ \bibnamefont
  {Ford}}\ and\ \bibinfo {author} {\bibfnamefont {D.~J.}\ \bibnamefont
  {Toms}},\ }\bibfield  {title} {\bibinfo {title} {Dynamical symmetry breaking
  due to radiative corrections in cosmology},\ }\href
  {https://doi.org/10.1103/PhysRevD.25.1510} {\bibfield  {journal} {\bibinfo
  {journal} {Phys. Rev. D}\ }\textbf {\bibinfo {volume} {25}},\ \bibinfo
  {pages} {1510} (\bibinfo {year} {1982})}\BibitemShut {NoStop}%
\bibitem [{\citenamefont {Parker}\ and\ \citenamefont
  {Toms}(1984)}]{Parker1984}%
  \BibitemOpen
  \bibfield  {author} {\bibinfo {author} {\bibfnamefont {L.}~\bibnamefont
  {Parker}}\ and\ \bibinfo {author} {\bibfnamefont {D.~J.}\ \bibnamefont
  {Toms}},\ }\bibfield  {title} {\bibinfo {title} {Renormalization-group
  analysis of grand unified theories in curved spacetime},\ }\href
  {https://doi.org/10.1103/PhysRevD.29.1584} {\bibfield  {journal} {\bibinfo
  {journal} {Phys. Rev. D}\ }\textbf {\bibinfo {volume} {29}},\ \bibinfo
  {pages} {1584} (\bibinfo {year} {1984})}\BibitemShut {NoStop}%
\bibitem [{\citenamefont {Parker}\ and\ \citenamefont
  {Toms}(1985)}]{Parker1985}%
  \BibitemOpen
  \bibfield  {author} {\bibinfo {author} {\bibfnamefont {L.}~\bibnamefont
  {Parker}}\ and\ \bibinfo {author} {\bibfnamefont {D.~J.}\ \bibnamefont
  {Toms}},\ }\bibfield  {title} {\bibinfo {title} {Renormalization group and
  nonlocal terms in the curved-spacetime effective action: Weak-field
  results},\ }\href {https://doi.org/10.1103/PhysRevD.32.1409} {\bibfield
  {journal} {\bibinfo  {journal} {Phys. Rev. D}\ }\textbf {\bibinfo {volume}
  {32}},\ \bibinfo {pages} {1409} (\bibinfo {year} {1985})}\BibitemShut
  {NoStop}%
\bibitem [{\citenamefont {Buchbinder}\ \emph {et~al.}(2017)\citenamefont
  {Buchbinder}, \citenamefont {Odintsov},\ and\ \citenamefont
  {Shapiro}}]{Buchbinder2017}%
  \BibitemOpen
  \bibfield  {author} {\bibinfo {author} {\bibfnamefont {I.~L.}\ \bibnamefont
  {Buchbinder}}, \bibinfo {author} {\bibfnamefont {S.~D.}\ \bibnamefont
  {Odintsov}},\ and\ \bibinfo {author} {\bibfnamefont {I.~L.}\ \bibnamefont
  {Shapiro}},\ }\href {https://doi.org/10.1201/9780203758922} {\emph {\bibinfo
  {title} {Effective Action in Quantum Gravity}}}\ (\bibinfo  {publisher}
  {Routledge},\ \bibinfo {year} {2017})\BibitemShut {NoStop}%
\bibitem [{\citenamefont {Faraoni}(2004)}]{faraoni2004cosmology}%
  \BibitemOpen
  \bibfield  {author} {\bibinfo {author} {\bibfnamefont {V.}~\bibnamefont
  {Faraoni}},\ }\href {https://books.google.com.au/books?id=lDT0BwAAQBAJ}
  {\emph {\bibinfo {title} {Cosmology in Scalar-Tensor Gravity}}},\ Fundamental
  Theories of Physics\ (\bibinfo  {publisher} {Springer Netherlands},\ \bibinfo
  {year} {2004})\BibitemShut {NoStop}%
\bibitem [{\citenamefont {Lee}\ \emph {et~al.}(2022)\citenamefont {Lee},
  \citenamefont {Modak}, \citenamefont {Oda},\ and\ \citenamefont
  {Takahashi}}]{Lee:2021rzy}%
  \BibitemOpen
  \bibfield  {author} {\bibinfo {author} {\bibfnamefont {S.~M.}\ \bibnamefont
  {Lee}}, \bibinfo {author} {\bibfnamefont {T.}~\bibnamefont {Modak}}, \bibinfo
  {author} {\bibfnamefont {K.-y.}\ \bibnamefont {Oda}},\ and\ \bibinfo {author}
  {\bibfnamefont {T.}~\bibnamefont {Takahashi}},\ }\bibfield  {title} {\bibinfo
  {title} {{The $R^2$-Higgs inflation with two Higgs doublets}},\ }\href
  {https://doi.org/10.1140/epjc/s10052-021-09978-w} {\bibfield  {journal}
  {\bibinfo  {journal} {Eur. Phys. J. C}\ }\textbf {\bibinfo {volume} {82}},\
  \bibinfo {pages} {18} (\bibinfo {year} {2022})},\ \Eprint
  {https://arxiv.org/abs/2108.02383} {arXiv:2108.02383 [hep-ph]} \BibitemShut
  {NoStop}%
\bibitem [{\citenamefont {Kaiser}(2010)}]{Kaiser2010}%
  \BibitemOpen
  \bibfield  {author} {\bibinfo {author} {\bibfnamefont {D.~I.}\ \bibnamefont
  {Kaiser}},\ }\bibfield  {title} {\bibinfo {title} {Conformal transformations
  with multiple scalar fields},\ }\href
  {https://doi.org/10.1103/PhysRevD.81.084044} {\bibfield  {journal} {\bibinfo
  {journal} {Phys. Rev. D}\ }\textbf {\bibinfo {volume} {81}},\ \bibinfo
  {pages} {084044} (\bibinfo {year} {2010})}\BibitemShut {NoStop}%
\bibitem [{\citenamefont {Kaiser}\ \emph {et~al.}(2013)\citenamefont {Kaiser},
  \citenamefont {Mazenc},\ and\ \citenamefont {Sfakianakis}}]{Kaiser:2012ak}%
  \BibitemOpen
  \bibfield  {author} {\bibinfo {author} {\bibfnamefont {D.~I.}\ \bibnamefont
  {Kaiser}}, \bibinfo {author} {\bibfnamefont {E.~A.}\ \bibnamefont {Mazenc}},\
  and\ \bibinfo {author} {\bibfnamefont {E.~I.}\ \bibnamefont {Sfakianakis}},\
  }\bibfield  {title} {\bibinfo {title} {{Primordial Bispectrum from Multifield
  Inflation with Nonminimal Couplings}},\ }\href
  {https://doi.org/10.1103/PhysRevD.87.064004} {\bibfield  {journal} {\bibinfo
  {journal} {Phys. Rev. D}\ }\textbf {\bibinfo {volume} {87}},\ \bibinfo
  {pages} {064004} (\bibinfo {year} {2013})},\ \Eprint
  {https://arxiv.org/abs/1210.7487} {arXiv:1210.7487 [astro-ph.CO]}
  \BibitemShut {NoStop}%
\bibitem [{\citenamefont {Clark}\ \emph {et~al.}(2009)\citenamefont {Clark},
  \citenamefont {Liu}, \citenamefont {Love},\ and\ \citenamefont {ter
  Veldhuis}}]{Clark:2009dc}%
  \BibitemOpen
  \bibfield  {author} {\bibinfo {author} {\bibfnamefont {T.~E.}\ \bibnamefont
  {Clark}}, \bibinfo {author} {\bibfnamefont {B.}~\bibnamefont {Liu}}, \bibinfo
  {author} {\bibfnamefont {S.~T.}\ \bibnamefont {Love}},\ and\ \bibinfo
  {author} {\bibfnamefont {T.}~\bibnamefont {ter Veldhuis}},\ }\bibfield
  {title} {\bibinfo {title} {{The Standard Model Higgs Boson-Inflaton and Dark
  Matter}},\ }\href {https://doi.org/10.1103/PhysRevD.80.075019} {\bibfield
  {journal} {\bibinfo  {journal} {Phys. Rev. D}\ }\textbf {\bibinfo {volume}
  {80}},\ \bibinfo {pages} {075019} (\bibinfo {year} {2009})},\ \Eprint
  {https://arxiv.org/abs/0906.5595} {arXiv:0906.5595 [hep-ph]} \BibitemShut
  {NoStop}%
\bibitem [{\citenamefont {Greenwood}\ \emph {et~al.}(2013)\citenamefont
  {Greenwood}, \citenamefont {Kaiser},\ and\ \citenamefont
  {Sfakianakis}}]{greenwood2013}%
  \BibitemOpen
  \bibfield  {author} {\bibinfo {author} {\bibfnamefont {R.~N.}\ \bibnamefont
  {Greenwood}}, \bibinfo {author} {\bibfnamefont {D.~I.}\ \bibnamefont
  {Kaiser}},\ and\ \bibinfo {author} {\bibfnamefont {E.~I.}\ \bibnamefont
  {Sfakianakis}},\ }\bibfield  {title} {\bibinfo {title} {Multifield dynamics
  of higgs inflation},\ }\href {https://doi.org/10.1103/PhysRevD.87.064021}
  {\bibfield  {journal} {\bibinfo  {journal} {Phys. Rev. D}\ }\textbf {\bibinfo
  {volume} {87}},\ \bibinfo {pages} {064021} (\bibinfo {year}
  {2013})}\BibitemShut {NoStop}%
\bibitem [{\citenamefont {Ade}\ \emph {et~al.}(2021)\citenamefont {Ade} \emph
  {et~al.}}]{BICEP:2021xfz}%
  \BibitemOpen
  \bibfield  {author} {\bibinfo {author} {\bibfnamefont {P.~A.~R.}\
  \bibnamefont {Ade}} \emph {et~al.} (\bibinfo {collaboration} {BICEP, Keck}),\
  }\bibfield  {title} {\bibinfo {title} {{Improved Constraints on Primordial
  Gravitational Waves using Planck, WMAP, and BICEP/Keck Observations through
  the 2018 Observing Season}},\ }\href
  {https://doi.org/10.1103/PhysRevLett.127.151301} {\bibfield  {journal}
  {\bibinfo  {journal} {Phys. Rev. Lett.}\ }\textbf {\bibinfo {volume} {127}},\
  \bibinfo {pages} {151301} (\bibinfo {year} {2021})},\ \Eprint
  {https://arxiv.org/abs/2110.00483} {arXiv:2110.00483 [astro-ph.CO]}
  \BibitemShut {NoStop}%
\bibitem [{\citenamefont {Aghanim}\ \emph {et~al.}(2020)\citenamefont {Aghanim}
  \emph {et~al.}}]{PlanckCosParam}%
  \BibitemOpen
  \bibfield  {author} {\bibinfo {author} {\bibfnamefont {N.}~\bibnamefont
  {Aghanim}} \emph {et~al.} (\bibinfo {collaboration} {Planck}),\ }\bibfield
  {title} {\bibinfo {title} {{Planck 2018 results. VI. Cosmological
  parameters}},\ }\href {https://doi.org/10.1051/0004-6361/201833910}
  {\bibfield  {journal} {\bibinfo  {journal} {Astron. Astrophys.}\ }\textbf
  {\bibinfo {volume} {641}},\ \bibinfo {pages} {A6} (\bibinfo {year} {2020})},\
  \bibinfo {note} {[Erratum: Astron.Astrophys. 652, C4 (2021)]},\ \Eprint
  {https://arxiv.org/abs/1807.06209} {arXiv:1807.06209 [astro-ph.CO]}
  \BibitemShut {NoStop}%
\bibitem [{\citenamefont {Garcia-Bellido}\ \emph {et~al.}(2009)\citenamefont
  {Garcia-Bellido}, \citenamefont {Figueroa},\ and\ \citenamefont
  {Rubio}}]{Garcia-Bellido:2008ycs}%
  \BibitemOpen
  \bibfield  {author} {\bibinfo {author} {\bibfnamefont {J.}~\bibnamefont
  {Garcia-Bellido}}, \bibinfo {author} {\bibfnamefont {D.~G.}\ \bibnamefont
  {Figueroa}},\ and\ \bibinfo {author} {\bibfnamefont {J.}~\bibnamefont
  {Rubio}},\ }\bibfield  {title} {\bibinfo {title} {{Preheating in the Standard
  Model with the Higgs-Inflaton coupled to gravity}},\ }\href
  {https://doi.org/10.1103/PhysRevD.79.063531} {\bibfield  {journal} {\bibinfo
  {journal} {Phys. Rev. D}\ }\textbf {\bibinfo {volume} {79}},\ \bibinfo
  {pages} {063531} (\bibinfo {year} {2009})},\ \Eprint
  {https://arxiv.org/abs/0812.4624} {arXiv:0812.4624 [hep-ph]} \BibitemShut
  {NoStop}%
\bibitem [{\citenamefont {Fairbairn}\ \emph {et~al.}(2015)\citenamefont
  {Fairbairn}, \citenamefont {Hogan},\ and\ \citenamefont
  {Marsh}}]{Fairbairn:2014zta}%
  \BibitemOpen
  \bibfield  {author} {\bibinfo {author} {\bibfnamefont {M.}~\bibnamefont
  {Fairbairn}}, \bibinfo {author} {\bibfnamefont {R.}~\bibnamefont {Hogan}},\
  and\ \bibinfo {author} {\bibfnamefont {D.~J.~E.}\ \bibnamefont {Marsh}},\
  }\bibfield  {title} {\bibinfo {title} {{Unifying inflation and dark matter
  with the Peccei-Quinn field: observable axions and observable tensors}},\
  }\href {https://doi.org/10.1103/PhysRevD.91.023509} {\bibfield  {journal}
  {\bibinfo  {journal} {Phys. Rev. D}\ }\textbf {\bibinfo {volume} {91}},\
  \bibinfo {pages} {023509} (\bibinfo {year} {2015})},\ \Eprint
  {https://arxiv.org/abs/1410.1752} {arXiv:1410.1752 [hep-ph]} \BibitemShut
  {NoStop}%
\bibitem [{\citenamefont {Gibbons}\ and\ \citenamefont
  {Hawking}(1977)}]{Gibbons1977}%
  \BibitemOpen
  \bibfield  {author} {\bibinfo {author} {\bibfnamefont {G.~W.}\ \bibnamefont
  {Gibbons}}\ and\ \bibinfo {author} {\bibfnamefont {S.~W.}\ \bibnamefont
  {Hawking}},\ }\bibfield  {title} {\bibinfo {title} {Cosmological event
  horizons, thermodynamics, and particle creation},\ }\href
  {https://doi.org/10.1103/PhysRevD.15.2738} {\bibfield  {journal} {\bibinfo
  {journal} {Phys. Rev. D}\ }\textbf {\bibinfo {volume} {15}},\ \bibinfo
  {pages} {2738} (\bibinfo {year} {1977})}\BibitemShut {NoStop}%
\bibitem [{\citenamefont {Hertzberg}\ \emph {et~al.}(2008)\citenamefont
  {Hertzberg}, \citenamefont {Tegmark},\ and\ \citenamefont
  {Wilczek}}]{Hertzberg:2008wr}%
  \BibitemOpen
  \bibfield  {author} {\bibinfo {author} {\bibfnamefont {M.~P.}\ \bibnamefont
  {Hertzberg}}, \bibinfo {author} {\bibfnamefont {M.}~\bibnamefont {Tegmark}},\
  and\ \bibinfo {author} {\bibfnamefont {F.}~\bibnamefont {Wilczek}},\
  }\bibfield  {title} {\bibinfo {title} {{Axion Cosmology and the Energy Scale
  of Inflation}},\ }\href {https://doi.org/10.1103/PhysRevD.78.083507}
  {\bibfield  {journal} {\bibinfo  {journal} {Phys. Rev. D}\ }\textbf {\bibinfo
  {volume} {78}},\ \bibinfo {pages} {083507} (\bibinfo {year} {2008})},\
  \Eprint {https://arxiv.org/abs/0807.1726} {arXiv:0807.1726 [astro-ph]}
  \BibitemShut {NoStop}%
\bibitem [{\citenamefont {Linde}(1991)}]{LINDE199138}%
  \BibitemOpen
  \bibfield  {author} {\bibinfo {author} {\bibfnamefont {A.}~\bibnamefont
  {Linde}},\ }\bibfield  {title} {\bibinfo {title} {Axions in inflationary
  cosmology},\ }\href
  {https://doi.org/https://doi.org/10.1016/0370-2693(91)90130-I} {\bibfield
  {journal} {\bibinfo  {journal} {Physics Letters B}\ }\textbf {\bibinfo
  {volume} {259}},\ \bibinfo {pages} {38} (\bibinfo {year} {1991})}\BibitemShut
  {NoStop}%
\bibitem [{\citenamefont {Kofman}\ \emph {et~al.}(1994)\citenamefont {Kofman},
  \citenamefont {Linde},\ and\ \citenamefont {Starobinsky}}]{Kofman:1994rk}%
  \BibitemOpen
  \bibfield  {author} {\bibinfo {author} {\bibfnamefont {L.}~\bibnamefont
  {Kofman}}, \bibinfo {author} {\bibfnamefont {A.~D.}\ \bibnamefont {Linde}},\
  and\ \bibinfo {author} {\bibfnamefont {A.~A.}\ \bibnamefont {Starobinsky}},\
  }\bibfield  {title} {\bibinfo {title} {{Reheating after inflation}},\ }\href
  {https://doi.org/10.1103/PhysRevLett.73.3195} {\bibfield  {journal} {\bibinfo
   {journal} {Phys. Rev. Lett.}\ }\textbf {\bibinfo {volume} {73}},\ \bibinfo
  {pages} {3195} (\bibinfo {year} {1994})},\ \Eprint
  {https://arxiv.org/abs/hep-th/9405187} {arXiv:hep-th/9405187} \BibitemShut
  {NoStop}%
\bibitem [{\citenamefont {Kofman}\ \emph {et~al.}(1997)\citenamefont {Kofman},
  \citenamefont {Linde},\ and\ \citenamefont {Starobinsky}}]{Kofman:1997yn}%
  \BibitemOpen
  \bibfield  {author} {\bibinfo {author} {\bibfnamefont {L.}~\bibnamefont
  {Kofman}}, \bibinfo {author} {\bibfnamefont {A.~D.}\ \bibnamefont {Linde}},\
  and\ \bibinfo {author} {\bibfnamefont {A.~A.}\ \bibnamefont {Starobinsky}},\
  }\bibfield  {title} {\bibinfo {title} {{Towards the theory of reheating after
  inflation}},\ }\href {https://doi.org/10.1103/PhysRevD.56.3258} {\bibfield
  {journal} {\bibinfo  {journal} {Phys. Rev. D}\ }\textbf {\bibinfo {volume}
  {56}},\ \bibinfo {pages} {3258} (\bibinfo {year} {1997})},\ \Eprint
  {https://arxiv.org/abs/hep-ph/9704452} {arXiv:hep-ph/9704452} \BibitemShut
  {NoStop}%
\bibitem [{\citenamefont {DeCross}\ \emph
  {et~al.}(2018{\natexlab{b}})\citenamefont {DeCross}, \citenamefont {Kaiser},
  \citenamefont {Prabhu}, \citenamefont {Prescod-Weinstein},\ and\
  \citenamefont {Sfakianakis}}]{DeCross:2016fdz}%
  \BibitemOpen
  \bibfield  {author} {\bibinfo {author} {\bibfnamefont {M.~P.}\ \bibnamefont
  {DeCross}}, \bibinfo {author} {\bibfnamefont {D.~I.}\ \bibnamefont {Kaiser}},
  \bibinfo {author} {\bibfnamefont {A.}~\bibnamefont {Prabhu}}, \bibinfo
  {author} {\bibfnamefont {C.}~\bibnamefont {Prescod-Weinstein}},\ and\
  \bibinfo {author} {\bibfnamefont {E.~I.}\ \bibnamefont {Sfakianakis}},\
  }\bibfield  {title} {\bibinfo {title} {{Preheating after multifield inflation
  with nonminimal couplings, II: Resonance Structure}},\ }\href
  {https://doi.org/10.1103/PhysRevD.97.023527} {\bibfield  {journal} {\bibinfo
  {journal} {Phys. Rev. D}\ }\textbf {\bibinfo {volume} {97}},\ \bibinfo
  {pages} {023527} (\bibinfo {year} {2018}{\natexlab{b}})},\ \Eprint
  {https://arxiv.org/abs/1610.08868} {arXiv:1610.08868 [astro-ph.CO]}
  \BibitemShut {NoStop}%
\bibitem [{\citenamefont {Tkachev}\ \emph {et~al.}(1998)\citenamefont
  {Tkachev}, \citenamefont {Khlebnikov}, \citenamefont {Kofman},\ and\
  \citenamefont {Linde}}]{Tkachev:1998dc}%
  \BibitemOpen
  \bibfield  {author} {\bibinfo {author} {\bibfnamefont {I.}~\bibnamefont
  {Tkachev}}, \bibinfo {author} {\bibfnamefont {S.}~\bibnamefont {Khlebnikov}},
  \bibinfo {author} {\bibfnamefont {L.}~\bibnamefont {Kofman}},\ and\ \bibinfo
  {author} {\bibfnamefont {A.~D.}\ \bibnamefont {Linde}},\ }\bibfield  {title}
  {\bibinfo {title} {{Cosmic strings from preheating}},\ }\href
  {https://doi.org/10.1016/S0370-2693(98)01094-6} {\bibfield  {journal}
  {\bibinfo  {journal} {Phys. Lett. B}\ }\textbf {\bibinfo {volume} {440}},\
  \bibinfo {pages} {262} (\bibinfo {year} {1998})},\ \Eprint
  {https://arxiv.org/abs/hep-ph/9805209} {arXiv:hep-ph/9805209} \BibitemShut
  {NoStop}%
\bibitem [{\citenamefont {Ballesteros}\ \emph {et~al.}(2021)\citenamefont
  {Ballesteros}, \citenamefont {Ringwald}, \citenamefont {Tamarit},\ and\
  \citenamefont {Welling}}]{Ballesteros:2021bee}%
  \BibitemOpen
  \bibfield  {author} {\bibinfo {author} {\bibfnamefont {G.}~\bibnamefont
  {Ballesteros}}, \bibinfo {author} {\bibfnamefont {A.}~\bibnamefont
  {Ringwald}}, \bibinfo {author} {\bibfnamefont {C.}~\bibnamefont {Tamarit}},\
  and\ \bibinfo {author} {\bibfnamefont {Y.}~\bibnamefont {Welling}},\
  }\bibfield  {title} {\bibinfo {title} {{Revisiting isocurvature bounds in
  models unifying the axion with the inflaton}},\ }\href
  {https://doi.org/10.1088/1475-7516/2021/09/036} {\bibfield  {journal}
  {\bibinfo  {journal} {JCAP}\ }\textbf {\bibinfo {volume} {09}},\ \bibinfo
  {pages} {036}},\ \Eprint {https://arxiv.org/abs/2104.13847} {arXiv:2104.13847
  [hep-ph]} \BibitemShut {NoStop}%
\bibitem [{\citenamefont {Kawasaki}\ \emph {et~al.}(2013)\citenamefont
  {Kawasaki}, \citenamefont {Yanagida},\ and\ \citenamefont
  {Yoshino}}]{Kawasaki:2013iha}%
  \BibitemOpen
  \bibfield  {author} {\bibinfo {author} {\bibfnamefont {M.}~\bibnamefont
  {Kawasaki}}, \bibinfo {author} {\bibfnamefont {T.~T.}\ \bibnamefont
  {Yanagida}},\ and\ \bibinfo {author} {\bibfnamefont {K.}~\bibnamefont
  {Yoshino}},\ }\bibfield  {title} {\bibinfo {title} {{Domain wall and
  isocurvature perturbation problems in axion models}},\ }\href
  {https://doi.org/10.1088/1475-7516/2013/11/030} {\bibfield  {journal}
  {\bibinfo  {journal} {JCAP}\ }\textbf {\bibinfo {volume} {11}},\ \bibinfo
  {pages} {030}},\ \Eprint {https://arxiv.org/abs/1305.5338} {arXiv:1305.5338
  [hep-ph]} \BibitemShut {NoStop}%
\bibitem [{\citenamefont {Kearney}\ \emph {et~al.}(2016)\citenamefont
  {Kearney}, \citenamefont {Orlofsky},\ and\ \citenamefont
  {Pierce}}]{Kearney:2016vqw}%
  \BibitemOpen
  \bibfield  {author} {\bibinfo {author} {\bibfnamefont {J.}~\bibnamefont
  {Kearney}}, \bibinfo {author} {\bibfnamefont {N.}~\bibnamefont {Orlofsky}},\
  and\ \bibinfo {author} {\bibfnamefont {A.}~\bibnamefont {Pierce}},\
  }\bibfield  {title} {\bibinfo {title} {{High-Scale Axions without
  Isocurvature from Inflationary Dynamics}},\ }\href
  {https://doi.org/10.1103/PhysRevD.93.095026} {\bibfield  {journal} {\bibinfo
  {journal} {Phys. Rev. D}\ }\textbf {\bibinfo {volume} {93}},\ \bibinfo
  {pages} {095026} (\bibinfo {year} {2016})},\ \Eprint
  {https://arxiv.org/abs/1601.03049} {arXiv:1601.03049 [hep-ph]} \BibitemShut
  {NoStop}%
\bibitem [{\citenamefont {Chung}\ \emph {et~al.}(1999)\citenamefont {Chung},
  \citenamefont {Kolb},\ and\ \citenamefont {Riotto}}]{chung1999}%
  \BibitemOpen
  \bibfield  {author} {\bibinfo {author} {\bibfnamefont {D.~J.~H.}\
  \bibnamefont {Chung}}, \bibinfo {author} {\bibfnamefont {E.~W.}\ \bibnamefont
  {Kolb}},\ and\ \bibinfo {author} {\bibfnamefont {A.}~\bibnamefont {Riotto}},\
  }\bibfield  {title} {\bibinfo {title} {Production of massive particles during
  reheating},\ }\href {https://doi.org/10.1103/PhysRevD.60.063504} {\bibfield
  {journal} {\bibinfo  {journal} {Phys. Rev. D}\ }\textbf {\bibinfo {volume}
  {60}},\ \bibinfo {pages} {063504} (\bibinfo {year} {1999})}\BibitemShut
  {NoStop}%
\bibitem [{\citenamefont {Bezrukov}\ \emph {et~al.}(2009)\citenamefont
  {Bezrukov}, \citenamefont {Gorbunov},\ and\ \citenamefont
  {Shaposhnikov}}]{Bezrukov:2008utReheat}%
  \BibitemOpen
  \bibfield  {author} {\bibinfo {author} {\bibfnamefont {F.}~\bibnamefont
  {Bezrukov}}, \bibinfo {author} {\bibfnamefont {D.}~\bibnamefont {Gorbunov}},\
  and\ \bibinfo {author} {\bibfnamefont {M.}~\bibnamefont {Shaposhnikov}},\
  }\bibfield  {title} {\bibinfo {title} {{On initial conditions for the Hot Big
  Bang}},\ }\href {https://doi.org/10.1088/1475-7516/2009/06/029} {\bibfield
  {journal} {\bibinfo  {journal} {JCAP}\ }\textbf {\bibinfo {volume} {06}},\
  \bibinfo {pages} {029}},\ \Eprint {https://arxiv.org/abs/0812.3622}
  {arXiv:0812.3622 [hep-ph]} \BibitemShut {NoStop}%
\bibitem [{\citenamefont {Lerner}\ and\ \citenamefont
  {McDonald}(2011)}]{Lerner2011}%
  \BibitemOpen
  \bibfield  {author} {\bibinfo {author} {\bibfnamefont {R.~N.}\ \bibnamefont
  {Lerner}}\ and\ \bibinfo {author} {\bibfnamefont {J.}~\bibnamefont
  {McDonald}},\ }\bibfield  {title} {\bibinfo {title} {{Distinguishing Higgs
  inflation and its variants}},\ }\href
  {https://doi.org/10.1103/PhysRevD.83.123522} {\bibfield  {journal} {\bibinfo
  {journal} {Phys. Rev. D}\ }\textbf {\bibinfo {volume} {83}},\ \bibinfo
  {pages} {123522} (\bibinfo {year} {2011})},\ \Eprint
  {https://arxiv.org/abs/1104.2468} {arXiv:1104.2468 [hep-ph]} \BibitemShut
  {NoStop}%
\bibitem [{\citenamefont {Masso}\ \emph {et~al.}(2002)\citenamefont {Masso},
  \citenamefont {Rota},\ and\ \citenamefont {Zsembinszki}}]{Masso:2002np}%
  \BibitemOpen
  \bibfield  {author} {\bibinfo {author} {\bibfnamefont {E.}~\bibnamefont
  {Masso}}, \bibinfo {author} {\bibfnamefont {F.}~\bibnamefont {Rota}},\ and\
  \bibinfo {author} {\bibfnamefont {G.}~\bibnamefont {Zsembinszki}},\
  }\bibfield  {title} {\bibinfo {title} {{On axion thermalization in the early
  universe}},\ }\href {https://doi.org/10.1103/PhysRevD.66.023004} {\bibfield
  {journal} {\bibinfo  {journal} {Phys. Rev. D}\ }\textbf {\bibinfo {volume}
  {66}},\ \bibinfo {pages} {023004} (\bibinfo {year} {2002})},\ \Eprint
  {https://arxiv.org/abs/hep-ph/0203221} {arXiv:hep-ph/0203221} \BibitemShut
  {NoStop}%
\bibitem [{\citenamefont {Figueroa}\ and\ \citenamefont
  {Byrnes}(2017)}]{Figueroa:2016dsc}%
  \BibitemOpen
  \bibfield  {author} {\bibinfo {author} {\bibfnamefont {D.~G.}\ \bibnamefont
  {Figueroa}}\ and\ \bibinfo {author} {\bibfnamefont {C.~T.}\ \bibnamefont
  {Byrnes}},\ }\bibfield  {title} {\bibinfo {title} {{The Standard Model Higgs
  as the origin of the hot Big Bang}},\ }\href
  {https://doi.org/10.1016/j.physletb.2017.01.059} {\bibfield  {journal}
  {\bibinfo  {journal} {Phys. Lett. B}\ }\textbf {\bibinfo {volume} {767}},\
  \bibinfo {pages} {272} (\bibinfo {year} {2017})},\ \Eprint
  {https://arxiv.org/abs/1604.03905} {arXiv:1604.03905 [hep-ph]} \BibitemShut
  {NoStop}%
\bibitem [{\citenamefont {Gorghetto}\ \emph {et~al.}(2021)\citenamefont
  {Gorghetto}, \citenamefont {Hardy},\ and\ \citenamefont
  {Villadoro}}]{GorghettoAxions}%
  \BibitemOpen
  \bibfield  {author} {\bibinfo {author} {\bibfnamefont {M.}~\bibnamefont
  {Gorghetto}}, \bibinfo {author} {\bibfnamefont {E.}~\bibnamefont {Hardy}},\
  and\ \bibinfo {author} {\bibfnamefont {G.}~\bibnamefont {Villadoro}},\
  }\bibfield  {title} {\bibinfo {title} {{More Axions from Strings}},\ }\href
  {https://doi.org/10.21468/SciPostPhys.10.2.050} {\bibfield  {journal}
  {\bibinfo  {journal} {SciPost Phys.}\ }\textbf {\bibinfo {volume} {10}},\
  \bibinfo {pages} {50} (\bibinfo {year} {2021})}\BibitemShut {NoStop}%
\bibitem [{\citenamefont {Preskill}\ \emph {et~al.}(1983)\citenamefont
  {Preskill}, \citenamefont {Wise},\ and\ \citenamefont {Wilczek}}]{MisAlign1}%
  \BibitemOpen
  \bibfield  {author} {\bibinfo {author} {\bibfnamefont {J.}~\bibnamefont
  {Preskill}}, \bibinfo {author} {\bibfnamefont {M.~B.}\ \bibnamefont {Wise}},\
  and\ \bibinfo {author} {\bibfnamefont {F.}~\bibnamefont {Wilczek}},\
  }\bibfield  {title} {\bibinfo {title} {Cosmology of the invisible axion},\
  }\href {https://doi.org/https://doi.org/10.1016/0370-2693(83)90637-8}
  {\bibfield  {journal} {\bibinfo  {journal} {Physics Letters B}\ }\textbf
  {\bibinfo {volume} {120}},\ \bibinfo {pages} {127} (\bibinfo {year}
  {1983})}\BibitemShut {NoStop}%
\bibitem [{\citenamefont {Abbott}\ and\ \citenamefont
  {Sikivie}(1983)}]{MisAlign2}%
  \BibitemOpen
  \bibfield  {author} {\bibinfo {author} {\bibfnamefont {L.}~\bibnamefont
  {Abbott}}\ and\ \bibinfo {author} {\bibfnamefont {P.}~\bibnamefont
  {Sikivie}},\ }\bibfield  {title} {\bibinfo {title} {A cosmological bound on
  the invisible axion},\ }\href
  {https://doi.org/https://doi.org/10.1016/0370-2693(83)90638-X} {\bibfield
  {journal} {\bibinfo  {journal} {Physics Letters B}\ }\textbf {\bibinfo
  {volume} {120}},\ \bibinfo {pages} {133} (\bibinfo {year}
  {1983})}\BibitemShut {NoStop}%
\bibitem [{\citenamefont {Dine}\ and\ \citenamefont
  {Fischler}(1983)}]{MisAlign3}%
  \BibitemOpen
  \bibfield  {author} {\bibinfo {author} {\bibfnamefont {M.}~\bibnamefont
  {Dine}}\ and\ \bibinfo {author} {\bibfnamefont {W.}~\bibnamefont
  {Fischler}},\ }\bibfield  {title} {\bibinfo {title} {The not-so-harmless
  axion},\ }\href
  {https://doi.org/https://doi.org/10.1016/0370-2693(83)90639-1} {\bibfield
  {journal} {\bibinfo  {journal} {Physics Letters B}\ }\textbf {\bibinfo
  {volume} {120}},\ \bibinfo {pages} {137} (\bibinfo {year}
  {1983})}\BibitemShut {NoStop}%
\bibitem [{\citenamefont {Ringwald}(2018)}]{RingwaldPostInf}%
  \BibitemOpen
  \bibfield  {author} {\bibinfo {author} {\bibfnamefont {A.}~\bibnamefont
  {Ringwald}},\ }\bibfield  {title} {\bibinfo {title} {{Axion mass in the case
  of post-inflationary Peccei-Quinn symmetry breaking}},\ }in\ \href@noop {}
  {\emph {\bibinfo {booktitle} {{53rd Rencontres de Moriond on Cosmology}}}}\
  (\bibinfo {year} {2018})\ pp.\ \bibinfo {pages} {241--244},\ \Eprint
  {https://arxiv.org/abs/1805.09618} {arXiv:1805.09618 [hep-ph]} \BibitemShut
  {NoStop}%
\bibitem [{\citenamefont {Bors{\'a}nyi}\ \emph {et~al.}(2016)\citenamefont
  {Bors{\'a}nyi}, \citenamefont {Fodor}, \citenamefont {Guenther},
  \citenamefont {Kampert}, \citenamefont {Katz}, \citenamefont {Kawanai},
  \citenamefont {Kov{\'a}cs}, \citenamefont {Mages}, \citenamefont
  {P{\'a}sztor}, \citenamefont {Pittler}, \citenamefont {Redondo},
  \citenamefont {Ringwald},\ and\ \citenamefont
  {Szab{\'o}}}]{Borsnyi2016CalculationOT}%
  \BibitemOpen
  \bibfield  {author} {\bibinfo {author} {\bibfnamefont {S.}~\bibnamefont
  {Bors{\'a}nyi}}, \bibinfo {author} {\bibfnamefont {Z.}~\bibnamefont {Fodor}},
  \bibinfo {author} {\bibfnamefont {J.~N.}\ \bibnamefont {Guenther}}, \bibinfo
  {author} {\bibfnamefont {K.-H.}\ \bibnamefont {Kampert}}, \bibinfo {author}
  {\bibfnamefont {S.~D.}\ \bibnamefont {Katz}}, \bibinfo {author}
  {\bibfnamefont {T.}~\bibnamefont {Kawanai}}, \bibinfo {author} {\bibfnamefont
  {T.}~\bibnamefont {Kov{\'a}cs}}, \bibinfo {author} {\bibfnamefont
  {S.}~\bibnamefont {Mages}}, \bibinfo {author} {\bibfnamefont
  {A.}~\bibnamefont {P{\'a}sztor}}, \bibinfo {author} {\bibfnamefont
  {F.}~\bibnamefont {Pittler}}, \bibinfo {author} {\bibfnamefont
  {J.}~\bibnamefont {Redondo}}, \bibinfo {author} {\bibfnamefont
  {A.}~\bibnamefont {Ringwald}},\ and\ \bibinfo {author} {\bibfnamefont
  {K.~K.}\ \bibnamefont {Szab{\'o}}},\ }\bibfield  {title} {\bibinfo {title}
  {Calculation of the axion mass based on high-temperature lattice quantum
  chromodynamics},\ }\href@noop {} {\bibfield  {journal} {\bibinfo  {journal}
  {Nature}\ }\textbf {\bibinfo {volume} {539}},\ \bibinfo {pages} {69}
  (\bibinfo {year} {2016})}\BibitemShut {NoStop}%
\bibitem [{\citenamefont {Davis}(1986)}]{Davis}%
  \BibitemOpen
  \bibfield  {author} {\bibinfo {author} {\bibfnamefont {R.}~\bibnamefont
  {Davis}},\ }\bibfield  {title} {\bibinfo {title} {Cosmic axions from cosmic
  strings},\ }\href
  {https://doi.org/https://doi.org/10.1016/0370-2693(86)90300-X} {\bibfield
  {journal} {\bibinfo  {journal} {Physics Letters B}\ }\textbf {\bibinfo
  {volume} {180}},\ \bibinfo {pages} {225} (\bibinfo {year}
  {1986})}\BibitemShut {NoStop}%
\bibitem [{\citenamefont {Gorghetto}\ and\ \citenamefont
  {Villadoro}(2019)}]{GorghettoTopSus}%
  \BibitemOpen
  \bibfield  {author} {\bibinfo {author} {\bibfnamefont {M.}~\bibnamefont
  {Gorghetto}}\ and\ \bibinfo {author} {\bibfnamefont {G.}~\bibnamefont
  {Villadoro}},\ }\bibfield  {title} {\bibinfo {title} {{Topological
  Susceptibility and QCD Axion Mass: QED and NNLO corrections}},\ }\href
  {https://doi.org/10.1007/JHEP03(2019)033} {\bibfield  {journal} {\bibinfo
  {journal} {JHEP}\ }\textbf {\bibinfo {volume} {03}},\ \bibinfo {pages}
  {033}},\ \Eprint {https://arxiv.org/abs/1812.01008} {arXiv:1812.01008
  [hep-ph]} \BibitemShut {NoStop}%
\bibitem [{\citenamefont {Kawasaki}\ and\ \citenamefont
  {Nakayama}(2013)}]{Kawasaki2013}%
  \BibitemOpen
  \bibfield  {author} {\bibinfo {author} {\bibfnamefont {M.}~\bibnamefont
  {Kawasaki}}\ and\ \bibinfo {author} {\bibfnamefont {K.}~\bibnamefont
  {Nakayama}},\ }\bibfield  {title} {\bibinfo {title} {{Axions: Theory and
  Cosmological Role}},\ }\href
  {https://doi.org/10.1146/annurev-nucl-102212-170536} {\bibfield  {journal}
  {\bibinfo  {journal} {Ann. Rev. Nucl. Part. Sci.}\ }\textbf {\bibinfo
  {volume} {63}},\ \bibinfo {pages} {69} (\bibinfo {year} {2013})},\ \Eprint
  {https://arxiv.org/abs/1301.1123} {arXiv:1301.1123 [hep-ph]} \BibitemShut
  {NoStop}%
\bibitem [{\citenamefont {Vilenkin}\ and\ \citenamefont
  {Everett}(1982)}]{Vilenkin:1982ks}%
  \BibitemOpen
  \bibfield  {author} {\bibinfo {author} {\bibfnamefont {A.}~\bibnamefont
  {Vilenkin}}\ and\ \bibinfo {author} {\bibfnamefont {A.~E.}\ \bibnamefont
  {Everett}},\ }\bibfield  {title} {\bibinfo {title} {{Cosmic Strings and
  Domain Walls in Models with Goldstone and PseudoGoldstone Bosons}},\ }\href
  {https://doi.org/10.1103/PhysRevLett.48.1867} {\bibfield  {journal} {\bibinfo
   {journal} {Phys. Rev. Lett.}\ }\textbf {\bibinfo {volume} {48}},\ \bibinfo
  {pages} {1867} (\bibinfo {year} {1982})}\BibitemShut {NoStop}%
\bibitem [{\citenamefont {Hiramatsu}\ \emph {et~al.}(2012)\citenamefont
  {Hiramatsu}, \citenamefont {Kawasaki}, \citenamefont {Saikawa},\ and\
  \citenamefont {Sekiguchi}}]{Hiramatsu2012}%
  \BibitemOpen
  \bibfield  {author} {\bibinfo {author} {\bibfnamefont {T.}~\bibnamefont
  {Hiramatsu}}, \bibinfo {author} {\bibfnamefont {M.}~\bibnamefont {Kawasaki}},
  \bibinfo {author} {\bibfnamefont {K.}~\bibnamefont {Saikawa}},\ and\ \bibinfo
  {author} {\bibfnamefont {T.}~\bibnamefont {Sekiguchi}},\ }\bibfield  {title}
  {\bibinfo {title} {{Production of dark matter axions from collapse of
  string-wall systems}},\ }\href {https://doi.org/10.1103/PhysRevD.85.105020}
  {\bibfield  {journal} {\bibinfo  {journal} {Phys. Rev. D}\ }\textbf {\bibinfo
  {volume} {85}},\ \bibinfo {pages} {105020} (\bibinfo {year} {2012})},\
  \bibinfo {note} {[Erratum: Phys.Rev.D 86, 089902 (2012)]},\ \Eprint
  {https://arxiv.org/abs/1202.5851} {arXiv:1202.5851 [hep-ph]} \BibitemShut
  {NoStop}%
\bibitem [{\citenamefont {Hoof}\ \emph {et~al.}(2022)\citenamefont {Hoof},
  \citenamefont {Riess},\ and\ \citenamefont {Marsh}}]{HoofDomWalls}%
  \BibitemOpen
  \bibfield  {author} {\bibinfo {author} {\bibfnamefont {S.}~\bibnamefont
  {Hoof}}, \bibinfo {author} {\bibfnamefont {J.}~\bibnamefont {Riess}},\ and\
  \bibinfo {author} {\bibfnamefont {D.~J.}\ \bibnamefont {Marsh}},\ }\bibfield
  {title} {\bibinfo {title} {Statistical uncertainties of the {$N_{\text{DW}} =
  1$} {QCD} axion mass window from topological defects},\ }\bibfield  {journal}
  {\bibinfo  {journal} {The Open Journal of Astrophysics}\ }\textbf {\bibinfo
  {volume} {5}},\ \href {https://doi.org/10.21105/astro.2108.09563}
  {10.21105/astro.2108.09563} (\bibinfo {year} {2022})\BibitemShut {NoStop}%
\bibitem [{\citenamefont {Buschmann}\ \emph {et~al.}(2022)\citenamefont
  {Buschmann}, \citenamefont {Foster}, \citenamefont {Hook}, \citenamefont
  {Peterson}, \citenamefont {Willcox}, \citenamefont {Zhang},\ and\
  \citenamefont {Safdi}}]{Buschmann2021}%
  \BibitemOpen
  \bibfield  {author} {\bibinfo {author} {\bibfnamefont {M.}~\bibnamefont
  {Buschmann}}, \bibinfo {author} {\bibfnamefont {J.~W.}\ \bibnamefont
  {Foster}}, \bibinfo {author} {\bibfnamefont {A.}~\bibnamefont {Hook}},
  \bibinfo {author} {\bibfnamefont {A.}~\bibnamefont {Peterson}}, \bibinfo
  {author} {\bibfnamefont {D.~E.}\ \bibnamefont {Willcox}}, \bibinfo {author}
  {\bibfnamefont {W.}~\bibnamefont {Zhang}},\ and\ \bibinfo {author}
  {\bibfnamefont {B.~R.}\ \bibnamefont {Safdi}},\ }\bibfield  {title} {\bibinfo
  {title} {{Dark matter from axion strings with adaptive mesh refinement}},\
  }\href {https://doi.org/10.1038/s41467-022-28669-y} {\bibfield  {journal}
  {\bibinfo  {journal} {Nature Commun.}\ }\textbf {\bibinfo {volume} {13}},\
  \bibinfo {pages} {1049} (\bibinfo {year} {2022})},\ \Eprint
  {https://arxiv.org/abs/2108.05368} {arXiv:2108.05368 [hep-ph]} \BibitemShut
  {NoStop}%
\bibitem [{\citenamefont {Saikawa}\ and\ \citenamefont
  {Yanagida}(2020)}]{Saikawa:2019lng}%
  \BibitemOpen
  \bibfield  {author} {\bibinfo {author} {\bibfnamefont {K.}~\bibnamefont
  {Saikawa}}\ and\ \bibinfo {author} {\bibfnamefont {T.~T.}\ \bibnamefont
  {Yanagida}},\ }\bibfield  {title} {\bibinfo {title} {{Stellar cooling
  anomalies and variant axion models}},\ }\href
  {https://doi.org/10.1088/1475-7516/2020/03/007} {\bibfield  {journal}
  {\bibinfo  {journal} {JCAP}\ }\textbf {\bibinfo {volume} {03}},\ \bibinfo
  {pages} {007}},\ \Eprint {https://arxiv.org/abs/1907.07662} {arXiv:1907.07662
  [hep-ph]} \BibitemShut {NoStop}%
\bibitem [{\citenamefont {Di~Luzio}\ \emph {et~al.}(2022)\citenamefont
  {Di~Luzio}, \citenamefont {Fedele}, \citenamefont {Giannotti}, \citenamefont
  {Mescia},\ and\ \citenamefont {Nardi}}]{DiLuzio:2021ysg}%
  \BibitemOpen
  \bibfield  {author} {\bibinfo {author} {\bibfnamefont {L.}~\bibnamefont
  {Di~Luzio}}, \bibinfo {author} {\bibfnamefont {M.}~\bibnamefont {Fedele}},
  \bibinfo {author} {\bibfnamefont {M.}~\bibnamefont {Giannotti}}, \bibinfo
  {author} {\bibfnamefont {F.}~\bibnamefont {Mescia}},\ and\ \bibinfo {author}
  {\bibfnamefont {E.}~\bibnamefont {Nardi}},\ }\bibfield  {title} {\bibinfo
  {title} {{Stellar evolution confronts axion models}},\ }\href
  {https://doi.org/10.1088/1475-7516/2022/02/035} {\bibfield  {journal}
  {\bibinfo  {journal} {JCAP}\ }\textbf {\bibinfo {volume} {02}},\ \bibinfo
  {pages} {035}},\ \Eprint {https://arxiv.org/abs/2109.10368} {arXiv:2109.10368
  [hep-ph]} \BibitemShut {NoStop}%
\bibitem [{\citenamefont {Miller~Bertolami}\ \emph {et~al.}(2014)\citenamefont
  {Miller~Bertolami}, \citenamefont {Melendez}, \citenamefont {Althaus},\ and\
  \citenamefont {Isern}}]{MillerBertolami}%
  \BibitemOpen
  \bibfield  {author} {\bibinfo {author} {\bibfnamefont {M.~M.}\ \bibnamefont
  {Miller~Bertolami}}, \bibinfo {author} {\bibfnamefont {B.~E.}\ \bibnamefont
  {Melendez}}, \bibinfo {author} {\bibfnamefont {L.~G.}\ \bibnamefont
  {Althaus}},\ and\ \bibinfo {author} {\bibfnamefont {J.}~\bibnamefont
  {Isern}},\ }\bibfield  {title} {\bibinfo {title} {{Revisiting the axion
  bounds from the Galactic white dwarf luminosity function}},\ }\href
  {https://doi.org/10.1088/1475-7516/2014/10/069} {\bibfield  {journal}
  {\bibinfo  {journal} {JCAP}\ }\textbf {\bibinfo {volume} {10}},\ \bibinfo
  {pages} {069}},\ \Eprint {https://arxiv.org/abs/1406.7712} {arXiv:1406.7712
  [hep-ph]} \BibitemShut {NoStop}%
\bibitem [{\citenamefont {Straniero}\ \emph {et~al.}(2020)\citenamefont
  {Straniero}, \citenamefont {Pallanca}, \citenamefont {Dalessandro},
  \citenamefont {Dominguez}, \citenamefont {Ferraro}, \citenamefont
  {Giannotti}, \citenamefont {Mirizzi},\ and\ \citenamefont
  {Piersanti}}]{Straniero:2020iyi}%
  \BibitemOpen
  \bibfield  {author} {\bibinfo {author} {\bibfnamefont {O.}~\bibnamefont
  {Straniero}}, \bibinfo {author} {\bibfnamefont {C.}~\bibnamefont {Pallanca}},
  \bibinfo {author} {\bibfnamefont {E.}~\bibnamefont {Dalessandro}}, \bibinfo
  {author} {\bibfnamefont {I.}~\bibnamefont {Dominguez}}, \bibinfo {author}
  {\bibfnamefont {F.~R.}\ \bibnamefont {Ferraro}}, \bibinfo {author}
  {\bibfnamefont {M.}~\bibnamefont {Giannotti}}, \bibinfo {author}
  {\bibfnamefont {A.}~\bibnamefont {Mirizzi}},\ and\ \bibinfo {author}
  {\bibfnamefont {L.}~\bibnamefont {Piersanti}},\ }\bibfield  {title} {\bibinfo
  {title} {{The RGB tip of galactic globular clusters and the revision of the
  axion-electron coupling bound}},\ }\href
  {https://doi.org/10.1051/0004-6361/202038775} {\bibfield  {journal} {\bibinfo
   {journal} {Astron. Astrophys.}\ }\textbf {\bibinfo {volume} {644}},\
  \bibinfo {pages} {A166} (\bibinfo {year} {2020})},\ \Eprint
  {https://arxiv.org/abs/2010.03833} {arXiv:2010.03833 [astro-ph.SR]}
  \BibitemShut {NoStop}%
\bibitem [{\citenamefont {Capozzi}\ and\ \citenamefont
  {Raffelt}(2020)}]{Capozzi2020}%
  \BibitemOpen
  \bibfield  {author} {\bibinfo {author} {\bibfnamefont {F.}~\bibnamefont
  {Capozzi}}\ and\ \bibinfo {author} {\bibfnamefont {G.}~\bibnamefont
  {Raffelt}},\ }\bibfield  {title} {\bibinfo {title} {{Axion and neutrino
  bounds improved with new calibrations of the tip of the red-giant branch
  using geometric distance determinations}},\ }\href
  {https://doi.org/10.1103/PhysRevD.102.083007} {\bibfield  {journal} {\bibinfo
   {journal} {Phys. Rev. D}\ }\textbf {\bibinfo {volume} {102}},\ \bibinfo
  {pages} {083007} (\bibinfo {year} {2020})},\ \Eprint
  {https://arxiv.org/abs/2007.03694} {arXiv:2007.03694 [astro-ph.SR]}
  \BibitemShut {NoStop}%
\bibitem [{\citenamefont {Giannotti}\ \emph {et~al.}(2017)\citenamefont
  {Giannotti}, \citenamefont {Irastorza}, \citenamefont {Redondo},
  \citenamefont {Ringwald},\ and\ \citenamefont {Saikawa}}]{Giannotti:2017hny}%
  \BibitemOpen
  \bibfield  {author} {\bibinfo {author} {\bibfnamefont {M.}~\bibnamefont
  {Giannotti}}, \bibinfo {author} {\bibfnamefont {I.~G.}\ \bibnamefont
  {Irastorza}}, \bibinfo {author} {\bibfnamefont {J.}~\bibnamefont {Redondo}},
  \bibinfo {author} {\bibfnamefont {A.}~\bibnamefont {Ringwald}},\ and\
  \bibinfo {author} {\bibfnamefont {K.}~\bibnamefont {Saikawa}},\ }\bibfield
  {title} {\bibinfo {title} {{Stellar Recipes for Axion Hunters}},\ }\href
  {https://doi.org/10.1088/1475-7516/2017/10/010} {\bibfield  {journal}
  {\bibinfo  {journal} {JCAP}\ }\textbf {\bibinfo {volume} {10}},\ \bibinfo
  {pages} {010}},\ \Eprint {https://arxiv.org/abs/1708.02111} {arXiv:1708.02111
  [hep-ph]} \BibitemShut {NoStop}%
\bibitem [{\citenamefont {Dolan}\ \emph {et~al.}(2022)\citenamefont {Dolan},
  \citenamefont {Hiskens},\ and\ \citenamefont {Volkas}}]{Dolan:2022kul}%
  \BibitemOpen
  \bibfield  {author} {\bibinfo {author} {\bibfnamefont {M.~J.}\ \bibnamefont
  {Dolan}}, \bibinfo {author} {\bibfnamefont {F.~J.}\ \bibnamefont {Hiskens}},\
  and\ \bibinfo {author} {\bibfnamefont {R.~R.}\ \bibnamefont {Volkas}},\
  }\bibfield  {title} {\bibinfo {title} {{Advancing globular cluster
  constraints on the axion-photon coupling}},\ }\href
  {https://doi.org/10.1088/1475-7516/2022/10/096} {\bibfield  {journal}
  {\bibinfo  {journal} {JCAP}\ }\textbf {\bibinfo {volume} {10}},\ \bibinfo
  {pages} {096}},\ \Eprint {https://arxiv.org/abs/2207.03102} {arXiv:2207.03102
  [hep-ph]} \BibitemShut {NoStop}%
\bibitem [{\citenamefont {Du}\ \emph {et~al.}(2018)\citenamefont {Du} \emph
  {et~al.}}]{ADMX:2018gho}%
  \BibitemOpen
  \bibfield  {author} {\bibinfo {author} {\bibfnamefont {N.}~\bibnamefont {Du}}
  \emph {et~al.} (\bibinfo {collaboration} {ADMX}),\ }\bibfield  {title}
  {\bibinfo {title} {{A Search for Invisible Axion Dark Matter with the Axion
  Dark Matter Experiment}},\ }\href
  {https://doi.org/10.1103/PhysRevLett.120.151301} {\bibfield  {journal}
  {\bibinfo  {journal} {Phys. Rev. Lett.}\ }\textbf {\bibinfo {volume} {120}},\
  \bibinfo {pages} {151301} (\bibinfo {year} {2018})},\ \Eprint
  {https://arxiv.org/abs/1804.05750} {arXiv:1804.05750 [hep-ex]} \BibitemShut
  {NoStop}%
\bibitem [{\citenamefont {Braine}\ \emph {et~al.}(2020)\citenamefont {Braine}
  \emph {et~al.}}]{ADMX:2019uok}%
  \BibitemOpen
  \bibfield  {author} {\bibinfo {author} {\bibfnamefont {T.}~\bibnamefont
  {Braine}} \emph {et~al.} (\bibinfo {collaboration} {ADMX}),\ }\bibfield
  {title} {\bibinfo {title} {{Extended Search for the Invisible Axion with the
  Axion Dark Matter Experiment}},\ }\href
  {https://doi.org/10.1103/PhysRevLett.124.101303} {\bibfield  {journal}
  {\bibinfo  {journal} {Phys. Rev. Lett.}\ }\textbf {\bibinfo {volume} {124}},\
  \bibinfo {pages} {101303} (\bibinfo {year} {2020})},\ \Eprint
  {https://arxiv.org/abs/1910.08638} {arXiv:1910.08638 [hep-ex]} \BibitemShut
  {NoStop}%
\bibitem [{\citenamefont {Bartram}\ \emph {et~al.}(2021)\citenamefont {Bartram}
  \emph {et~al.}}]{ADMX:2021nhd}%
  \BibitemOpen
  \bibfield  {author} {\bibinfo {author} {\bibfnamefont {C.}~\bibnamefont
  {Bartram}} \emph {et~al.} (\bibinfo {collaboration} {ADMX}),\ }\bibfield
  {title} {\bibinfo {title} {{Search for Invisible Axion Dark Matter in the
  3.3\textendash{}4.2\,\,\ensuremath{\mu}eV Mass Range}},\ }\href
  {https://doi.org/10.1103/PhysRevLett.127.261803} {\bibfield  {journal}
  {\bibinfo  {journal} {Phys. Rev. Lett.}\ }\textbf {\bibinfo {volume} {127}},\
  \bibinfo {pages} {261803} (\bibinfo {year} {2021})},\ \Eprint
  {https://arxiv.org/abs/2110.06096} {arXiv:2110.06096 [hep-ex]} \BibitemShut
  {NoStop}%
\bibitem [{\citenamefont {Caldwell}\ \emph {et~al.}(2017)\citenamefont
  {Caldwell}, \citenamefont {Dvali}, \citenamefont {Majorovits}, \citenamefont
  {Millar}, \citenamefont {Raffelt}, \citenamefont {Redondo}, \citenamefont
  {Reimann}, \citenamefont {Simon},\ and\ \citenamefont
  {Steffen}}]{Caldwell2017}%
  \BibitemOpen
  \bibfield  {author} {\bibinfo {author} {\bibfnamefont {A.}~\bibnamefont
  {Caldwell}}, \bibinfo {author} {\bibfnamefont {G.}~\bibnamefont {Dvali}},
  \bibinfo {author} {\bibfnamefont {B.}~\bibnamefont {Majorovits}}, \bibinfo
  {author} {\bibfnamefont {A.}~\bibnamefont {Millar}}, \bibinfo {author}
  {\bibfnamefont {G.}~\bibnamefont {Raffelt}}, \bibinfo {author} {\bibfnamefont
  {J.}~\bibnamefont {Redondo}}, \bibinfo {author} {\bibfnamefont
  {O.}~\bibnamefont {Reimann}}, \bibinfo {author} {\bibfnamefont
  {F.}~\bibnamefont {Simon}},\ and\ \bibinfo {author} {\bibfnamefont
  {F.}~\bibnamefont {Steffen}} (\bibinfo {collaboration} {MADMAX Working
  Group}),\ }\bibfield  {title} {\bibinfo {title} {Dielectric haloscopes: A new
  way to detect axion dark matter},\ }\href
  {https://doi.org/10.1103/PhysRevLett.118.091801} {\bibfield  {journal}
  {\bibinfo  {journal} {Phys. Rev. Lett.}\ }\textbf {\bibinfo {volume} {118}},\
  \bibinfo {pages} {091801} (\bibinfo {year} {2017})}\BibitemShut {NoStop}%
\bibitem [{\citenamefont {Semertzidis}\ \emph {et~al.}(2019)\citenamefont
  {Semertzidis} \emph {et~al.}}]{Semertzidis:2019gkj}%
  \BibitemOpen
  \bibfield  {author} {\bibinfo {author} {\bibfnamefont {Y.~K.}\ \bibnamefont
  {Semertzidis}} \emph {et~al.},\ }\bibfield  {title} {\bibinfo {title} {{Axion
  Dark Matter Research with IBS/CAPP}},\ }\href@noop {} {\  (\bibinfo {year}
  {2019})},\ \Eprint {https://arxiv.org/abs/1910.11591} {arXiv:1910.11591
  [physics.ins-det]} \BibitemShut {NoStop}%
\bibitem [{\citenamefont {Backes}\ \emph {et~al.}(2021)\citenamefont {Backes}
  \emph {et~al.}}]{HAYSTAC:2020kwv}%
  \BibitemOpen
  \bibfield  {author} {\bibinfo {author} {\bibfnamefont {K.~M.}\ \bibnamefont
  {Backes}} \emph {et~al.} (\bibinfo {collaboration} {HAYSTAC}),\ }\bibfield
  {title} {\bibinfo {title} {{A quantum-enhanced search for dark matter
  axions}},\ }\href {https://doi.org/10.1038/s41586-021-03226-7} {\bibfield
  {journal} {\bibinfo  {journal} {Nature}\ }\textbf {\bibinfo {volume} {590}},\
  \bibinfo {pages} {238} (\bibinfo {year} {2021})},\ \Eprint
  {https://arxiv.org/abs/2008.01853} {arXiv:2008.01853 [quant-ph]} \BibitemShut
  {NoStop}%
\bibitem [{\citenamefont {Lawson}\ \emph {et~al.}(2019)\citenamefont {Lawson},
  \citenamefont {Millar}, \citenamefont {Pancaldi}, \citenamefont
  {Vitagliano},\ and\ \citenamefont {Wilczek}}]{Lawson2019}%
  \BibitemOpen
  \bibfield  {author} {\bibinfo {author} {\bibfnamefont {M.}~\bibnamefont
  {Lawson}}, \bibinfo {author} {\bibfnamefont {A.~J.}\ \bibnamefont {Millar}},
  \bibinfo {author} {\bibfnamefont {M.}~\bibnamefont {Pancaldi}}, \bibinfo
  {author} {\bibfnamefont {E.}~\bibnamefont {Vitagliano}},\ and\ \bibinfo
  {author} {\bibfnamefont {F.}~\bibnamefont {Wilczek}},\ }\bibfield  {title}
  {\bibinfo {title} {Tunable axion plasma haloscopes},\ }\href
  {https://doi.org/10.1103/PhysRevLett.123.141802} {\bibfield  {journal}
  {\bibinfo  {journal} {Phys. Rev. Lett.}\ }\textbf {\bibinfo {volume} {123}},\
  \bibinfo {pages} {141802} (\bibinfo {year} {2019})}\BibitemShut {NoStop}%
\bibitem [{\citenamefont {Quiskamp}\ \emph {et~al.}(2022)\citenamefont
  {Quiskamp}, \citenamefont {McAllister}, \citenamefont {Altin}, \citenamefont
  {Ivanov}, \citenamefont {Goryachev},\ and\ \citenamefont
  {Tobar}}]{Quiskamp:2022pks}%
  \BibitemOpen
  \bibfield  {author} {\bibinfo {author} {\bibfnamefont {A.~P.}\ \bibnamefont
  {Quiskamp}}, \bibinfo {author} {\bibfnamefont {B.~T.}\ \bibnamefont
  {McAllister}}, \bibinfo {author} {\bibfnamefont {P.}~\bibnamefont {Altin}},
  \bibinfo {author} {\bibfnamefont {E.~N.}\ \bibnamefont {Ivanov}}, \bibinfo
  {author} {\bibfnamefont {M.}~\bibnamefont {Goryachev}},\ and\ \bibinfo
  {author} {\bibfnamefont {M.~E.}\ \bibnamefont {Tobar}},\ }\bibfield  {title}
  {\bibinfo {title} {{Direct search for dark matter axions excluding ALP
  cogenesis in the 63- to 67-\ensuremath{\mu}eV range with the ORGAN
  experiment}},\ }\href {https://doi.org/10.1126/sciadv.abq3765} {\bibfield
  {journal} {\bibinfo  {journal} {Sci. Adv.}\ }\textbf {\bibinfo {volume}
  {8}},\ \bibinfo {pages} {abq3765} (\bibinfo {year} {2022})},\ \Eprint
  {https://arxiv.org/abs/2203.12152} {arXiv:2203.12152 [hep-ex]} \BibitemShut
  {NoStop}%
\bibitem [{\citenamefont {Armengaud}\ \emph {et~al.}(2019)\citenamefont
  {Armengaud} \emph {et~al.}}]{IAXO:2019mpb}%
  \BibitemOpen
  \bibfield  {author} {\bibinfo {author} {\bibfnamefont {E.}~\bibnamefont
  {Armengaud}} \emph {et~al.} (\bibinfo {collaboration} {IAXO}),\ }\bibfield
  {title} {\bibinfo {title} {{Physics potential of the International Axion
  Observatory (IAXO)}},\ }\href {https://doi.org/10.1088/1475-7516/2019/06/047}
  {\bibfield  {journal} {\bibinfo  {journal} {JCAP}\ }\textbf {\bibinfo
  {volume} {06}},\ \bibinfo {pages} {047}},\ \Eprint
  {https://arxiv.org/abs/1904.09155} {arXiv:1904.09155 [hep-ph]} \BibitemShut
  {NoStop}%
\bibitem [{\citenamefont {Boucenna}\ and\ \citenamefont
  {Shafi}(2018)}]{Boucenna:2017fna}%
  \BibitemOpen
  \bibfield  {author} {\bibinfo {author} {\bibfnamefont {S.~M.}\ \bibnamefont
  {Boucenna}}\ and\ \bibinfo {author} {\bibfnamefont {Q.}~\bibnamefont
  {Shafi}},\ }\bibfield  {title} {\bibinfo {title} {{Axion inflation, proton
  decay, and leptogenesis in $SU(5)\times U(1)_{PQ}$}},\ }\href
  {https://doi.org/10.1103/PhysRevD.97.075012} {\bibfield  {journal} {\bibinfo
  {journal} {Phys. Rev. D}\ }\textbf {\bibinfo {volume} {97}},\ \bibinfo
  {pages} {075012} (\bibinfo {year} {2018})},\ \Eprint
  {https://arxiv.org/abs/1712.06526} {arXiv:1712.06526 [hep-ph]} \BibitemShut
  {NoStop}%
\bibitem [{\citenamefont {Zyla}\ \emph {et~al.}(2020)\citenamefont {Zyla} \emph
  {et~al.}}]{ParticleDataGroup:2020ssz}%
  \BibitemOpen
  \bibfield  {author} {\bibinfo {author} {\bibfnamefont {P.~A.}\ \bibnamefont
  {Zyla}} \emph {et~al.} (\bibinfo {collaboration} {Particle Data Group}),\
  }\bibfield  {title} {\bibinfo {title} {{Review of Particle Physics}},\ }\href
  {https://doi.org/10.1093/ptep/ptaa104} {\bibfield  {journal} {\bibinfo
  {journal} {PTEP}\ }\textbf {\bibinfo {volume} {2020}},\ \bibinfo {pages}
  {083C01} (\bibinfo {year} {2020})}\BibitemShut {NoStop}%
\bibitem [{\citenamefont {'t~Hooft}(1980)}]{tHooft:1979rat}%
  \BibitemOpen
  \bibfield  {author} {\bibinfo {author} {\bibfnamefont {G.}~\bibnamefont
  {'t~Hooft}},\ }\bibfield  {title} {\bibinfo {title} {{Naturalness, chiral
  symmetry, and spontaneous chiral symmetry breaking}},\ }\href
  {https://doi.org/10.1007/978-1-4684-7571-5_9} {\bibfield  {journal} {\bibinfo
   {journal} {NATO Sci. Ser. B}\ }\textbf {\bibinfo {volume} {59}},\ \bibinfo
  {pages} {135} (\bibinfo {year} {1980})}\BibitemShut {NoStop}%
\bibitem [{\citenamefont {Buchmuller}\ \emph {et~al.}(2002)\citenamefont
  {Buchmuller}, \citenamefont {Di~Bari},\ and\ \citenamefont
  {Plumacher}}]{Buchmuller:2002rq}%
  \BibitemOpen
  \bibfield  {author} {\bibinfo {author} {\bibfnamefont {W.}~\bibnamefont
  {Buchmuller}}, \bibinfo {author} {\bibfnamefont {P.}~\bibnamefont
  {Di~Bari}},\ and\ \bibinfo {author} {\bibfnamefont {M.}~\bibnamefont
  {Plumacher}},\ }\bibfield  {title} {\bibinfo {title} {{Cosmic microwave
  background, matter - antimatter asymmetry and neutrino masses}},\ }\href
  {https://doi.org/10.1016/S0550-3213(02)00737-X} {\bibfield  {journal}
  {\bibinfo  {journal} {Nucl. Phys. B}\ }\textbf {\bibinfo {volume} {643}},\
  \bibinfo {pages} {367} (\bibinfo {year} {2002})},\ \bibinfo {note} {[Erratum:
  Nucl.Phys.B 793, 362 (2008)]},\ \Eprint
  {https://arxiv.org/abs/hep-ph/0205349} {arXiv:hep-ph/0205349} \BibitemShut
  {NoStop}%
\bibitem [{\citenamefont {Pilaftsis}\ and\ \citenamefont
  {Underwood}(2004)}]{Pilaftsis:2003gt}%
  \BibitemOpen
  \bibfield  {author} {\bibinfo {author} {\bibfnamefont {A.}~\bibnamefont
  {Pilaftsis}}\ and\ \bibinfo {author} {\bibfnamefont {T.~E.~J.}\ \bibnamefont
  {Underwood}},\ }\bibfield  {title} {\bibinfo {title} {{Resonant
  leptogenesis}},\ }\href {https://doi.org/10.1016/j.nuclphysb.2004.05.029}
  {\bibfield  {journal} {\bibinfo  {journal} {Nucl. Phys. B}\ }\textbf
  {\bibinfo {volume} {692}},\ \bibinfo {pages} {303} (\bibinfo {year}
  {2004})},\ \Eprint {https://arxiv.org/abs/hep-ph/0309342}
  {arXiv:hep-ph/0309342} \BibitemShut {NoStop}%
\bibitem [{\citenamefont {Coleman}\ and\ \citenamefont
  {Weinberg}(1973)}]{Coleman:1973jx}%
  \BibitemOpen
  \bibfield  {author} {\bibinfo {author} {\bibfnamefont {S.~R.}\ \bibnamefont
  {Coleman}}\ and\ \bibinfo {author} {\bibfnamefont {E.~J.}\ \bibnamefont
  {Weinberg}},\ }\bibfield  {title} {\bibinfo {title} {{Radiative Corrections
  as the Origin of Spontaneous Symmetry Breaking}},\ }\href
  {https://doi.org/10.1103/PhysRevD.7.1888} {\bibfield  {journal} {\bibinfo
  {journal} {Phys. Rev. D}\ }\textbf {\bibinfo {volume} {7}},\ \bibinfo {pages}
  {1888} (\bibinfo {year} {1973})}\BibitemShut {NoStop}%
\bibitem [{\citenamefont {Degrassi}\ \emph {et~al.}(2012)\citenamefont
  {Degrassi}, \citenamefont {Di~Vita}, \citenamefont {Elias-Miro},
  \citenamefont {Espinosa}, \citenamefont {Giudice}, \citenamefont {Isidori},\
  and\ \citenamefont {Strumia}}]{Degrassi:2012ry}%
  \BibitemOpen
  \bibfield  {author} {\bibinfo {author} {\bibfnamefont {G.}~\bibnamefont
  {Degrassi}}, \bibinfo {author} {\bibfnamefont {S.}~\bibnamefont {Di~Vita}},
  \bibinfo {author} {\bibfnamefont {J.}~\bibnamefont {Elias-Miro}}, \bibinfo
  {author} {\bibfnamefont {J.~R.}\ \bibnamefont {Espinosa}}, \bibinfo {author}
  {\bibfnamefont {G.~F.}\ \bibnamefont {Giudice}}, \bibinfo {author}
  {\bibfnamefont {G.}~\bibnamefont {Isidori}},\ and\ \bibinfo {author}
  {\bibfnamefont {A.}~\bibnamefont {Strumia}},\ }\bibfield  {title} {\bibinfo
  {title} {{Higgs mass and vacuum stability in the Standard Model at NNLO}},\
  }\href {https://doi.org/10.1007/JHEP08(2012)098} {\bibfield  {journal}
  {\bibinfo  {journal} {JHEP}\ }\textbf {\bibinfo {volume} {08}},\ \bibinfo
  {pages} {098}},\ \Eprint {https://arxiv.org/abs/1205.6497} {arXiv:1205.6497
  [hep-ph]} \BibitemShut {NoStop}%
\bibitem [{\citenamefont {Delle~Rose}\ \emph {et~al.}(2016)\citenamefont
  {Delle~Rose}, \citenamefont {Marzo},\ and\ \citenamefont
  {Urbano}}]{DelleRose:2015bpo}%
  \BibitemOpen
  \bibfield  {author} {\bibinfo {author} {\bibfnamefont {L.}~\bibnamefont
  {Delle~Rose}}, \bibinfo {author} {\bibfnamefont {C.}~\bibnamefont {Marzo}},\
  and\ \bibinfo {author} {\bibfnamefont {A.}~\bibnamefont {Urbano}},\
  }\bibfield  {title} {\bibinfo {title} {{On the fate of the Standard Model at
  finite temperature}},\ }\href {https://doi.org/10.1007/JHEP05(2016)050}
  {\bibfield  {journal} {\bibinfo  {journal} {JHEP}\ }\textbf {\bibinfo
  {volume} {05}},\ \bibinfo {pages} {050}},\ \Eprint
  {https://arxiv.org/abs/1507.06912} {arXiv:1507.06912 [hep-ph]} \BibitemShut
  {NoStop}%
\bibitem [{\citenamefont {Lebedev}(2012)}]{Lebedev:2012zw}%
  \BibitemOpen
  \bibfield  {author} {\bibinfo {author} {\bibfnamefont {O.}~\bibnamefont
  {Lebedev}},\ }\bibfield  {title} {\bibinfo {title} {{On Stability of the
  Electroweak Vacuum and the Higgs Portal}},\ }\href
  {https://doi.org/10.1140/epjc/s10052-012-2058-2} {\bibfield  {journal}
  {\bibinfo  {journal} {Eur. Phys. J. C}\ }\textbf {\bibinfo {volume} {72}},\
  \bibinfo {pages} {2058} (\bibinfo {year} {2012})},\ \Eprint
  {https://arxiv.org/abs/1203.0156} {arXiv:1203.0156 [hep-ph]} \BibitemShut
  {NoStop}%
\bibitem [{\citenamefont {Elias-Miro}\ \emph {et~al.}(2012)\citenamefont
  {Elias-Miro}, \citenamefont {Espinosa}, \citenamefont {Giudice},
  \citenamefont {Lee},\ and\ \citenamefont {Strumia}}]{Elias-Miro:2012eoi}%
  \BibitemOpen
  \bibfield  {author} {\bibinfo {author} {\bibfnamefont {J.}~\bibnamefont
  {Elias-Miro}}, \bibinfo {author} {\bibfnamefont {J.~R.}\ \bibnamefont
  {Espinosa}}, \bibinfo {author} {\bibfnamefont {G.~F.}\ \bibnamefont
  {Giudice}}, \bibinfo {author} {\bibfnamefont {H.~M.}\ \bibnamefont {Lee}},\
  and\ \bibinfo {author} {\bibfnamefont {A.}~\bibnamefont {Strumia}},\
  }\bibfield  {title} {\bibinfo {title} {{Stabilization of the Electroweak
  Vacuum by a Scalar Threshold Effect}},\ }\href
  {https://doi.org/10.1007/JHEP06(2012)031} {\bibfield  {journal} {\bibinfo
  {journal} {JHEP}\ }\textbf {\bibinfo {volume} {06}},\ \bibinfo {pages}
  {031}},\ \Eprint {https://arxiv.org/abs/1203.0237} {arXiv:1203.0237 [hep-ph]}
  \BibitemShut {NoStop}%
\bibitem [{\citenamefont {Chakrabarty}\ and\ \citenamefont
  {Mukhopadhyaya}(2017{\natexlab{a}})}]{Chakrabarty:2016smc}%
  \BibitemOpen
  \bibfield  {author} {\bibinfo {author} {\bibfnamefont {N.}~\bibnamefont
  {Chakrabarty}}\ and\ \bibinfo {author} {\bibfnamefont {B.}~\bibnamefont
  {Mukhopadhyaya}},\ }\bibfield  {title} {\bibinfo {title} {{High-scale
  validity of a two Higgs doublet scenario: metastability included}},\ }\href
  {https://doi.org/10.1140/epjc/s10052-017-4705-0} {\bibfield  {journal}
  {\bibinfo  {journal} {Eur. Phys. J. C}\ }\textbf {\bibinfo {volume} {77}},\
  \bibinfo {pages} {153} (\bibinfo {year} {2017}{\natexlab{a}})},\ \Eprint
  {https://arxiv.org/abs/1603.05883} {arXiv:1603.05883 [hep-ph]} \BibitemShut
  {NoStop}%
\bibitem [{\citenamefont {Chakrabarty}\ and\ \citenamefont
  {Mukhopadhyaya}(2017{\natexlab{b}})}]{Chakrabarty:2017qkh}%
  \BibitemOpen
  \bibfield  {author} {\bibinfo {author} {\bibfnamefont {N.}~\bibnamefont
  {Chakrabarty}}\ and\ \bibinfo {author} {\bibfnamefont {B.}~\bibnamefont
  {Mukhopadhyaya}},\ }\bibfield  {title} {\bibinfo {title} {{High-scale
  validity of a two Higgs doublet scenario: predicting collider signals}},\
  }\href {https://doi.org/10.1103/PhysRevD.96.035028} {\bibfield  {journal}
  {\bibinfo  {journal} {Phys. Rev. D}\ }\textbf {\bibinfo {volume} {96}},\
  \bibinfo {pages} {035028} (\bibinfo {year} {2017}{\natexlab{b}})},\ \Eprint
  {https://arxiv.org/abs/1702.08268} {arXiv:1702.08268 [hep-ph]} \BibitemShut
  {NoStop}%
\bibitem [{\citenamefont {Branchina}\ \emph {et~al.}(2018)\citenamefont
  {Branchina}, \citenamefont {Contino},\ and\ \citenamefont
  {Ferreira}}]{Branchina:2018qlf}%
  \BibitemOpen
  \bibfield  {author} {\bibinfo {author} {\bibfnamefont {V.}~\bibnamefont
  {Branchina}}, \bibinfo {author} {\bibfnamefont {F.}~\bibnamefont {Contino}},\
  and\ \bibinfo {author} {\bibfnamefont {P.~M.}\ \bibnamefont {Ferreira}},\
  }\bibfield  {title} {\bibinfo {title} {{Electroweak vacuum lifetime in two
  Higgs doublet models}},\ }\href {https://doi.org/10.1007/JHEP11(2018)107}
  {\bibfield  {journal} {\bibinfo  {journal} {JHEP}\ }\textbf {\bibinfo
  {volume} {11}},\ \bibinfo {pages} {107}},\ \Eprint
  {https://arxiv.org/abs/1807.10802} {arXiv:1807.10802 [hep-ph]} \BibitemShut
  {NoStop}%
\bibitem [{\citenamefont {Krauss}\ \emph {et~al.}(2018)\citenamefont {Krauss},
  \citenamefont {Opferkuch},\ and\ \citenamefont {Staub}}]{Krauss:2018thf}%
  \BibitemOpen
  \bibfield  {author} {\bibinfo {author} {\bibfnamefont {M.~E.}\ \bibnamefont
  {Krauss}}, \bibinfo {author} {\bibfnamefont {T.}~\bibnamefont {Opferkuch}},\
  and\ \bibinfo {author} {\bibfnamefont {F.}~\bibnamefont {Staub}},\ }\bibfield
   {title} {\bibinfo {title} {{The Ultraviolet Landscape of Two-Higgs Doublet
  Models}},\ }\href {https://doi.org/10.1140/epjc/s10052-018-6489-2} {\bibfield
   {journal} {\bibinfo  {journal} {Eur. Phys. J. C}\ }\textbf {\bibinfo
  {volume} {78}},\ \bibinfo {pages} {1020} (\bibinfo {year} {2018})},\ \Eprint
  {https://arxiv.org/abs/1807.07581} {arXiv:1807.07581 [hep-ph]} \BibitemShut
  {NoStop}%
\bibitem [{\citenamefont {Basler}\ \emph {et~al.}(2018)\citenamefont {Basler},
  \citenamefont {Ferreira}, \citenamefont {M\"uhlleitner},\ and\ \citenamefont
  {Santos}}]{Basler:2017nzu}%
  \BibitemOpen
  \bibfield  {author} {\bibinfo {author} {\bibfnamefont {P.}~\bibnamefont
  {Basler}}, \bibinfo {author} {\bibfnamefont {P.~M.}\ \bibnamefont
  {Ferreira}}, \bibinfo {author} {\bibfnamefont {M.}~\bibnamefont
  {M\"uhlleitner}},\ and\ \bibinfo {author} {\bibfnamefont {R.}~\bibnamefont
  {Santos}},\ }\bibfield  {title} {\bibinfo {title} {{High scale impact in
  alignment and decoupling in two-Higgs doublet models}},\ }\href
  {https://doi.org/10.1103/PhysRevD.97.095024} {\bibfield  {journal} {\bibinfo
  {journal} {Phys. Rev. D}\ }\textbf {\bibinfo {volume} {97}},\ \bibinfo
  {pages} {095024} (\bibinfo {year} {2018})},\ \Eprint
  {https://arxiv.org/abs/1710.10410} {arXiv:1710.10410 [hep-ph]} \BibitemShut
  {NoStop}%
\bibitem [{\citenamefont {Barbon}\ and\ \citenamefont
  {Espinosa}(2009)}]{Barbon:2009ya}%
  \BibitemOpen
  \bibfield  {author} {\bibinfo {author} {\bibfnamefont {J.~L.~F.}\
  \bibnamefont {Barbon}}\ and\ \bibinfo {author} {\bibfnamefont {J.~R.}\
  \bibnamefont {Espinosa}},\ }\bibfield  {title} {\bibinfo {title} {{On the
  Naturalness of Higgs Inflation}},\ }\href
  {https://doi.org/10.1103/PhysRevD.79.081302} {\bibfield  {journal} {\bibinfo
  {journal} {Phys. Rev. D}\ }\textbf {\bibinfo {volume} {79}},\ \bibinfo
  {pages} {081302} (\bibinfo {year} {2009})},\ \Eprint
  {https://arxiv.org/abs/0903.0355} {arXiv:0903.0355 [hep-ph]} \BibitemShut
  {NoStop}%
\bibitem [{\citenamefont {Burgess}\ \emph {et~al.}(2009)\citenamefont
  {Burgess}, \citenamefont {Lee},\ and\ \citenamefont
  {Trott}}]{Burgess:2009ea}%
  \BibitemOpen
  \bibfield  {author} {\bibinfo {author} {\bibfnamefont {C.~P.}\ \bibnamefont
  {Burgess}}, \bibinfo {author} {\bibfnamefont {H.~M.}\ \bibnamefont {Lee}},\
  and\ \bibinfo {author} {\bibfnamefont {M.}~\bibnamefont {Trott}},\ }\bibfield
   {title} {\bibinfo {title} {{Power-counting and the Validity of the Classical
  Approximation During Inflation}},\ }\href
  {https://doi.org/10.1088/1126-6708/2009/09/103} {\bibfield  {journal}
  {\bibinfo  {journal} {JHEP}\ }\textbf {\bibinfo {volume} {09}},\ \bibinfo
  {pages} {103}},\ \Eprint {https://arxiv.org/abs/0902.4465} {arXiv:0902.4465
  [hep-ph]} \BibitemShut {NoStop}%
\bibitem [{\citenamefont {Bezrukov}(2013)}]{Bezrukov_2013}%
  \BibitemOpen
  \bibfield  {author} {\bibinfo {author} {\bibfnamefont {F.}~\bibnamefont
  {Bezrukov}},\ }\bibfield  {title} {\bibinfo {title} {The higgs field as an
  inflaton},\ }\href {https://doi.org/10.1088/0264-9381/30/21/214001}
  {\bibfield  {journal} {\bibinfo  {journal} {Classical and Quantum Gravity}\
  }\textbf {\bibinfo {volume} {30}},\ \bibinfo {pages} {214001} (\bibinfo
  {year} {2013})}\BibitemShut {NoStop}%
\bibitem [{\citenamefont {Bezrukov}\ \emph {et~al.}(2011)\citenamefont
  {Bezrukov}, \citenamefont {Magnin}, \citenamefont {Shaposhnikov},\ and\
  \citenamefont {Sibiryakov}}]{Bezrukov:2010jz}%
  \BibitemOpen
  \bibfield  {author} {\bibinfo {author} {\bibfnamefont {F.}~\bibnamefont
  {Bezrukov}}, \bibinfo {author} {\bibfnamefont {A.}~\bibnamefont {Magnin}},
  \bibinfo {author} {\bibfnamefont {M.}~\bibnamefont {Shaposhnikov}},\ and\
  \bibinfo {author} {\bibfnamefont {S.}~\bibnamefont {Sibiryakov}},\ }\bibfield
   {title} {\bibinfo {title} {{Higgs inflation: consistency and
  generalisations}},\ }\href {https://doi.org/10.1007/JHEP01(2011)016}
  {\bibfield  {journal} {\bibinfo  {journal} {JHEP}\ }\textbf {\bibinfo
  {volume} {01}},\ \bibinfo {pages} {016}},\ \Eprint
  {https://arxiv.org/abs/1008.5157} {arXiv:1008.5157 [hep-ph]} \BibitemShut
  {NoStop}%
\bibitem [{\citenamefont {Ema}\ \emph {et~al.}(2017)\citenamefont {Ema},
  \citenamefont {Jinno}, \citenamefont {Mukaida},\ and\ \citenamefont
  {Nakayama}}]{Ema_2017}%
  \BibitemOpen
  \bibfield  {author} {\bibinfo {author} {\bibfnamefont {Y.}~\bibnamefont
  {Ema}}, \bibinfo {author} {\bibfnamefont {R.}~\bibnamefont {Jinno}}, \bibinfo
  {author} {\bibfnamefont {K.}~\bibnamefont {Mukaida}},\ and\ \bibinfo {author}
  {\bibfnamefont {K.}~\bibnamefont {Nakayama}},\ }\bibfield  {title} {\bibinfo
  {title} {Violent preheating in inflation with nonminimal coupling},\ }\href
  {https://doi.org/10.1088/1475-7516/2017/02/045} {\bibfield  {journal}
  {\bibinfo  {journal} {Journal of Cosmology and Astroparticle Physics}\
  }\textbf {\bibinfo {volume} {2017}}\bibinfo  {number} { (02)},\ \bibinfo
  {pages} {045}}\BibitemShut {NoStop}%
\bibitem [{\citenamefont {DeCross}\ \emph
  {et~al.}(2018{\natexlab{c}})\citenamefont {DeCross}, \citenamefont {Kaiser},
  \citenamefont {Prabhu}, \citenamefont {Prescod-Weinstein},\ and\
  \citenamefont {Sfakianakis}}]{DeCross:2016cbs}%
  \BibitemOpen
\bibfield  {number} {  }\bibfield  {author} {\bibinfo {author} {\bibfnamefont
  {M.~P.}\ \bibnamefont {DeCross}}, \bibinfo {author} {\bibfnamefont {D.~I.}\
  \bibnamefont {Kaiser}}, \bibinfo {author} {\bibfnamefont {A.}~\bibnamefont
  {Prabhu}}, \bibinfo {author} {\bibfnamefont {C.}~\bibnamefont
  {Prescod-Weinstein}},\ and\ \bibinfo {author} {\bibfnamefont {E.~I.}\
  \bibnamefont {Sfakianakis}},\ }\bibfield  {title} {\bibinfo {title}
  {{Preheating after multifield inflation with nonminimal couplings, III:
  Dynamical spacetime results}},\ }\href
  {https://doi.org/10.1103/PhysRevD.97.023528} {\bibfield  {journal} {\bibinfo
  {journal} {Phys. Rev. D}\ }\textbf {\bibinfo {volume} {97}},\ \bibinfo
  {pages} {023528} (\bibinfo {year} {2018}{\natexlab{c}})},\ \Eprint
  {https://arxiv.org/abs/1610.08916} {arXiv:1610.08916 [astro-ph.CO]}
  \BibitemShut {NoStop}%
\bibitem [{\citenamefont {Sfakianakis}\ and\ \citenamefont {van~de
  Vis}(2019)}]{Sfakianakis2019}%
  \BibitemOpen
  \bibfield  {author} {\bibinfo {author} {\bibfnamefont {E.~I.}\ \bibnamefont
  {Sfakianakis}}\ and\ \bibinfo {author} {\bibfnamefont {J.}~\bibnamefont
  {van~de Vis}},\ }\bibfield  {title} {\bibinfo {title} {Preheating after higgs
  inflation: Self-resonance and gauge boson production},\ }\href
  {https://doi.org/10.1103/PhysRevD.99.083519} {\bibfield  {journal} {\bibinfo
  {journal} {Phys. Rev. D}\ }\textbf {\bibinfo {volume} {99}},\ \bibinfo
  {pages} {083519} (\bibinfo {year} {2019})}\BibitemShut {NoStop}%
\bibitem [{\citenamefont {Hamada}\ \emph {et~al.}(2021)\citenamefont {Hamada},
  \citenamefont {Kawana},\ and\ \citenamefont {Scherlis}}]{Hamada:2020kuy}%
  \BibitemOpen
  \bibfield  {author} {\bibinfo {author} {\bibfnamefont {Y.}~\bibnamefont
  {Hamada}}, \bibinfo {author} {\bibfnamefont {K.}~\bibnamefont {Kawana}},\
  and\ \bibinfo {author} {\bibfnamefont {A.}~\bibnamefont {Scherlis}},\
  }\bibfield  {title} {\bibinfo {title} {{On Preheating in Higgs Inflation}},\
  }\href {https://doi.org/10.1088/1475-7516/2021/03/062} {\bibfield  {journal}
  {\bibinfo  {journal} {JCAP}\ }\textbf {\bibinfo {volume} {03}},\ \bibinfo
  {pages} {062}},\ \Eprint {https://arxiv.org/abs/2007.04701} {arXiv:2007.04701
  [hep-ph]} \BibitemShut {NoStop}%
\bibitem [{\citenamefont {Ema}(2017)}]{EMA2017403}%
  \BibitemOpen
  \bibfield  {author} {\bibinfo {author} {\bibfnamefont {Y.}~\bibnamefont
  {Ema}},\ }\bibfield  {title} {\bibinfo {title} {Higgs scalaron mixed
  inflation},\ }\href
  {https://doi.org/https://doi.org/10.1016/j.physletb.2017.04.060} {\bibfield
  {journal} {\bibinfo  {journal} {Physics Letters B}\ }\textbf {\bibinfo
  {volume} {770}},\ \bibinfo {pages} {403} (\bibinfo {year}
  {2017})}\BibitemShut {NoStop}%
\bibitem [{\citenamefont {Rubio}\ and\ \citenamefont
  {Tomberg}(2019)}]{Rubio:2019ypq}%
  \BibitemOpen
  \bibfield  {author} {\bibinfo {author} {\bibfnamefont {J.}~\bibnamefont
  {Rubio}}\ and\ \bibinfo {author} {\bibfnamefont {E.~S.}\ \bibnamefont
  {Tomberg}},\ }\bibfield  {title} {\bibinfo {title} {{Preheating in Palatini
  Higgs inflation}},\ }\href {https://doi.org/10.1088/1475-7516/2019/04/021}
  {\bibfield  {journal} {\bibinfo  {journal} {JCAP}\ }\textbf {\bibinfo
  {volume} {04}},\ \bibinfo {pages} {021}},\ \Eprint
  {https://arxiv.org/abs/1902.10148} {arXiv:1902.10148 [hep-ph]} \BibitemShut
  {NoStop}%
\bibitem [{\citenamefont {Lebedev}\ and\ \citenamefont
  {Lee}(2011)}]{Lebedev:2011aq}%
  \BibitemOpen
  \bibfield  {author} {\bibinfo {author} {\bibfnamefont {O.}~\bibnamefont
  {Lebedev}}\ and\ \bibinfo {author} {\bibfnamefont {H.~M.}\ \bibnamefont
  {Lee}},\ }\bibfield  {title} {\bibinfo {title} {{Higgs Portal Inflation}},\
  }\href {https://doi.org/10.1140/epjc/s10052-011-1821-0} {\bibfield  {journal}
  {\bibinfo  {journal} {Eur. Phys. J. C}\ }\textbf {\bibinfo {volume} {71}},\
  \bibinfo {pages} {1821} (\bibinfo {year} {2011})},\ \Eprint
  {https://arxiv.org/abs/1105.2284} {arXiv:1105.2284 [hep-ph]} \BibitemShut
  {NoStop}%
\bibitem [{\citenamefont {Achucarro}\ \emph
  {et~al.}(2011{\natexlab{a}})\citenamefont {Achucarro}, \citenamefont {Gong},
  \citenamefont {Hardeman}, \citenamefont {Palma},\ and\ \citenamefont
  {Patil}}]{Achucarro:20101}%
  \BibitemOpen
  \bibfield  {author} {\bibinfo {author} {\bibfnamefont {A.}~\bibnamefont
  {Achucarro}}, \bibinfo {author} {\bibfnamefont {J.-O.}\ \bibnamefont {Gong}},
  \bibinfo {author} {\bibfnamefont {S.}~\bibnamefont {Hardeman}}, \bibinfo
  {author} {\bibfnamefont {G.~A.}\ \bibnamefont {Palma}},\ and\ \bibinfo
  {author} {\bibfnamefont {S.~P.}\ \bibnamefont {Patil}},\ }\bibfield  {title}
  {\bibinfo {title} {{Mass hierarchies and non-decoupling in multi-scalar field
  dynamics}},\ }\href {https://doi.org/10.1103/PhysRevD.84.043502} {\bibfield
  {journal} {\bibinfo  {journal} {Phys. Rev. D}\ }\textbf {\bibinfo {volume}
  {84}},\ \bibinfo {pages} {043502} (\bibinfo {year} {2011}{\natexlab{a}})},\
  \Eprint {https://arxiv.org/abs/1005.3848} {arXiv:1005.3848 [hep-th]}
  \BibitemShut {NoStop}%
\bibitem [{\citenamefont {Achucarro}\ \emph
  {et~al.}(2011{\natexlab{b}})\citenamefont {Achucarro}, \citenamefont {Gong},
  \citenamefont {Hardeman}, \citenamefont {Palma},\ and\ \citenamefont
  {Patil}}]{Achucarro:20102}%
  \BibitemOpen
  \bibfield  {author} {\bibinfo {author} {\bibfnamefont {A.}~\bibnamefont
  {Achucarro}}, \bibinfo {author} {\bibfnamefont {J.-O.}\ \bibnamefont {Gong}},
  \bibinfo {author} {\bibfnamefont {S.}~\bibnamefont {Hardeman}}, \bibinfo
  {author} {\bibfnamefont {G.~A.}\ \bibnamefont {Palma}},\ and\ \bibinfo
  {author} {\bibfnamefont {S.~P.}\ \bibnamefont {Patil}},\ }\bibfield  {title}
  {\bibinfo {title} {{Features of heavy physics in the CMB power spectrum}},\
  }\href {https://doi.org/10.1088/1475-7516/2011/01/030} {\bibfield  {journal}
  {\bibinfo  {journal} {JCAP}\ }\textbf {\bibinfo {volume} {01}},\ \bibinfo
  {pages} {030}},\ \Eprint {https://arxiv.org/abs/1010.3693} {arXiv:1010.3693
  [hep-ph]} \BibitemShut {NoStop}%
\bibitem [{\citenamefont {Cespedes}\ \emph {et~al.}(2012)\citenamefont
  {Cespedes}, \citenamefont {Atal},\ and\ \citenamefont
  {Palma}}]{Cespedes:2012hu}%
  \BibitemOpen
  \bibfield  {author} {\bibinfo {author} {\bibfnamefont {S.}~\bibnamefont
  {Cespedes}}, \bibinfo {author} {\bibfnamefont {V.}~\bibnamefont {Atal}},\
  and\ \bibinfo {author} {\bibfnamefont {G.~A.}\ \bibnamefont {Palma}},\
  }\bibfield  {title} {\bibinfo {title} {{On the importance of heavy fields
  during inflation}},\ }\href {https://doi.org/10.1088/1475-7516/2012/05/008}
  {\bibfield  {journal} {\bibinfo  {journal} {JCAP}\ }\textbf {\bibinfo
  {volume} {05}},\ \bibinfo {pages} {008}},\ \Eprint
  {https://arxiv.org/abs/1201.4848} {arXiv:1201.4848 [hep-th]} \BibitemShut
  {NoStop}%
\bibitem [{\citenamefont {Achucarro}\ \emph {et~al.}(2012)\citenamefont
  {Achucarro}, \citenamefont {Gong}, \citenamefont {Hardeman}, \citenamefont
  {Palma},\ and\ \citenamefont {Patil}}]{Achucarro:20123}%
  \BibitemOpen
  \bibfield  {author} {\bibinfo {author} {\bibfnamefont {A.}~\bibnamefont
  {Achucarro}}, \bibinfo {author} {\bibfnamefont {J.-O.}\ \bibnamefont {Gong}},
  \bibinfo {author} {\bibfnamefont {S.}~\bibnamefont {Hardeman}}, \bibinfo
  {author} {\bibfnamefont {G.~A.}\ \bibnamefont {Palma}},\ and\ \bibinfo
  {author} {\bibfnamefont {S.~P.}\ \bibnamefont {Patil}},\ }\bibfield  {title}
  {\bibinfo {title} {{Effective theories of single field inflation when heavy
  fields matter}},\ }\href {https://doi.org/10.1007/JHEP05(2012)066} {\bibfield
   {journal} {\bibinfo  {journal} {JHEP}\ }\textbf {\bibinfo {volume} {05}},\
  \bibinfo {pages} {066}},\ \Eprint {https://arxiv.org/abs/1201.6342}
  {arXiv:1201.6342 [hep-th]} \BibitemShut {NoStop}%
\bibitem [{\citenamefont {Sartore}\ and\ \citenamefont
  {Schienbein}(2021)}]{Sartore:2020gou}%
  \BibitemOpen
  \bibfield  {author} {\bibinfo {author} {\bibfnamefont {L.}~\bibnamefont
  {Sartore}}\ and\ \bibinfo {author} {\bibfnamefont {I.}~\bibnamefont
  {Schienbein}},\ }\bibfield  {title} {\bibinfo {title} {{PyR@TE 3}},\ }\href
  {https://doi.org/10.1016/j.cpc.2020.107819} {\bibfield  {journal} {\bibinfo
  {journal} {Comput. Phys. Commun.}\ }\textbf {\bibinfo {volume} {261}},\
  \bibinfo {pages} {107819} (\bibinfo {year} {2021})},\ \Eprint
  {https://arxiv.org/abs/2007.12700} {arXiv:2007.12700 [hep-ph]} \BibitemShut
  {NoStop}%
\bibitem [{\citenamefont {{Di Luzio}}\ \emph {et~al.}(2020)\citenamefont {{Di
  Luzio}}, \citenamefont {Giannotti}, \citenamefont {Nardi},\ and\
  \citenamefont {Visinelli}}]{DILUZIO20201}%
  \BibitemOpen
  \bibfield  {author} {\bibinfo {author} {\bibfnamefont {L.}~\bibnamefont {{Di
  Luzio}}}, \bibinfo {author} {\bibfnamefont {M.}~\bibnamefont {Giannotti}},
  \bibinfo {author} {\bibfnamefont {E.}~\bibnamefont {Nardi}},\ and\ \bibinfo
  {author} {\bibfnamefont {L.}~\bibnamefont {Visinelli}},\ }\bibfield  {title}
  {\bibinfo {title} {The landscape of {QCD} axion models},\ }\href
  {https://doi.org/https://doi.org/10.1016/j.physrep.2020.06.002} {\bibfield
  {journal} {\bibinfo  {journal} {Physics Reports}\ }\textbf {\bibinfo {volume}
  {870}},\ \bibinfo {pages} {1} (\bibinfo {year} {2020})},\ \bibinfo {note}
  {the landscape of QCD axion models}\BibitemShut {NoStop}%
\bibitem [{\citenamefont {Sun}\ and\ \citenamefont
  {He}(2020)}]{DFSZaxioncouplings}%
  \BibitemOpen
  \bibfield  {author} {\bibinfo {author} {\bibfnamefont {J.}~\bibnamefont
  {Sun}}\ and\ \bibinfo {author} {\bibfnamefont {X.-G.}\ \bibnamefont {He}},\
  }\bibfield  {title} {\bibinfo {title} {{DFSZ} axion couplings revisited},\
  }\href {https://doi.org/https://doi.org/10.1016/j.physletb.2020.135881}
  {\bibfield  {journal} {\bibinfo  {journal} {Physics Letters B}\ }\textbf
  {\bibinfo {volume} {811}},\ \bibinfo {pages} {135881} (\bibinfo {year}
  {2020})}\BibitemShut {NoStop}%
\end{thebibliography}%

\end{document}